\documentclass[letterpaper,11pt]{article}

\usepackage{jheppub}

\usepackage{jheppub}
\usepackage{subfig}
\usepackage{xspace}
\usepackage{graphicx}

\usepackage{color}

\newcommand{\fastjet}{\textsc{FastJet}}
\newcommand{\fjcontrib}{\textsc{FastJet Contrib}}

\newcommand{\abs}[1]{\lvert#1\rvert}

\newcommand{\Tau}{\mathcal{T}}

\newcommand{\GeV}{\text{GeV}}
\newcommand{\TeV}{\text{TeV}}

\allowdisplaybreaks[4]

\DeclareRobustCommand{\Sec}[1]{section~\ref{#1}}
\DeclareRobustCommand{\Secs}[2]{sections~\ref{#1} and \ref{#2}}

\DeclareRobustCommand{\App}[1]{appendix~\ref{#1}}

\DeclareRobustCommand{\Tabs}[2]{tables~\ref{#1} and \ref{#2}}
\DeclareRobustCommand{\Fig}[1]{figure~\ref{#1}}
\DeclareRobustCommand{\Figs}[2]{figures~\ref{#1} and \ref{#2}}

\DeclareRobustCommand{\Eq}[1]{eq.~(\ref{#1})}

\DeclareRobustCommand{\Ref}[1]{ref.~\cite{#1}}
\DeclareRobustCommand{\Refs}[1]{refs.~\cite{#1}}

\newcommand{\be}{\begin{equation}}
\newcommand{\ee}{\end{equation}}

\newcommand{\jet}{{\text{jet}}}
\newcommand{\beam}{{\text{beam}}}

\newcommand{\Rnaught}{R}

\preprint{ 
\begin{flushright}
MIT--CTP 4675
 \end{flushright}}
 
\title{Resolving Boosted Jets with XCone}

\author{Jesse~Thaler}
\author{and Thomas~F.~Wilkason}

\affiliation{Center for Theoretical Physics, Massachusetts Institute of Technology, Cambridge, MA 02139, USA}

\emailAdd{jthaler@mit.edu}
\emailAdd{tjwilk@mit.edu}

\abstract{We show how the recently proposed XCone jet algorithm \cite{Stewart:2015waa} smoothly interpolates between resolved and boosted kinematics.  When using standard jet algorithms to reconstruct the decays of hadronic resonances like top quarks and Higgs bosons, one typically needs separate analysis strategies to handle the resolved regime of well-separated jets and the boosted regime of fat jets with substructure.  XCone, by contrast, is an exclusive cone jet algorithm that always returns a fixed number of jets, so jet regions remain resolved even when (sub)jets are overlapping in the boosted regime.  In this paper, we perform three LHC case studies---dijet resonances, Higgs decays to bottom quarks, and all-hadronic top pairs---that demonstrate the physics applications of XCone over a wide kinematic range.
}

\begin{document}

\maketitle

\section{Introduction}
\label{sec:intro}

The field of jet substructure has matured significantly over the past five years \cite{Abdesselam:2010pt,Altheimer:2012mn,Altheimer:2013yza,Adams:2015hiv}, with a variety of techniques in active use at the Large Hadron Collider (LHC) to tag boosted hadronic objects.  A prototypical example is the search for heavy new resonances that decay to pairs of top quarks \cite{Chatrchyan:2013lca,Aad:2012raa}; in the all-hadronic channel, each top quark obtains a large Lorentz boost from the heavy resonance decay, yielding a collimated fat jet with 3-prong substructure.  Since not all top quarks are produced in the boosted regime, one should also perform analyses in the resolved regime where the top decay products are well-separated and identified as individual jets.  Since new (and old) physics could show up at any energy scale, it is important to develop robust techniques to handle jets in both resolved and boosted kinematics.

In a companion paper, we introduced a new jet algorithm called XCone that blurs the boundary between resolved and boosted kinematics \cite{Stewart:2015waa}.  The name XCone refers to the fact that it is an \emph{exclusive cone} jet algorithm.  Like the exclusive $k_T$ algorithm \cite{Catani:1993hr}, XCone always returns a fixed number of jets $N$.  When jets are well-separated, XCone yields nearly conical jet regions with radius $\Rnaught$.  When jets are overlapping, XCone dynamically splits the jet regions into nearest neighbor partitions.  Thus, XCone smoothly interpolates between isolated conical jets and merged jets with substructure, making it ideally suited for studying the boosted and quasi-boosted regimes.

In this paper, we present three applications of XCone which are relevant for LHC physics in and beyond the standard model.  In \Sec{sec:dijet}, we study high-mass dijet resonances with isolated final state jets, showing that XCone has nearly identical performance to the popular anti-$k_T$ algorithm \cite{Cacciari:2008gp}.  In \Sec{sec:higgs}, we study associated Higgs boson production, showing that XCone can resolve $H \to b \bar{b}$ decays, even when the $R_{b \bar{b}}$ angle is less than the radius parameter $\Rnaught$, in contrast to anti-$k_T$.  In \Sec{sec:tops}, we study the classic example of boosted top quarks, showing how XCone can simultaneously identify jets and subjets in a high multiplicity final state, achieving higher signal efficiency than a traditional fat jet strategy.   These three case studies highlight the versatility of the XCone jet algorithm across a wide kinematic range and motivate the use of XCone as a viable alternative to anti-$k_T$.

It is worth noting that there have been other attempts to merge the resolved and boosted regimes into a single analysis, such as \Ref{Gouzevitch:2013qca} which combines different event topologies into a single search.  Cone algorithms like SISCone \cite{Salam:2007xv} have an overlap parameter that can be adjusted to achieve some of the desired jet splitting needed to resolve substructure.  More recently, the ``mass jump'' algorithm was introduced to avoid merging separated hard prongs \cite{Stoll:2014hsa,Hamaguchi:2015uqa}.  A key novelty of the XCone approach is that no explicit distinction is made between jets and subjets in the initial jet finding algorithm.  XCone can only partially replace a dedicated substructure analysis, especially since it is well-known that a fixed radius $\Rnaught$ no longer performs well in the hyper-boosted regime (see, e.g.\ \cite{Larkoski:2015yqa,Spannowsky:2015eba}).  We suspect that it will be advantageous to combine XCone jet finding with other jet substructure techniques, though we do not pursue that possibility in the present work. 

Note that exclusive clustering has long been part of the jet physics toolbox, though mainly in the context of sequential recombination algorithms.  Indeed, for reasons of computational efficiency, XCone uses $k_T$-style clustering internally as part of its jet finding procedure \cite{Stewart:2015waa}.  As shown in \App{app:exclusive}, exclusive $k_T$ clustering  \cite{Catani:1993hr} (with an $\Rnaught$ parameter) also successfully interpolates between the resolved and boosted regimes.  The key difference is that $k_T$-style jets have irregular boundaries and non-uniform active jet areas \cite{Cacciari:2007fd,Cacciari:2008gn}, while XCone jets are conical and uniform.  This turns out to give XCone a performance advantage over exclusive $k_T$, yielding better mass resolution for boosted Higgs bosons and top quarks.     

Before beginning our case studies, we briefly review the XCone jet algorithm \cite{Stewart:2015waa}.  XCone is based on minimizing the event shape $N$-jettiness \cite{Stewart:2010tn} using a measure inspired by the jet shape $N$-subjettiness \cite{Thaler:2010tr,Thaler:2011gf}.  A generic definition of $N$-jettiness is
\be
\label{eq:tauNdef}
\widetilde{\Tau}_N = \sum_i \min\left\{\rho_{\jet}(p_i,n_1), \ldots, \rho_{\jet}(p_i, n_N), \rho_{\beam}(p_i) \right\},
\ee
where $n_A = \{1, \hat{n}_A\}$ are $N$ light-like axes and $p_i$ are the particles in the event.  Based on a jet measure $\rho_\jet(p_i,n_A)$ and a beam measure $\rho_{\beam}(p_i)$, the minimum inside of $\widetilde{\Tau}_N$ partitions the event into $N$ jet regions and one unclustered beam region.  Ideally, one would find the global minimum of $\widetilde{\Tau}_N$ over all possible axes $n_A$,
\be
\label{eq:mincriteria}
\Tau_N = \min_{n_1, n_2, \ldots, n_N} \widetilde{\Tau}_N, 
\ee
though in practice, one uses iterative procedures to find a local $\Tau_N$ minimum starting from infrared and collinear safe seed axes.  A variety of $N$-jettiness measures have been proposed in the literature (most especially in \Refs{Jouttenus:2013hs,Thaler:2011gf}, see also \cite{Kim:2010uj,Thaler:2015uja}), but here we stick exclusively to the XCone recommended measure, namely the conical geometric measure with $\gamma = 1$ \cite{Stewart:2015waa}:
\be
\label{eq:dotproductmeasure}
\begin{aligned}
\rho_{\jet}(p_i,n_A) &= p_{T i} \left(\frac{2 n_A \cdot p_i}{n_{TA} \, p_{Ti}}\frac{1}{\Rnaught^2}\right)^{\beta/2} \approx p_{Ti} \left(\frac{R_{iA}}{\Rnaught} \right)^\beta,\\
\rho_{\beam}(p_i) &= p_{T i} .
\end{aligned}
\ee
As recommended in \Ref{Stewart:2015waa}, we consider two default values for the parameter $\beta$.  The XCone default is $\beta = 2$, which (approximately) aligns the jet axis with the jet momentum, as with standard cone algorithms \cite{Ellis:2001aa}.  A recoil-free default option is provided by $\beta = 1$, where the jet axis aligns with the hardest cluster in a jet \cite{Thaler:2011gf,Larkoski:2014uqa}, providing enhanced robustness against jet contamination \cite{Larkoski:2014bia}.  As described in \Ref{Stewart:2015waa}, we use a generalized $k_T$ clustering algorithm to define seed axes for one-pass $\Tau_N$ minimization.

\section{Dijet Resonances and Comparison to Anti-$k_T$}
\label{sec:dijet}

For our first case study, we compare the performance of XCone to anti-$k_T$ in the resolved regime of well-separated jets.  Inclusive jet algorithms like anti-$k_T$ identify a variable number of jets above some $p_T$ threshold, which is useful for classifying events into different jet multiplicity bins.  Exclusive jet algorithms like XCone always return a fixed number of jets $N$, which is useful if the number of desired jets is known in advance.  For widely separated \emph{cone} jets, however, the distinction between inclusive and exclusive cone jet algorithms is rather mild, since for typical $\Rnaught$ values, an exclusive cone jet algorithm will just return the $N$ hardest jets from an inclusive cone jet algorithm.  Since anti-$k_{T}$ acts like an idealized cone algorithm for well-separated jets \cite{Cacciari:2008gp}, XCone jets should be quite similar to the hardest $N$ anti-$k_T$ jets.  When we study overlapping jets in \Secs{sec:higgs}{sec:tops}, the inclusive/exclusive distinction will become much more important.

\begin{figure}
\centering
\subfloat[]{
\includegraphics[width = 0.45\columnwidth]{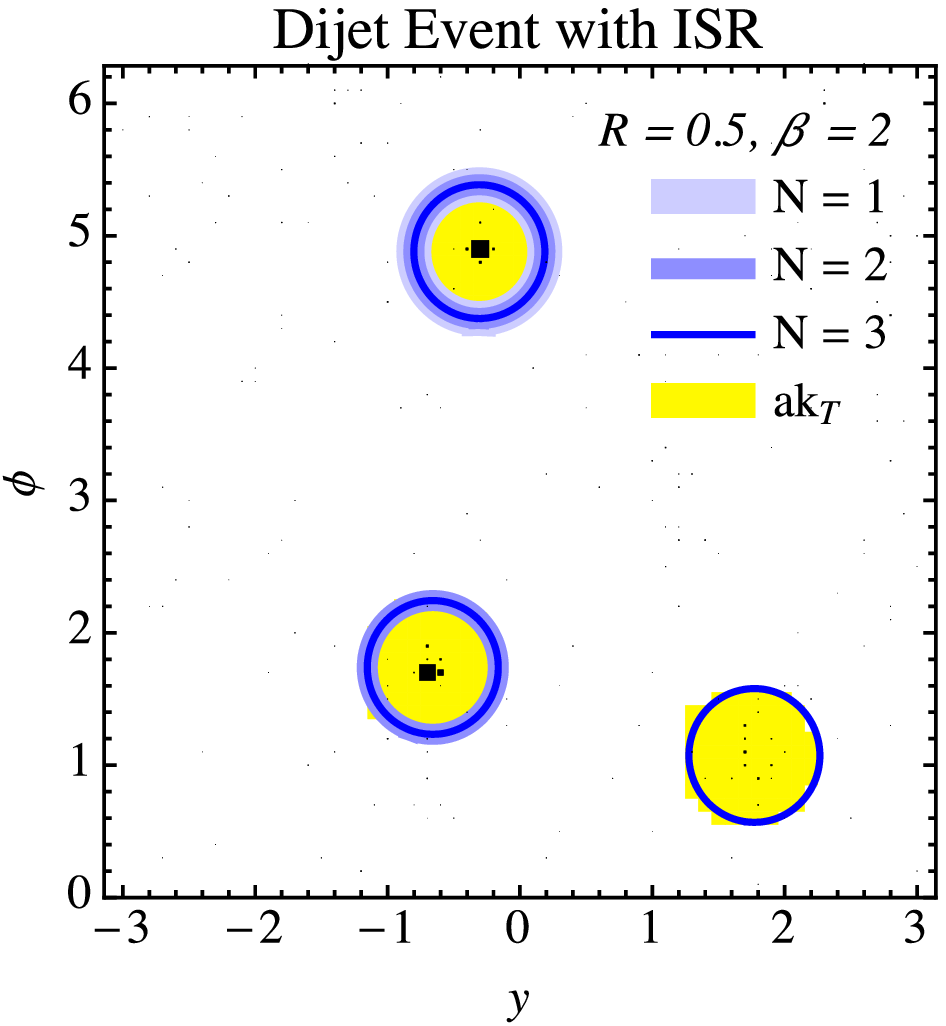}
\label{fig:dijet_display:a}}
$\qquad$
\subfloat[]{
\includegraphics[width = 0.45\columnwidth]{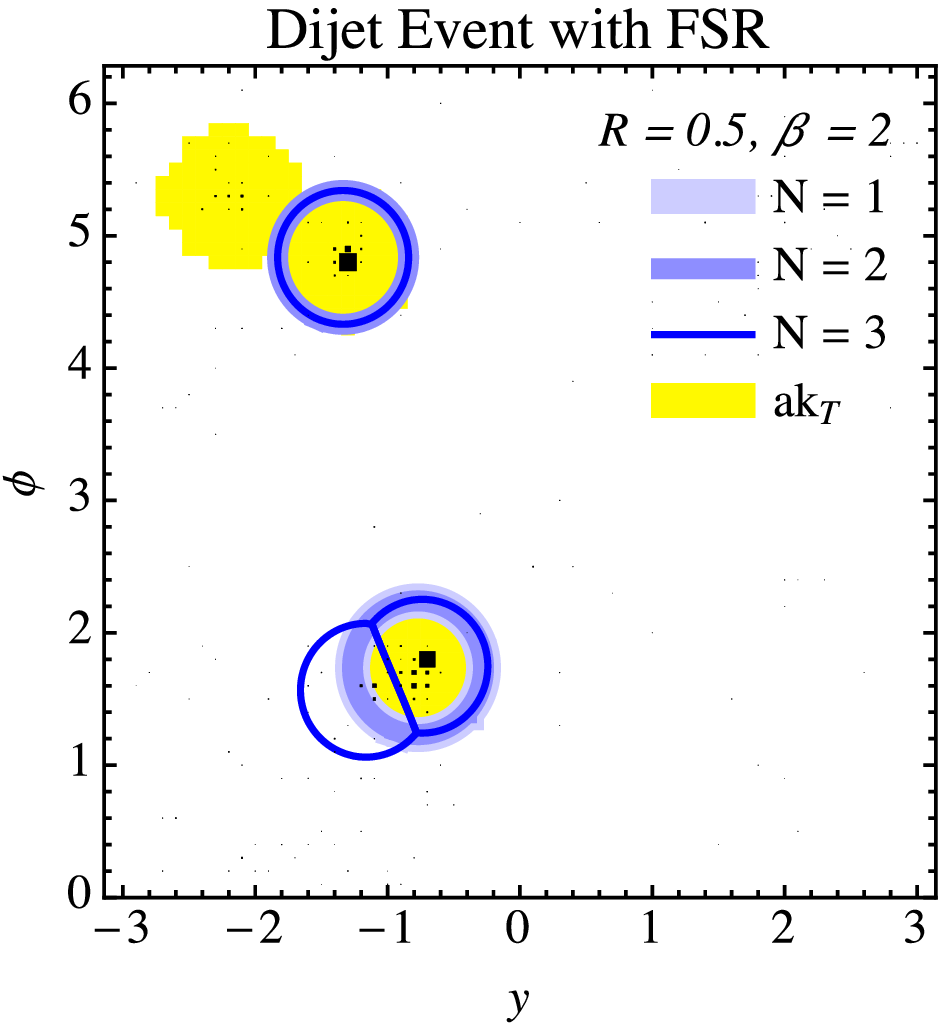}
\label{fig:dijet_display:b}}
\caption{Example XCone jet regions found with $N = 1$, $2$, and $3$.  These are dijet resonance events, using $\Rnaught = 0.5$ and $\beta = 2$.  (a) For widely separated jets, XCone and anti-$k_T$ identify nearly identical jet regions, including an additional jet from ISR when $N = 3$.  (b) XCone is able to identify jet substructure from FSR, even if the jet regions are closer than $\approx \Rnaught$.}
\label{fig:dijet_display}
\end{figure}

A good setting to study the resolved regime is a heavy resonance decay to dijets, where the two resulting jets are back-to-back and isolated.   Here, we consider the scenario 
\be
p p \to Z' \to q \bar{q},
\ee
where $Z'$ is a heavy boson with mass $m_{Z'}$ and $q$ is a $u$, $d$, or $s$ quark.   We start with $N = 1$ and show that XCone typically matches the hardest anti-$k_T$ jet, up to an expected two-fold ambiguity when the jets are nearly degenerate in $p_T$.  Going to $N = 2$, both XCone and anti-$k_T$ can successfully reconstruct the dijet resonance peak.  Even at $N=3$, the found jets are quite similar for typical choices of jet parameters, though XCone will identify final state jets with substructure.  Overall, XCone has essentially identical performance to anti-$k_{T}$ in this basic jet reconstruction scenario for $N = 2$, but can exhibit different behavior for $N = 3$ depending on the event topology. Example XCone jet regions are shown in \Fig{fig:dijet_display}.

In the following study, we use Pythia 8.176 \cite{Sjostrand:2006za,Sjostrand:2007gs} to simulate $Z'$ events at the $\sqrt{s} = 14$ TeV LHC.  We take $m_{Z'} = 1$ TeV and assume equal couplings to the three light quarks.   All of the final-state particles (except neutrinos) with $\abs{\eta} < 3.0$ are considered for analysis.  Anti-$k_T$ jets are found using \fastjet~3.1.2 \cite{Cacciari:2011ma} with standard $E$-scheme recombination.  XCone jets are found using \textsc{Nsubjettiness}~2.2.0 as part of \fjcontrib\ \cite{fjcontrib}, using the XCone default measure with $\beta = 2$ and $\beta = 1$.   For all algorithms, the jet radius parameter is $\Rnaught = 0.5$.

\subsection{N = 1 for Hardest Jet}

\begin{figure}
\centering
\subfloat[]{
\includegraphics[width =0.32\columnwidth]{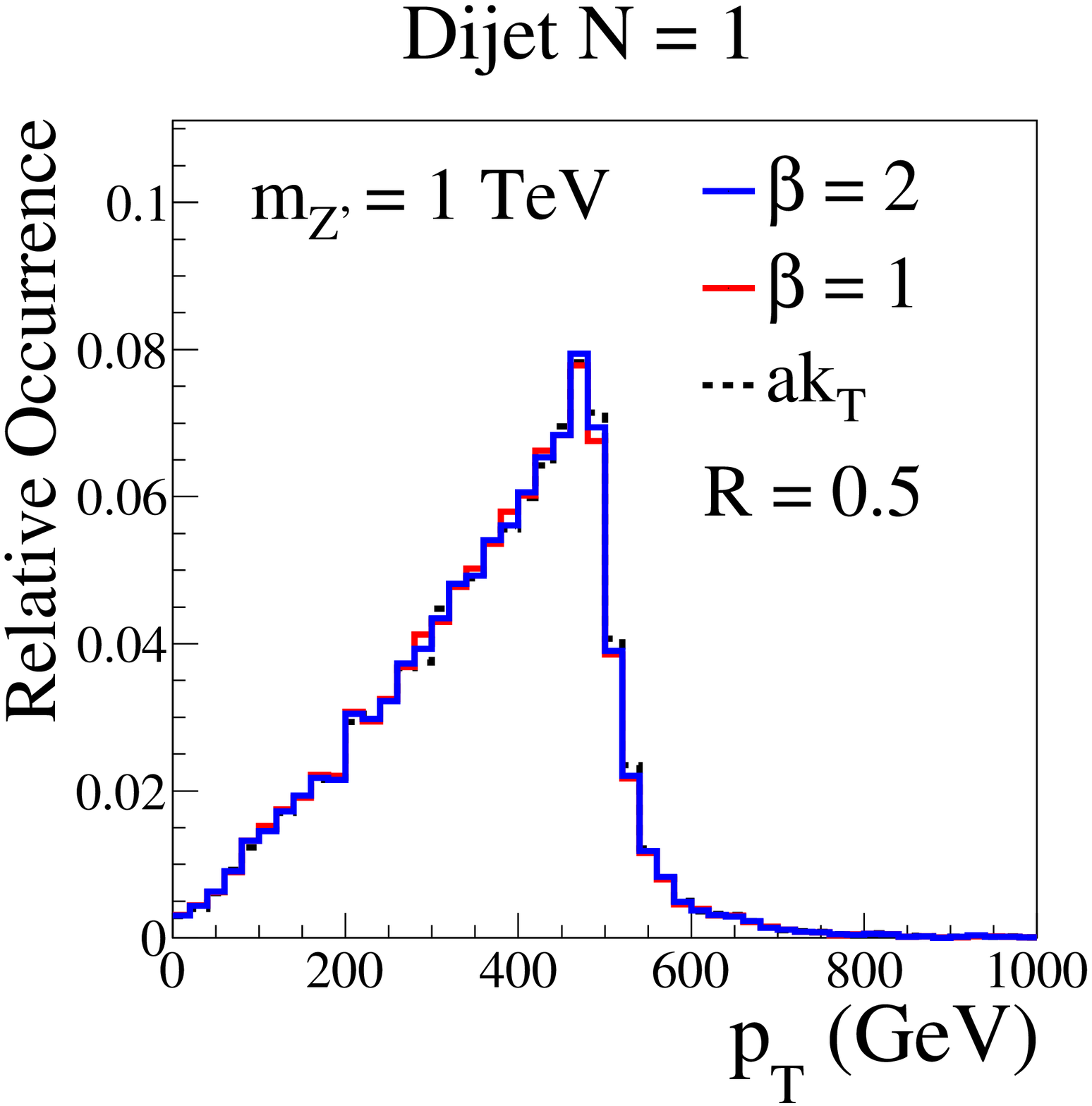}
\label{fig:dijet_1jet_perp}}
\subfloat[]{
\includegraphics[width = 0.32\columnwidth]{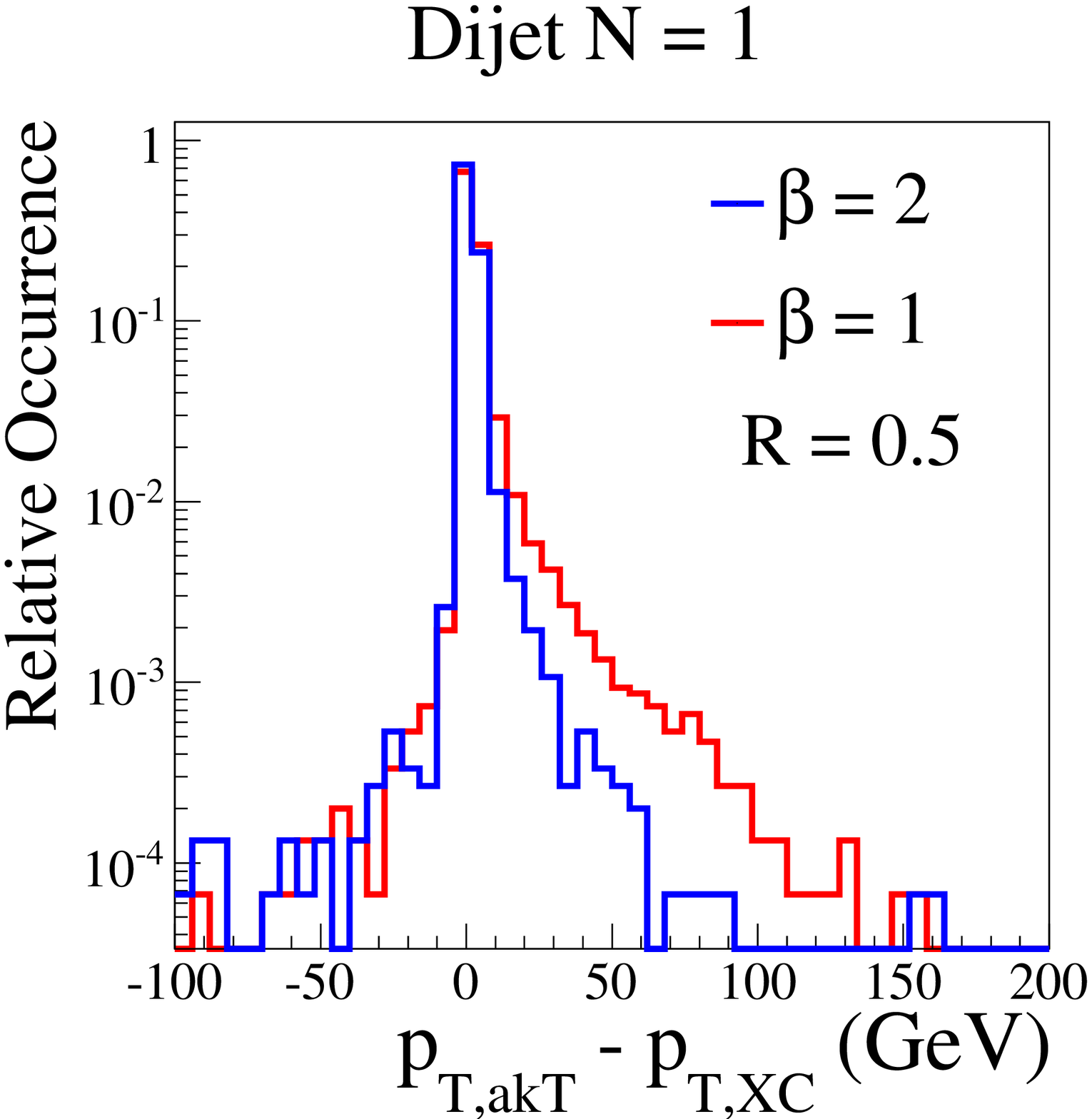}
\label{fig:dijet_1jet_perpdiff}}
\subfloat[]{
\includegraphics[width = 0.32\columnwidth]{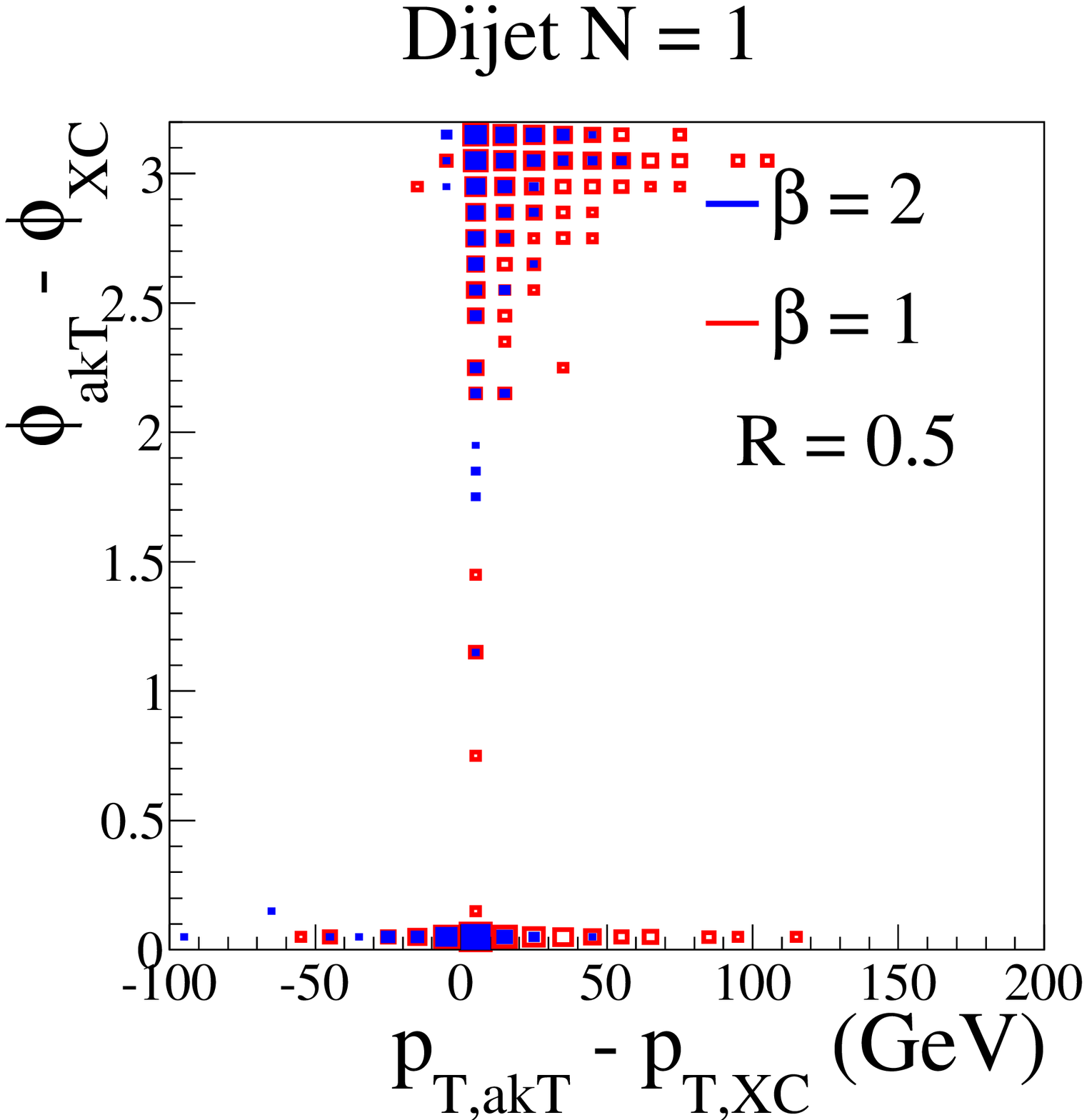}
\label{fig:dijet_1jet_perp_phidiff}}
\caption{Single jet kinematics of $N = 1$ XCone versus the hardest anti-$k_{T}$ jet, measured on the dijet resonance sample.  Shown are both the XCone default ($\beta = 2$) as well as the recoil-free variant ($\beta = 1$).   (a) Single jet $p_{T}$ spectrum. (b) Jet $p_T$ difference between XCone and anti-$k_{T}$ jet, showing that anti-$k_{T}$ jets are slightly harder on average.  (c)  Jet $p_T$ difference versus azimuth difference, showing the expected two-fold $\phi \leftrightarrow \phi + \pi$ ambiguity for dijets of comparable $p_T$.  Here and in \Fig{fig:dijet_3jet_perp_phidiff} below, the sizes of the boxes scale logarithmically with the number of entries, with solid blue boxes for $\beta = 2$ and empty red boxes for $\beta = 1$.}
\label{fig:singlejetkinematics}
\end{figure}

For $N = 1$, the XCone jet will tend to align with the hardest anti-$k_T$ jet in the event.  The reason is that the XCone measure in \Eq{eq:dotproductmeasure} penalizes unclustered $p_T$ by design.  In \Fig{fig:dijet_1jet_perp}, we see that anti-$k_T$ and XCone yield nearly identical single jet $p_T$ spectra, with the expected structure at $m_{Z'} / 2$ from a dijet resonance decay.  As shown in \Fig{fig:dijet_1jet_perp_phidiff}, there is a two-fold $\phi$ ambiguity in the found jets, as expected from dijet events where both jets have similar $p_T$ values.  Note that the box sizes are logarithmic in bin counts, and in the majority of cases, XCone and anti-$k_T$ find very similar jet regions.

\begin{figure}
\centering
\subfloat[]{
\includegraphics[width = 0.32\columnwidth]{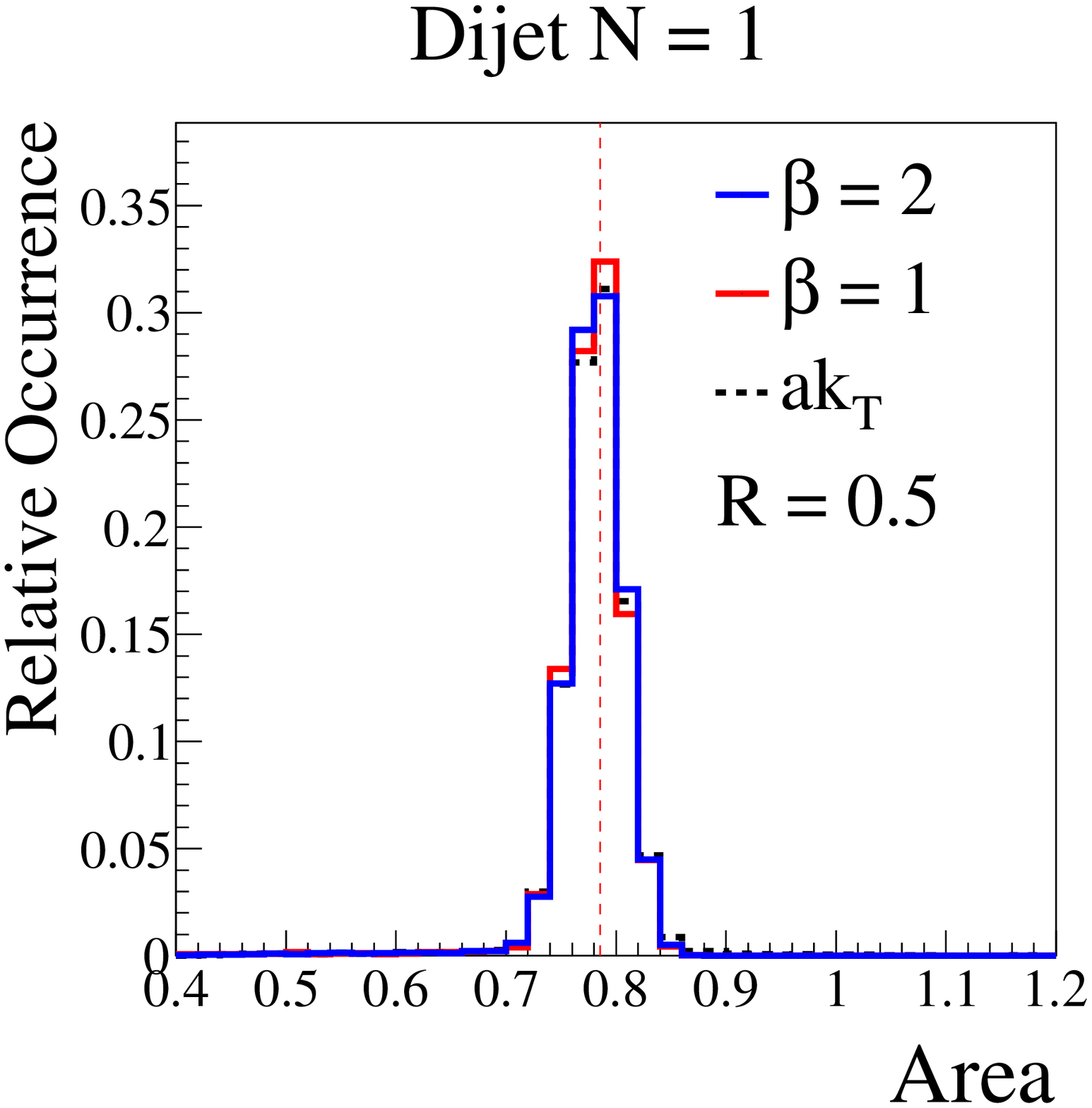}
\label{fig:dijet_1jet_area}}
$\qquad$
\subfloat[]{
\includegraphics[width = 0.32\columnwidth]{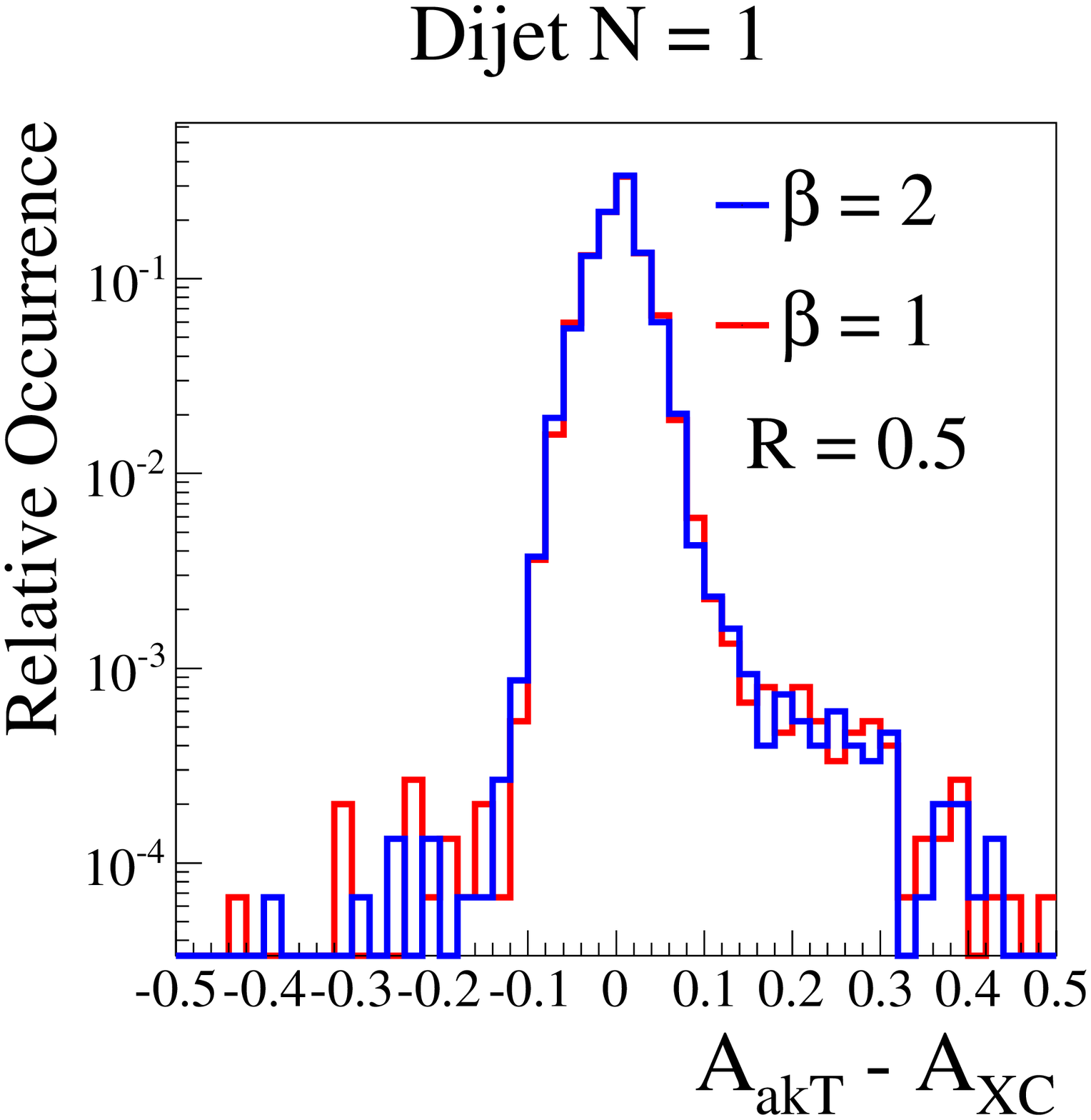}
\label{fig:dijet_1jet_areadiff}}
\caption{Same $N = 1$ comparison as \Fig{fig:singlejetkinematics}.  (a)  Single jet areas, showing the expected peak at $\pi \Rnaught^2$.  (b)  Jet area difference, showing that anti-$k_{T}$ jets occasionally have a higher area, explaining the $p_T$ asymmetry seen in \Fig{fig:dijet_1jet_perpdiff}.}
\label{fig:dijet_1jet_compare}
\end{figure}

In \Fig{fig:dijet_1jet_perpdiff}, we compare the found jet $p_T$ on an event-by-event basis, and find a sharp peak at $\Delta p_{T} = p_T^{\text{anti-$k_T$}} -  p_T^{\text{XCone}} = 0$.  On a logarithmic scale, one can see a small tail extending to $\mathcal{O}(50~\GeV)$ for $\beta = 2$, with larger deviations possible in the $\beta = 1$ case. It is interesting that the $\Delta p_{T}$ distribution is not symmetric, such that anti-$k_{T}$ jets tend to have a larger $p_{T}$ than XCone jets.  In \Fig{fig:dijet_1jet_area}, we plot the active jet area \cite{Cacciari:2007fd,Cacciari:2008gn},\footnote{We use the built-in \fastjet\ area determination routines using active ghosts.  For the general conical measure introduced in \Ref{Stewart:2015waa}, the active jet area can be determined analytically, though this is not as straightforward for the conical geometric measure used here.} which is quite similar between the two algorithms and peaked at the expected value of $\pi \Rnaught^2$.  On an event-by-event basis, though, there is a population of anti-$k_T$ jets that are systematically larger than XCone jets, as shown in \Fig{fig:dijet_1jet_areadiff}.  This occurs because anti-$k_T$ clustering can yield jets that extend beyond the conical boundary \cite{Cacciari:2008gp}.  The $\Delta p_{T}$ asymmetry then arises because these slightly bigger anti-$k_{T}$ jets contain more particles.

Comparing the performance of different $\beta$ values, $\beta = 2$ jets are more similar to anti-$k_T$ jets since both methods align the jet axis with the jet momentum.  The $\beta = 1$ jets are slightly softer, since they do not recoil away from the hard jet center to absorb soft radiation.  In the absence of pileup, however, the $\beta = 2$ and $\beta = 1$ performance is quite similar on single jet reconstruction.\footnote{In the presence of pileup, the directions of the $\beta = 1$ axes are more robust to pileup contamination \cite{Larkoski:2014bia}.}

\subsection{N = 2 for Dijet Reconstruction}

\begin{figure}
\centering
\subfloat[]{
\includegraphics[width = 0.45\columnwidth]{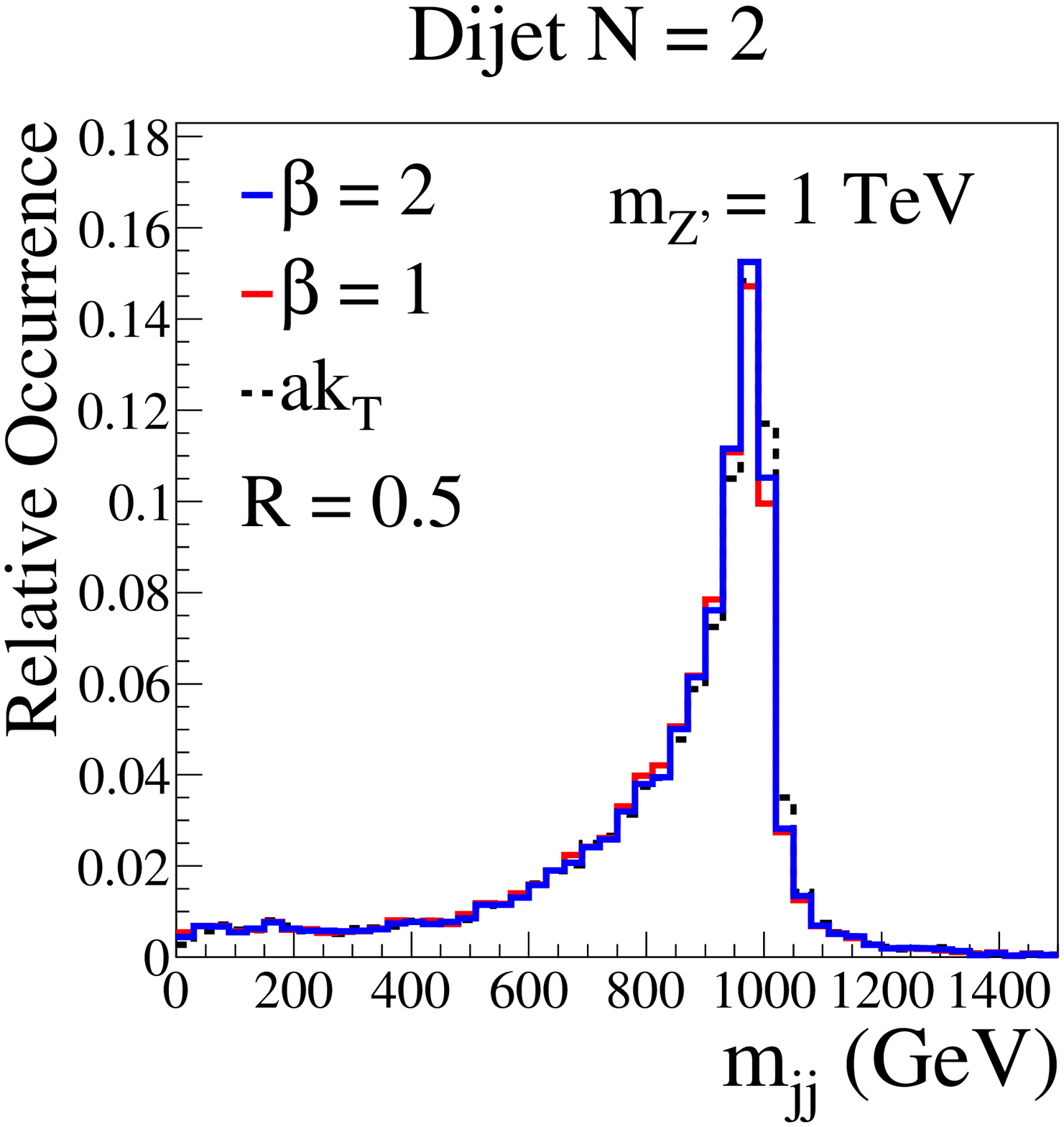}
\label{fig:dijet_2jet_mass}}
$\quad$
\subfloat[]{
\includegraphics[width = 0.45\columnwidth]{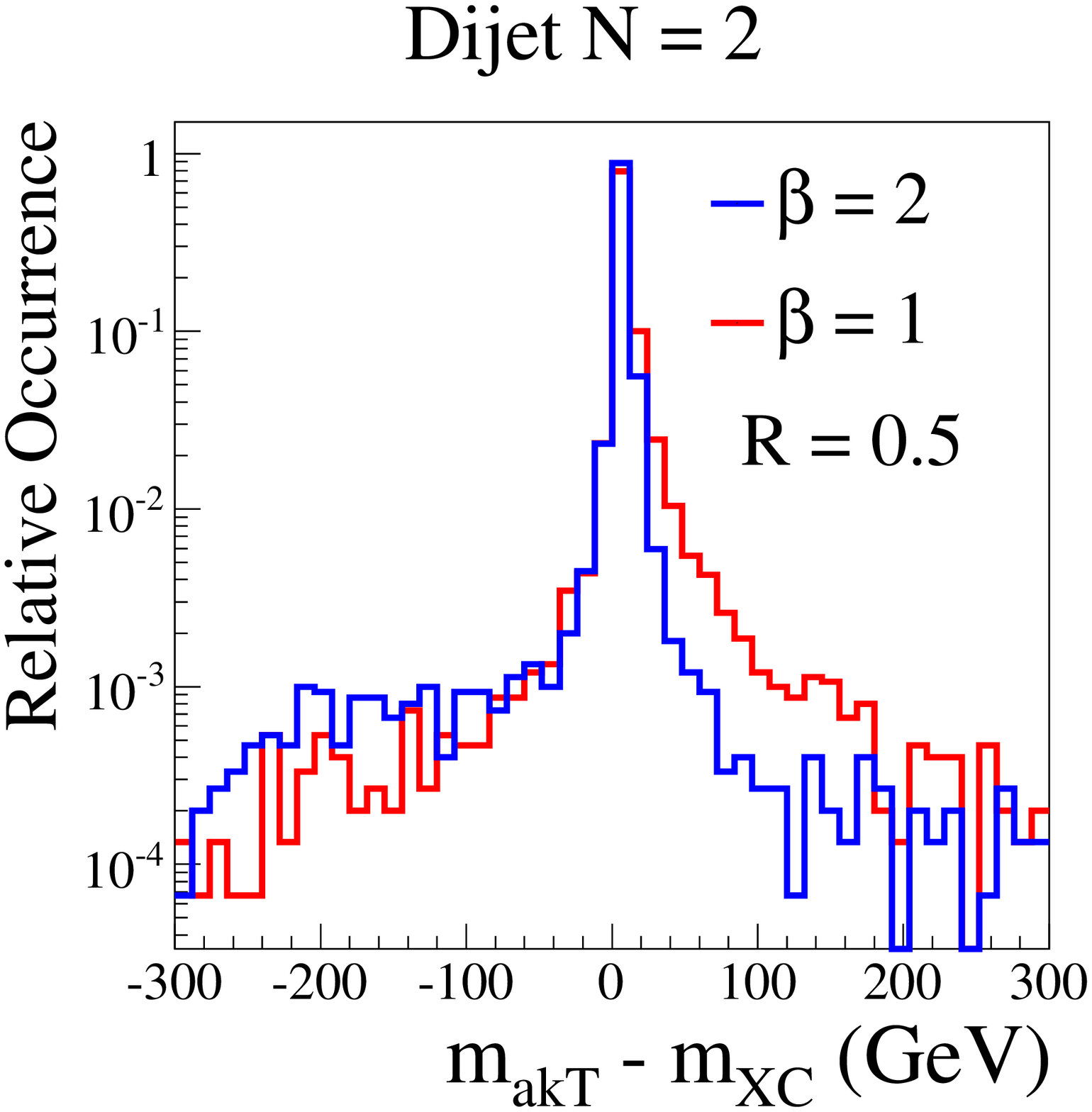}
\label{fig:dijet_2jet_massdiff}}
\caption{Dijet kinematics of $N = 2$ XCone versus the two hardest anti-$k_{T}$ jets, measured on the dijet resonance sample.  (a)  Dijet mass, showing the expected peak at $m_{Z'} = 1~\TeV$. (b) Dijet mass difference between XCone and anti-$k_{T}$ jet, showing comparable reconstruction.}
\label{fig:dijet_2jet_properties}
\end{figure}

For dijet resonance reconstruction, $N = 2$ is the most natural choice for running an exclusive cone jet algorithm.  As shown in \Figs{fig:dijet_2jet_mass}{fig:dijet_2jet_massdiff}, both anti-$k_T$ and XCone give a good reconstruction of the resonance peak, and they largely agree on the $m_{jj}$ value on an event-by-event basis, without much of an asymmetry in the $m_{jj}^{\text{anti-$k_T$}} -  m_{jj}^{\text{XCone}}$ distribution.  XCone can therefore act as a replacement for anti-$k_T$ for dijet resonance reconstruction, with comparable performance.\footnote{It is known that the $\Rnaught = 0.5$ cone size is typically too small to capture all of the dijet decay products \cite{Dasgupta:2007wa}.  While we did find that better performance could be obtained with somewhat larger $\Rnaught$, we wanted all of the plots in this paper to have a common cone size for ease of comparison.}

\subsection{N = 3 and ISR vs.\ FSR}

Thus far, XCone and anti-$k_T$ have exhibited very similar behavior, but differences start to appear when considering $N = 3$.  There are two main ways to achieve three jet configurations:  either there is sufficient initial state radiation (ISR) to form an additional widely-separated jet, or there is sufficient final state radiation (FSR) to give one of the primary jets some two-prong substructure.  In the ISR case (as in \Fig{fig:dijet_display:a}), anti-$k_T$ and XCone still give very similar results since the jets are non-overlapping.  In the FSR case (as in \Fig{fig:dijet_display:b}), XCone will often identify two separate prongs inside a fat clover jet, whereas anti-$k_T$ can only identify FSR if it is further away than $\Rnaught$ from the hard jet core.

\begin{figure}
\centering
\subfloat[]{
\includegraphics[width = 0.32\columnwidth]{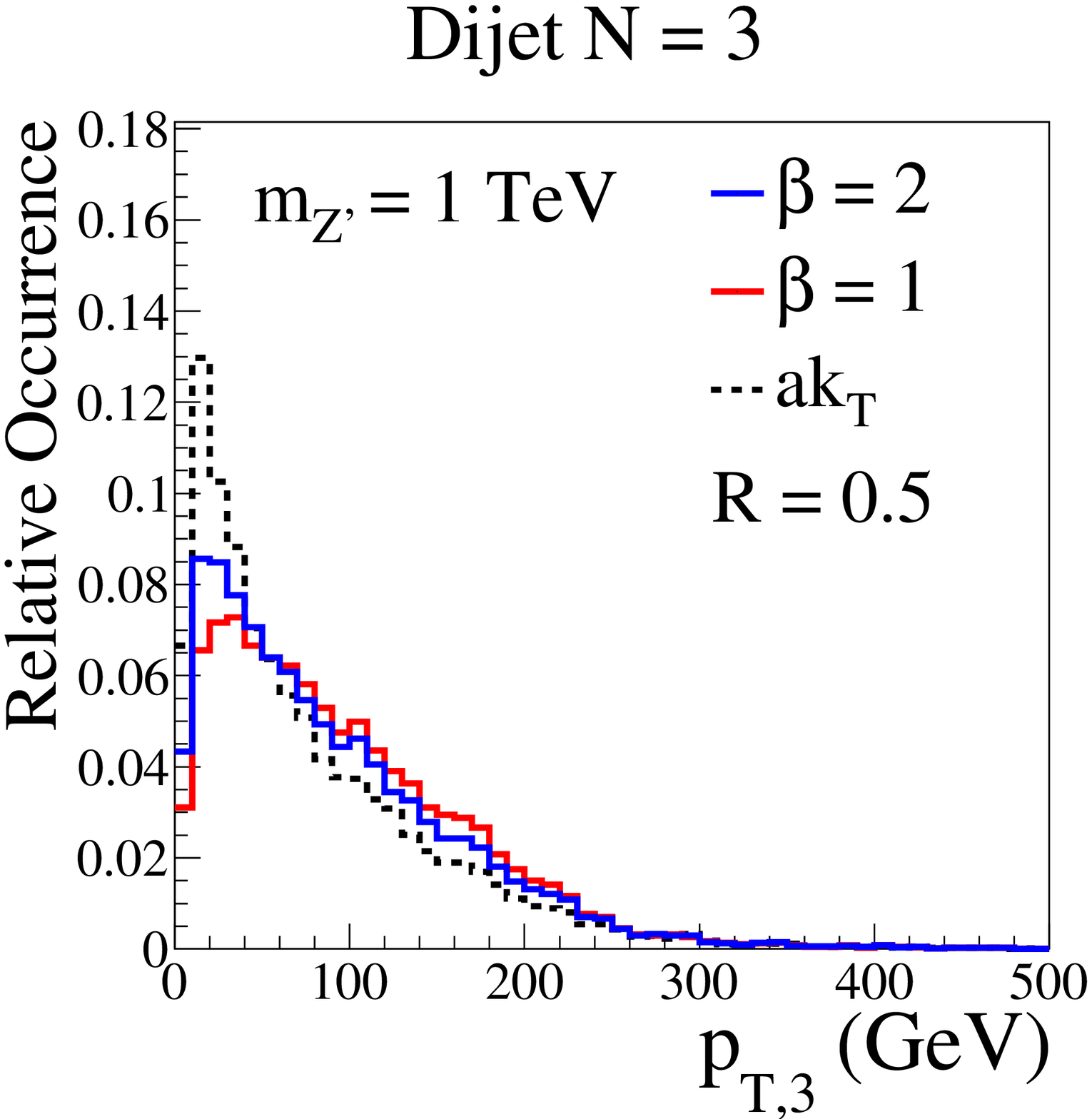}
\label{fig:dijet_3jet_perp}}
\subfloat[]{
\includegraphics[width = 0.32\columnwidth]{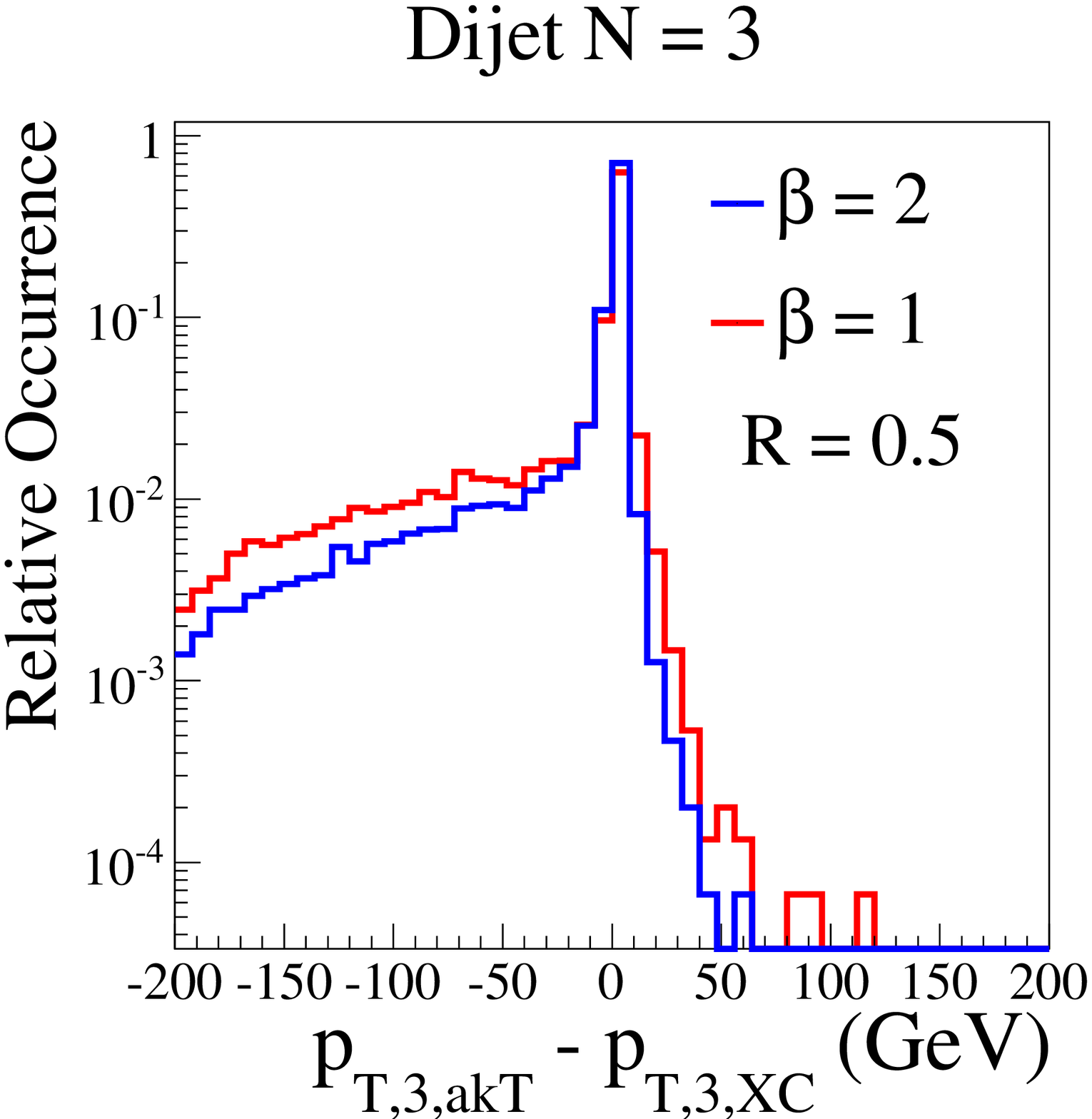}
\label{fig:dijet_3jet_perpdiff}}
\subfloat[]{
\includegraphics[width = 0.32\columnwidth]{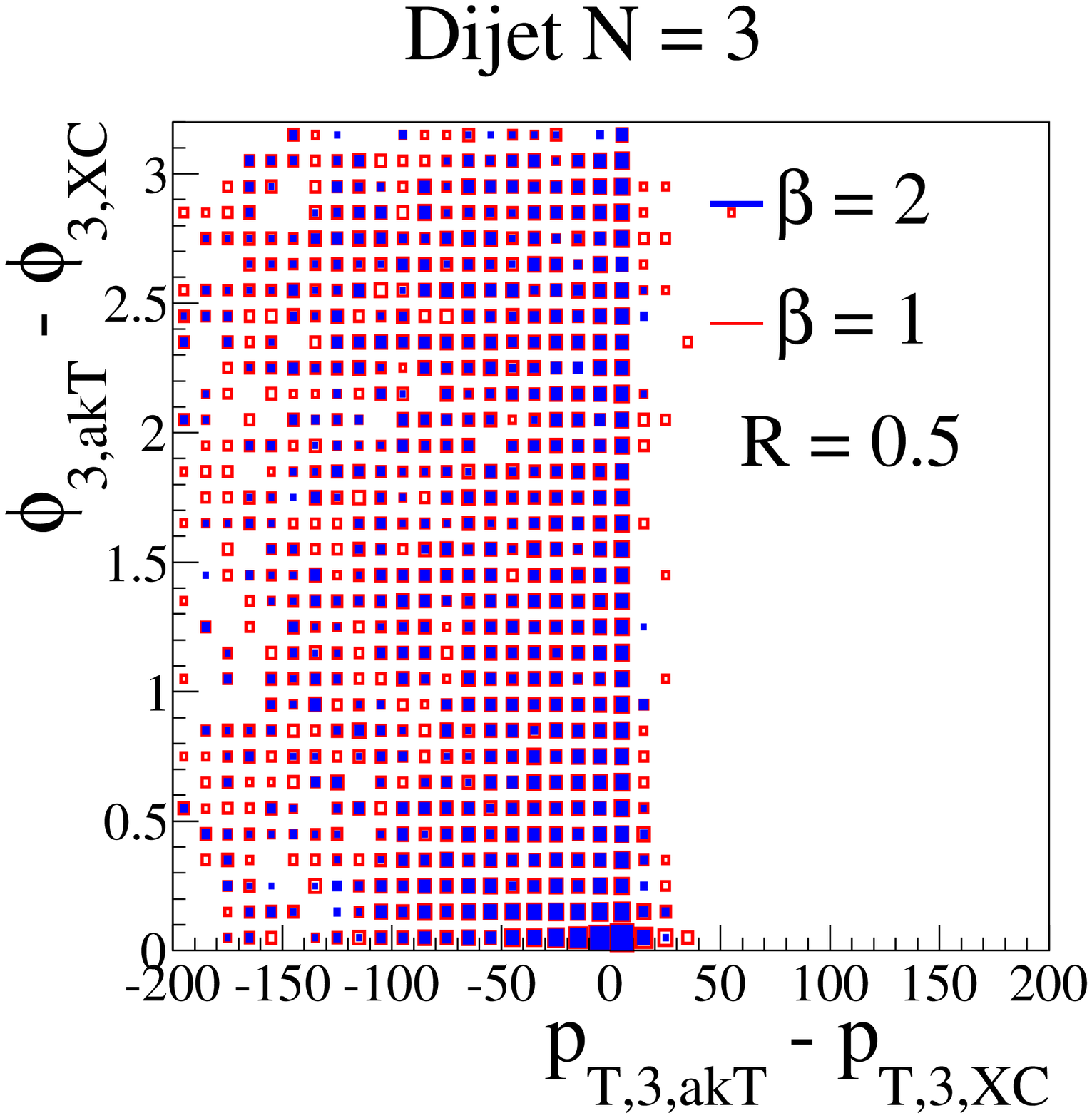}
\label{fig:dijet_3jet_perp_phidiff}}
\caption{Comparing the third hardest jet between XCone $N = 3$ and anti-$k_T$, measured on the dijet resonance sample.  (a)  Third jet $p_T$ spectrum. (b) Third jet $p_T$ difference between XCone and anti-$k_T$, showing that XCone has a somewhat harder spectrum due to its ability to identify FSR subjets.  (c)  Third jet $p_T$ difference versus azimuth difference, showing a population of events where the third jet kinematics are completely different between the algorithms. Again, the sizes of the boxes scale logarithmically with the number of entries.}
\label{fig:thirdjet}
\end{figure}

\begin{figure}[t]
\centering
\subfloat[]{
\includegraphics[width = 0.32\columnwidth]{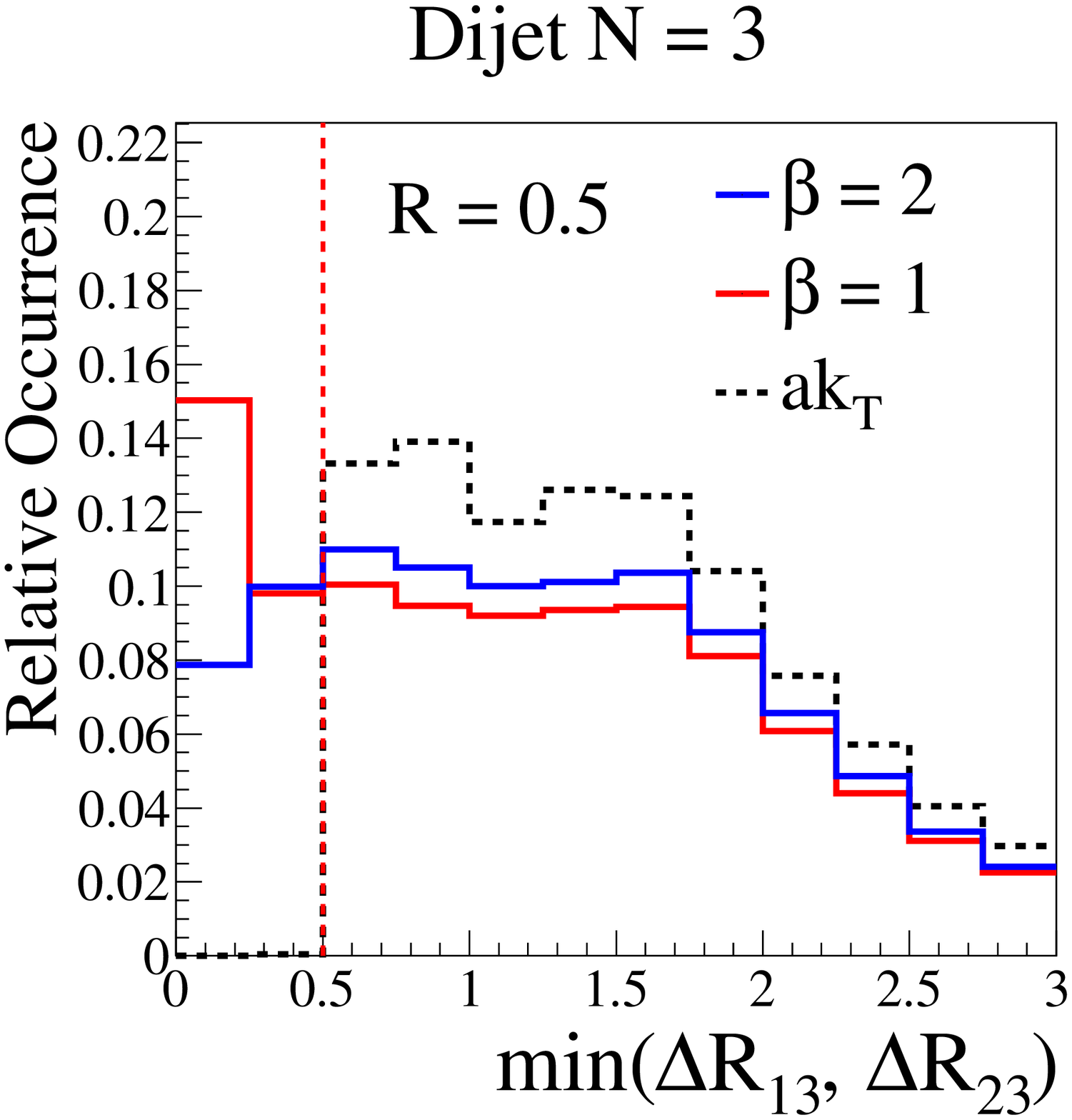}
\label{fig:dijet_3jet_mainjet_phidiff}}
\subfloat[]{
\includegraphics[width = 0.32\columnwidth]{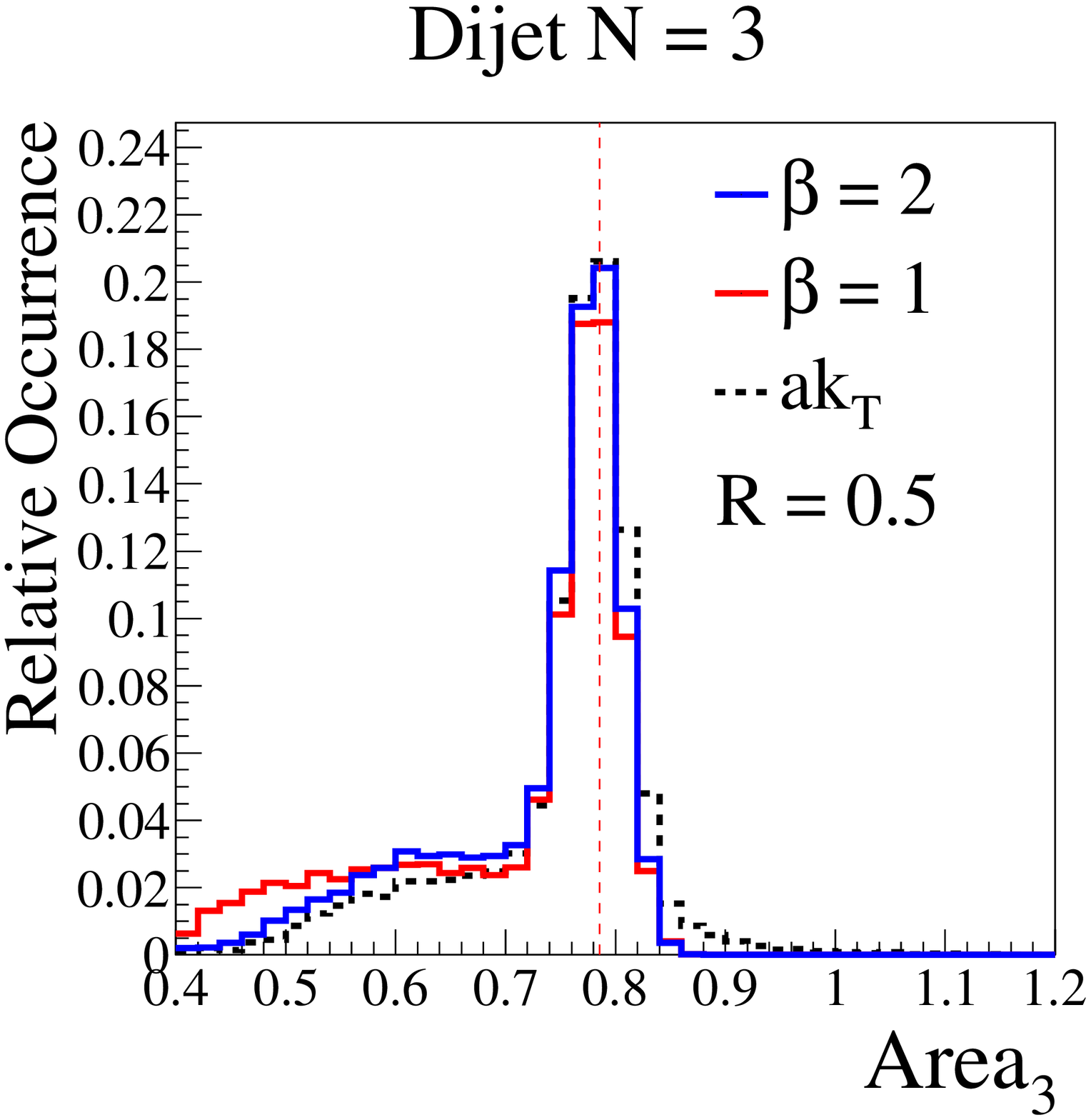}
\label{fig:dijet_3jet_area}}
\subfloat[]{
\includegraphics[width = 0.32\columnwidth]{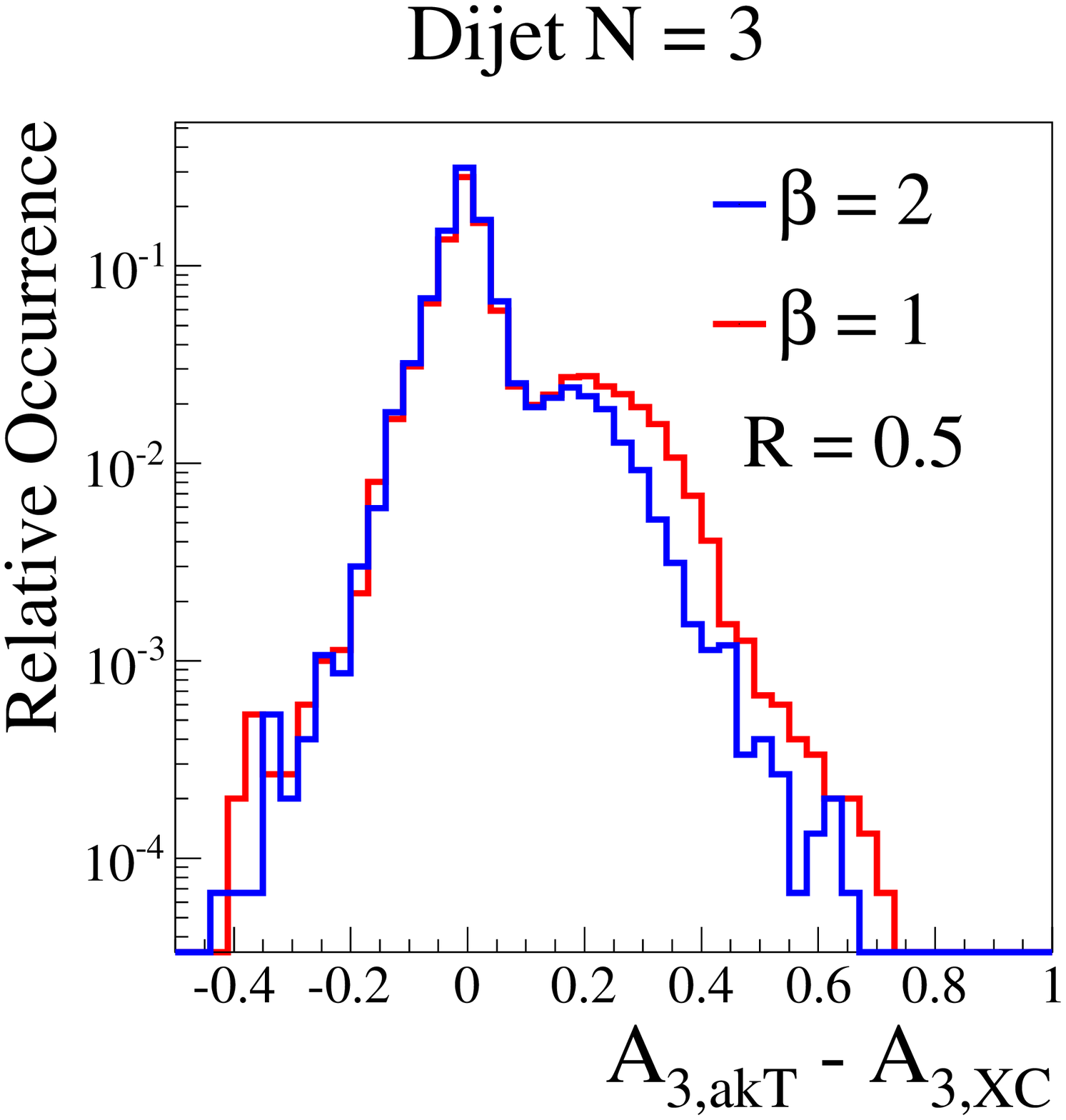}
\label{fig:dijet_3jet_areadiff}}
\caption{Same $N = 3$ comparison as \Fig{fig:thirdjet}.  (a) Angle of the third hardest jet to the nearest harder jet, showing that XCone jets can be located as close as $\Delta R = 0$ whereas anti-$k_{T}$ jets are forced to have $\Delta R > \Rnaught$.  (b)  Third jet area distribution, showing the expected peak at $\pi \Rnaught^2$, but with larger tails than in the $N = 1$ case in \Fig{fig:dijet_1jet_area}.  (c) Third jet area difference, showing a population of XCone jets with much smaller areas due to jet splitting.}
\label{fig:dijet_3jet_compare}
\end{figure}

\begin{table}[t]
\begin{center}
\subfloat[]{
\small
\begin{tabular}{cccc}
\hline \hline
$(\beta = 2)$ & XCone 1 & XCone 2 & XCone 3 \\
\hline\hline
AKT 1 & 0.925 & 0.077 & 0.031 \\ 
\hline
AKT 2 & 0.075 & 0.913 & 0.042 \\ 
\hline
AKT 3 & 0.000 & 0.006 & 0.795 \\ 
\hline
AKT 4 & 0.000 & 0.001 & 0.043 \\ 
\hline\hline
\end{tabular}
\label{tab:akt_njet_overlap_beta_2}}
\qquad
\subfloat[]{
\small
\begin{tabular}{cccc}
\hline \hline
$(\beta = 1)$ & XCone 1 & XCone 2 & XCone 3 \\
\hline\hline
AKT 1 & 0.884 & 0.120 & 0.070 \\ 
\hline
AKT 2 & 0.116 & 0.870 & 0.076 \\ 
\hline
AKT 3 & 0.000 & 0.007 & 0.720 \\ 
\hline
AKT 4 & 0.000 & 0.001 & 0.041 \\ 
\hline\hline
\end{tabular}
\label{tab:akt_njet_overlap_beta_1}}
\caption{Comparing XCone $N = 3$ to anti-$k_T$.  Shown is the fraction of events where the $n\textsuperscript{th}$ hardest XCone jet is within $\Rnaught/2 = 0.25$ of the $n\textsuperscript{th}$ hardest anti-$k_T$ jet.  (a)  The $\beta = 2$ default which behaves most similarly to anti-$k_T$.  (b) The $\beta = 1$ recoil-free variant where larger differences are possible.}
\end{center}
\end{table}

From the $p_T$ spectra in \Fig{fig:dijet_3jet_perp}, we see that the third jet is often softer in the anti-$k_T$ case, as expected if anti-$k_T$ tends to identify ISR jets. XCone, on the other hand, is able to find FSR radiation that lies close to one of the two original jets, and thus is more likely to find a hard third jet adjacent to the hard dijet structure. This is highlighted in \Figs{fig:dijet_3jet_perpdiff}{fig:dijet_3jet_perp_phidiff}, which shows that the third anti-$k_T$ jet can have completely uncorrelated kinematics from the third XCone jet.  We can gain further insight in \Fig{fig:dijet_3jet_mainjet_phidiff}, which shows the distance between the third jet and the closest harder jet.  In the anti-$k_T$ case, the third jet is forced to be further away than $\Rnaught$ from the jet core, whereas in XCone case, the third jet can go nearly to $\Delta R = 0$, corresponding to XCone finding substructure within a fat clover jet, as desired, instead of finding a separate ISR jet.

The same effects are visible in the area distributions in \Figs{fig:dijet_3jet_area}{fig:dijet_3jet_areadiff}.  While the overall area distributions are not so dissimilar (particularly in the $\beta = 2$ case), on an event-by-event basis, there is a population of events where the third XCone jet has substantially smaller jet area, indicative of jet overlap.  This is the flip side of the area distributions for $N = 1$ in \Fig{fig:dijet_1jet_areadiff}, where anti-$k_T$ jets could grow larger in size by incorporating a neighboring subjet.  In the XCone case, that subjet is separately identified as its own jet for $N = 3$.

Despite these differences, the overall jet reconstruction is still rather similar between XCone and anti-$k_T$.  In \Tabs{tab:akt_njet_overlap_beta_2}{tab:akt_njet_overlap_beta_1}, we show the fraction of events for which the $n\textsuperscript{th}$ hardest XCone jet is within $\Rnaught/2 = 0.25$ of the $n\textsuperscript{th}$ hardest anti-$k_{T}$ jet.  For $\beta = 2$, the three hardest jets are well aligned 80\% to 90\% of the time.  For $\beta = 1$, there are larger deviations, though often this is just because the first and second jets are reversed in $p_T$ ordering.  We conclude that the use of XCone is particularly advantageous for tagging small-angle FSR, but otherwise will have similar performance to anti-$k_{T}$.

Of course, there may be physics contexts where splitting jets by nearest-neighbor is undesirable and circular jet regions are preferred.  After all, the hardest anti-$k_T$ jet in an event tends to be circular, whereas proximate softer jets form crescent shaped regions, and this is often a desirable feature for jet calibrations and calculations.  In this context, note that XCone jet regions are fully determined by the locations of the corresponding jet axes, independent of the details of the jet constituents.  Therefore, it is straightforward to test for jet splitting by simply checking whether any two jet axes are closer than $2R$.  One could even imagine running XCone in a mode where $N$ is subsequently decreased until the distances between all axis pairs are larger than $2R$, forcing circular jet regions.   As we will see below, though, it is precisely the ability of XCone to split abutting jets which allows it to handle extreme kinematic circumstances where anti-$k_T$ reconstruction inevitability leads to jet merging.

\section{Boosted Higgs Bosons and Intelligent Jet Splitting}
\label{sec:higgs}

\begin{figure}[t]
\centering
\subfloat[]{
\includegraphics[width=0.32\columnwidth]{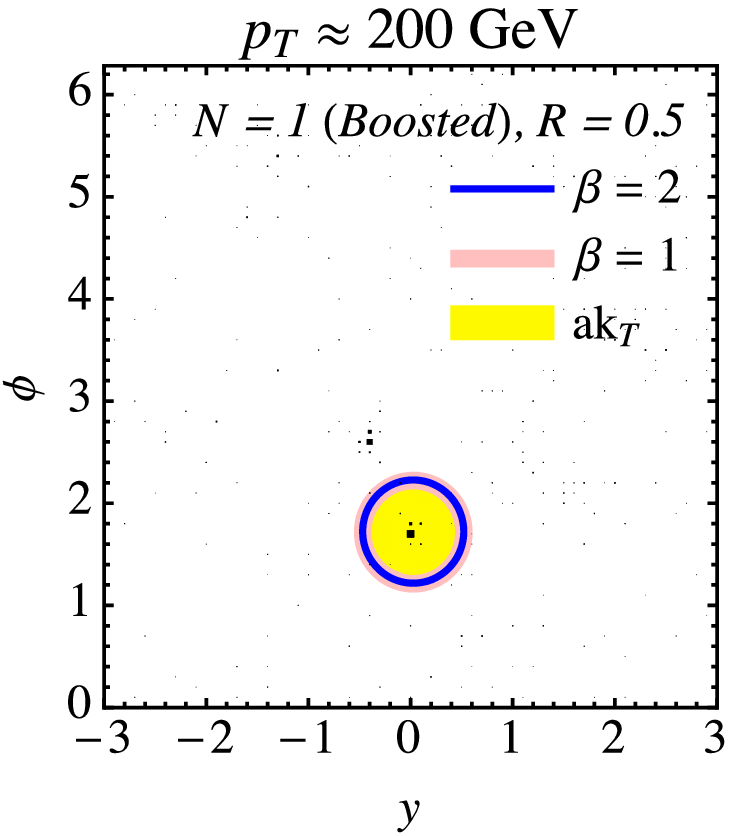}
\label{fig:higgs_1jet_display_200}
}
\subfloat[]{
\includegraphics[width=0.32\columnwidth]{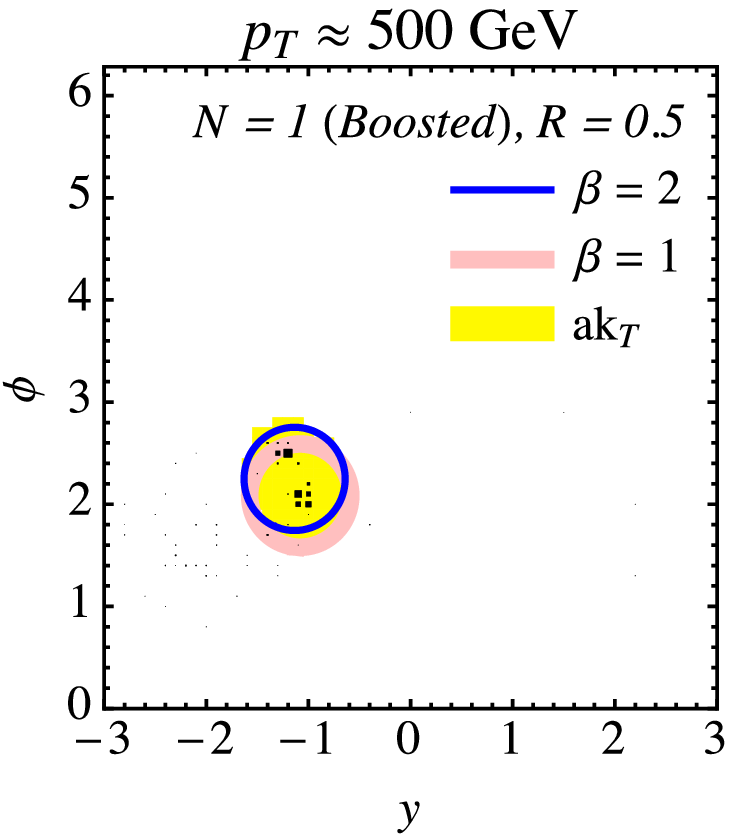}
\label{fig:higgs_1jet_display_500}
}
\subfloat[]{
\includegraphics[width=0.32\columnwidth]{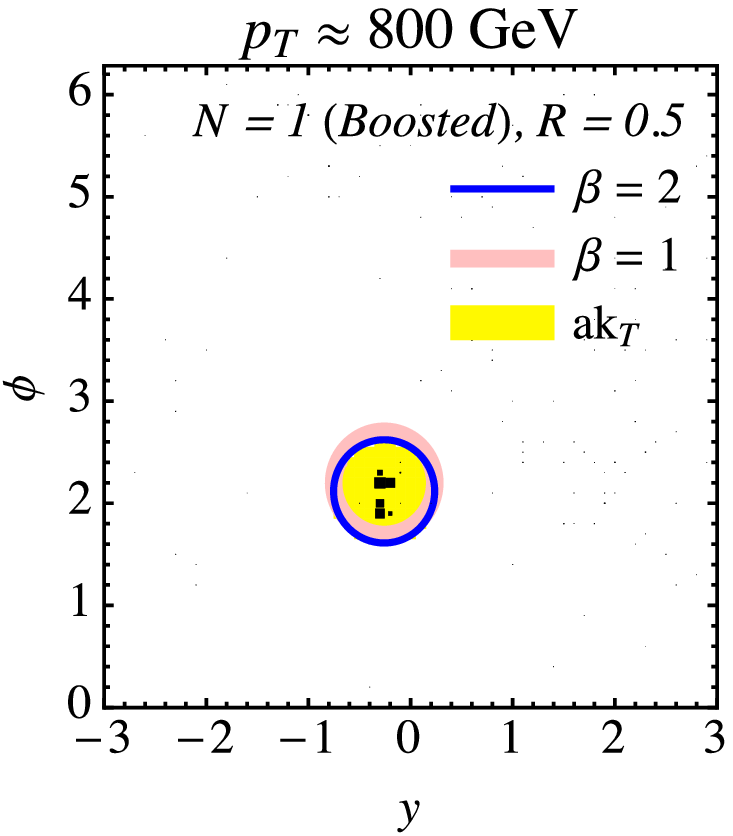}
\label{fig:higgs_1jet_display_800}
}

\subfloat[]{
\includegraphics[width=0.32\columnwidth]{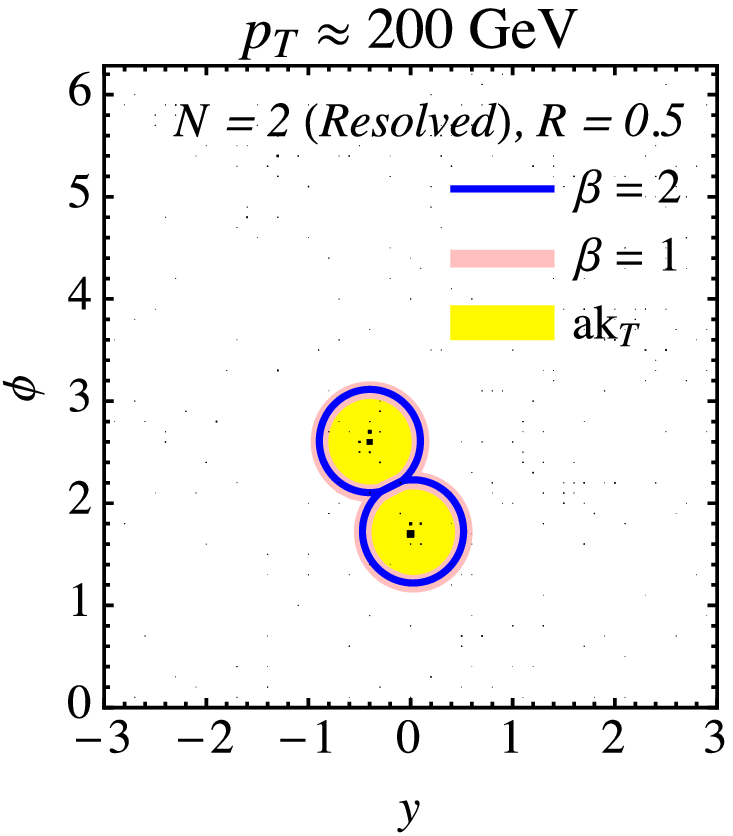}
\label{fig:higgs_2jet_display_200}
}
\subfloat[]{
\includegraphics[width=0.32\columnwidth]{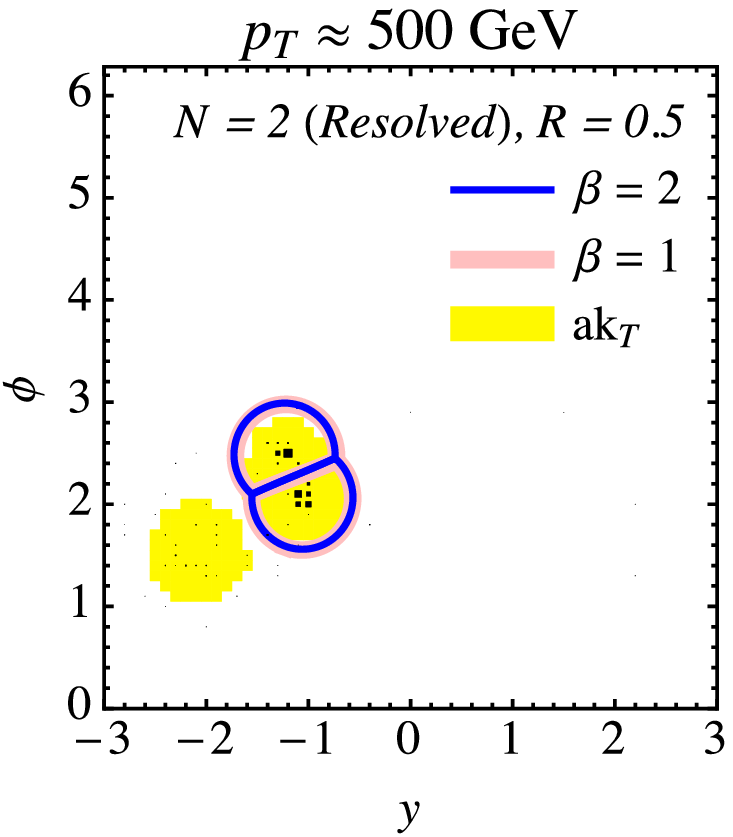}
\label{fig:higgs_2jet_display_500}
}
\subfloat[]{
\includegraphics[width=0.32\columnwidth]{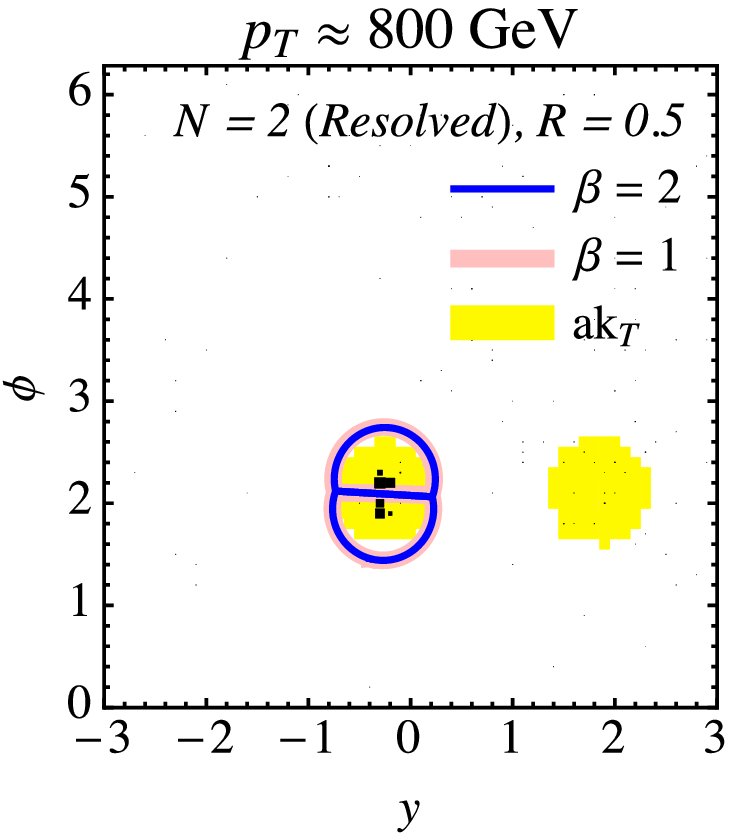}
\label{fig:higgs_2jet_display_800}
}
\caption{Example XCone jet regions for the Higgs analysis for $p_T \approx \{200, 500, 800\}~\GeV$, using $\Rnaught = 0.5$.  Top row:  A boosted Higgs analysis with $N = 1$, where XCone and anti-$k_T$ identify nearly the same jet regions.  Bottom row:  A resolved Higgs analysis with $N = 2$, where XCone separately identifies the $H \to b \overline{b}\,$ substructure while anti-$k_T$ often identifies ISR.  For each $p_T$ value, the same event appears for $N = 1$ and $N = 2$.
}
\label{fig:higgs_displays}
\end{figure}

To highlight the distinct advantages of XCone, we now consider physics situations where resolving substructure is a key element of the analysis.  Because XCone always identifies $N$ jets, it is well-suited to physics applications with a fixed number of expected (sub)jets.  This is particularly interesting for cases involving jet substructure, where traditional jet algorithms yield merged fat jets, but XCone can identify jets and subjets simultaneously.

As a well-motivated example at the LHC, consider associated Higgs boson production where the Higgs decays to bottom pairs \cite{Chatrchyan:2013zna,Aad:2014xzb}:
\be
p p \to H Z \to b \bar{b} \nu \bar{\nu}. 
\ee
Apart from possible ISR, the final state consists only of two $b$-jets.  To fight QCD backgrounds, the $p_T$ of the Higgs boson must be reasonably large \cite{Gallicchio:2010dq}.  However, in this regime, the Higgs decay products are more collimated, often resulting in jet merging.  Roughly speaking, the two $b$-jets will be merged into a single fat jet when the Higgs boson is at the scale
\be
\label{eq:higgsmergingscale}
p_T^{\rm merge} \simeq \frac{2 m_H}{\Rnaught}.
\ee
In order to counteract this effect, either $\Rnaught$ can be decreased until it is small enough to resolve two separate $b$-jets, or jet substructure techniques can be used \cite{Butterworth:2008iy}.

This (quasi-)boosted Higgs analysis is well-suited for XCone.  At minimum, $N = 1$ can identify a single fat jet, after which existing jet substructure techniques can be applied. Though we will not perform a detailed jet substructure analysis here, we will show that XCone with $N=1$ has nearly the same signal efficiency as anti-$k_T$ as a function of $p_T$, and therefore can be used as a suitable starting point for a full boosted Higgs analysis.

More intuitively, $N = 2$ can be used to identity the two hard $b$-jets in the event at all $p_{T}$ scales.  This $N=2$ strategy is very similar to what is already being done in existing ATLAS and CMS studies \cite{Chatrchyan:2013zna,Aad:2014xzb}, where anti-$k_T$ is used to resolve two separate $b$-jets.  For anti-$k_T$, the $b$-jets merge at high enough $p_T$, so this resolved technique is no longer efficient (see \cite{Butterworth:2015bya} for a recent discussion).  For XCone, we show that the corresponding $N=2$ resolved strategy can be pushed deep into the high $p_T$ regime while maintaining good signal efficiency.  For this reason, we advocate XCone as a promising approach to extrapolate resolved analyses into the boosted regime.

Example event displays using the $N=1$ and $N=2$ methods are shown in \Fig{fig:higgs_displays}.  In the text, we restrict our comparisons to anti-$k_T$, though in \App{app:exclusive_higgs}, we also show results for exclusive $k_T$, which has performance comparable to XCone, albeit with irregular jet boundaries.  A full accounting of background processes is beyond the scope of this work, though we do perform a sanity check in \App{app:boostedbackground} to verify that XCone does not unnecessarily sculpt the $Z + \text{jets}$ background.  Since the relative performance of the boosted and resolved strategies depends on the details of the background, we postpone a direct comparison of $N=1$ with $N=2$ to future work.

Like the previous dijet study, we use Pythia 8.176 \cite{Sjostrand:2006za,Sjostrand:2007gs} at the $\sqrt{s} = 14$ TeV LHC to simulate $pp \rightarrow HZ$.  For simplicity, we force the decays $Z \rightarrow \nu\bar{\nu}$ and $H \rightarrow b\bar{b}$. Like in the previous study, all of the final-state particles (except neutrinos) with $\abs{\eta} < 3.0$ are considered for analysis. In order to analyze the properties of the algorithm in different $p_{T}$ regimes, we place generator-level $p_{T}$ cuts on the Higgs boson between $200$ and $1000$ GeV.  We use the same $\Rnaught = 0.5$ jet radius for all analyses in this section, such that $p_T^{\rm merge} \simeq 500~\GeV$ is in the middle of our studied $p_T$ range.

\subsection{N = 1 for Boosted Analysis}
\label{sec:Higgs1}

\begin{figure}
\centering
\subfloat[]{
\includegraphics[width=0.32\columnwidth]{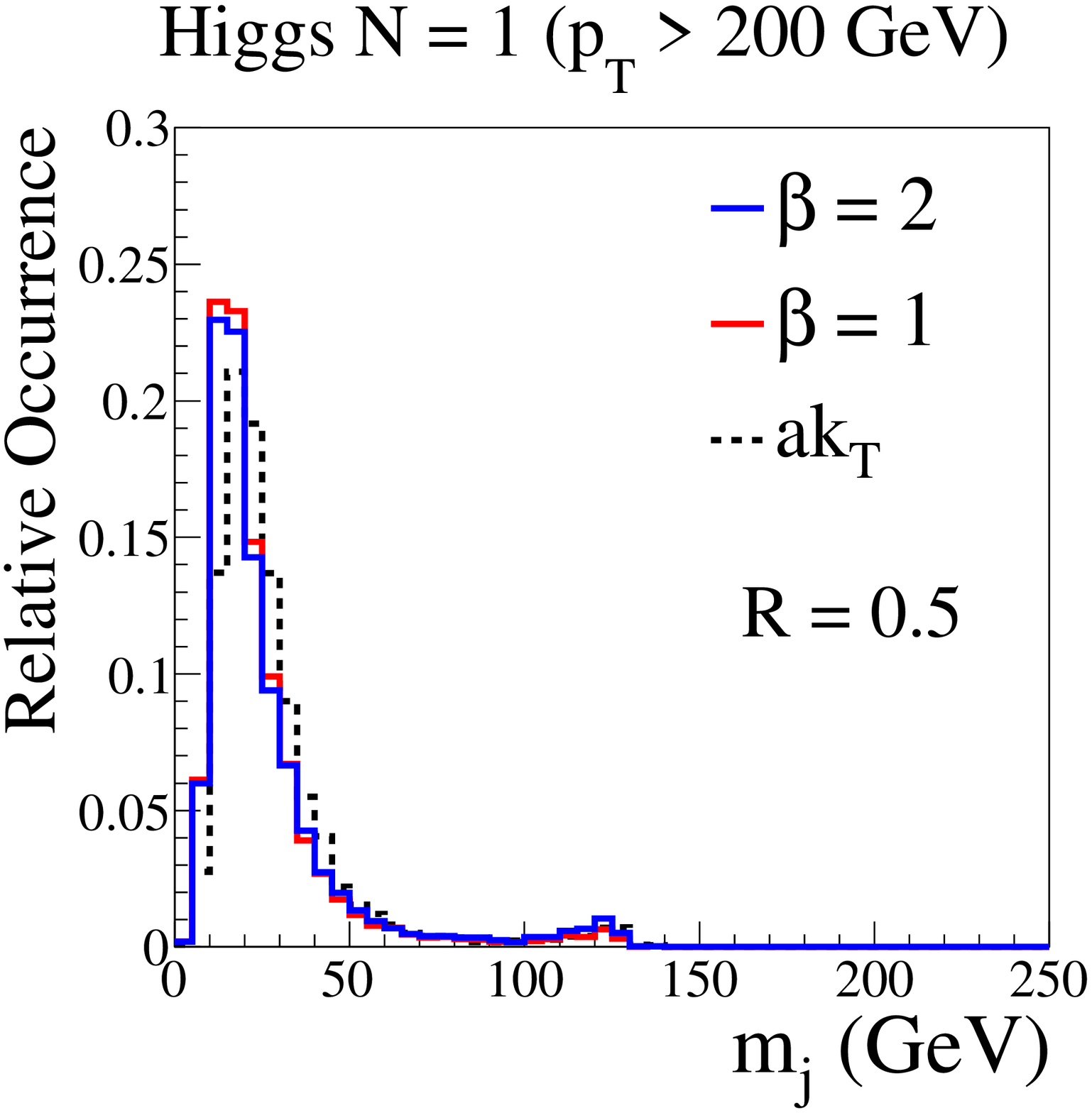}
\label{fig:higgs_1jet_mass_200}
}
\subfloat[]{
\includegraphics[width=0.32\columnwidth]{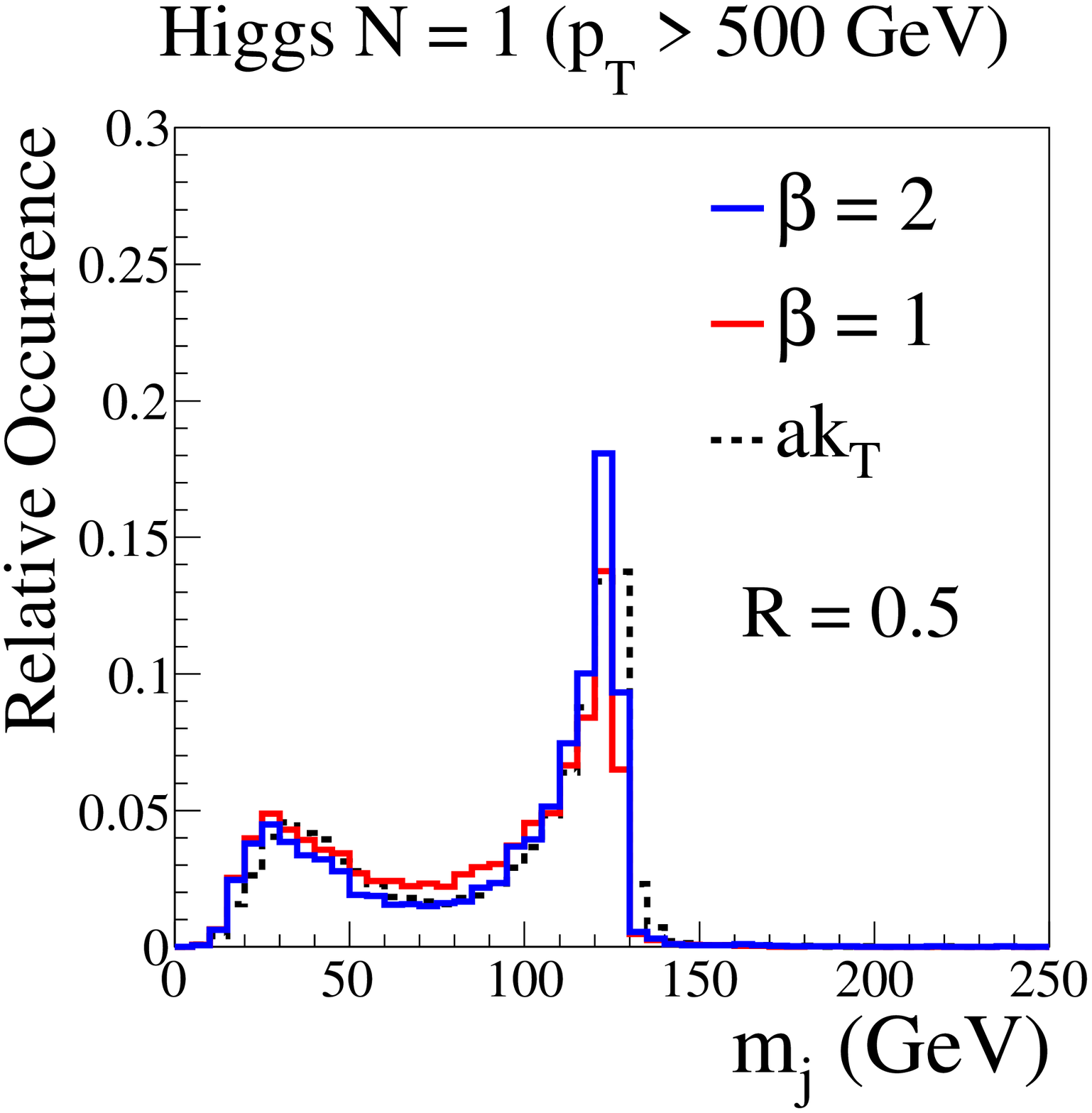}
\label{fig:higgs_1jet_mass_500}
}
\subfloat[]{
\includegraphics[width=0.32\columnwidth]{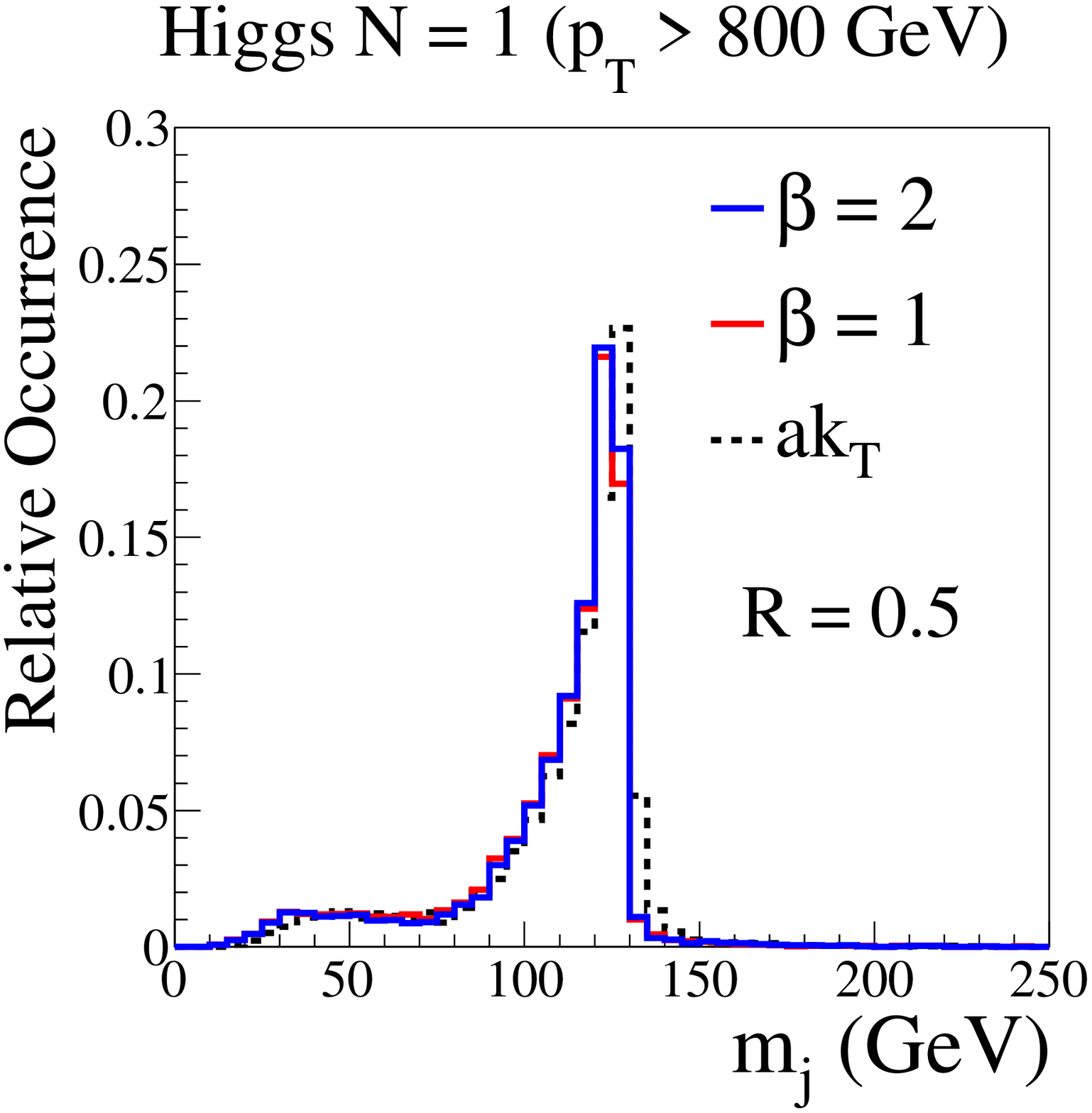}
\label{fig:higgs_1jet_mass_800}
}

\subfloat[]{
\includegraphics[width=0.32\columnwidth]{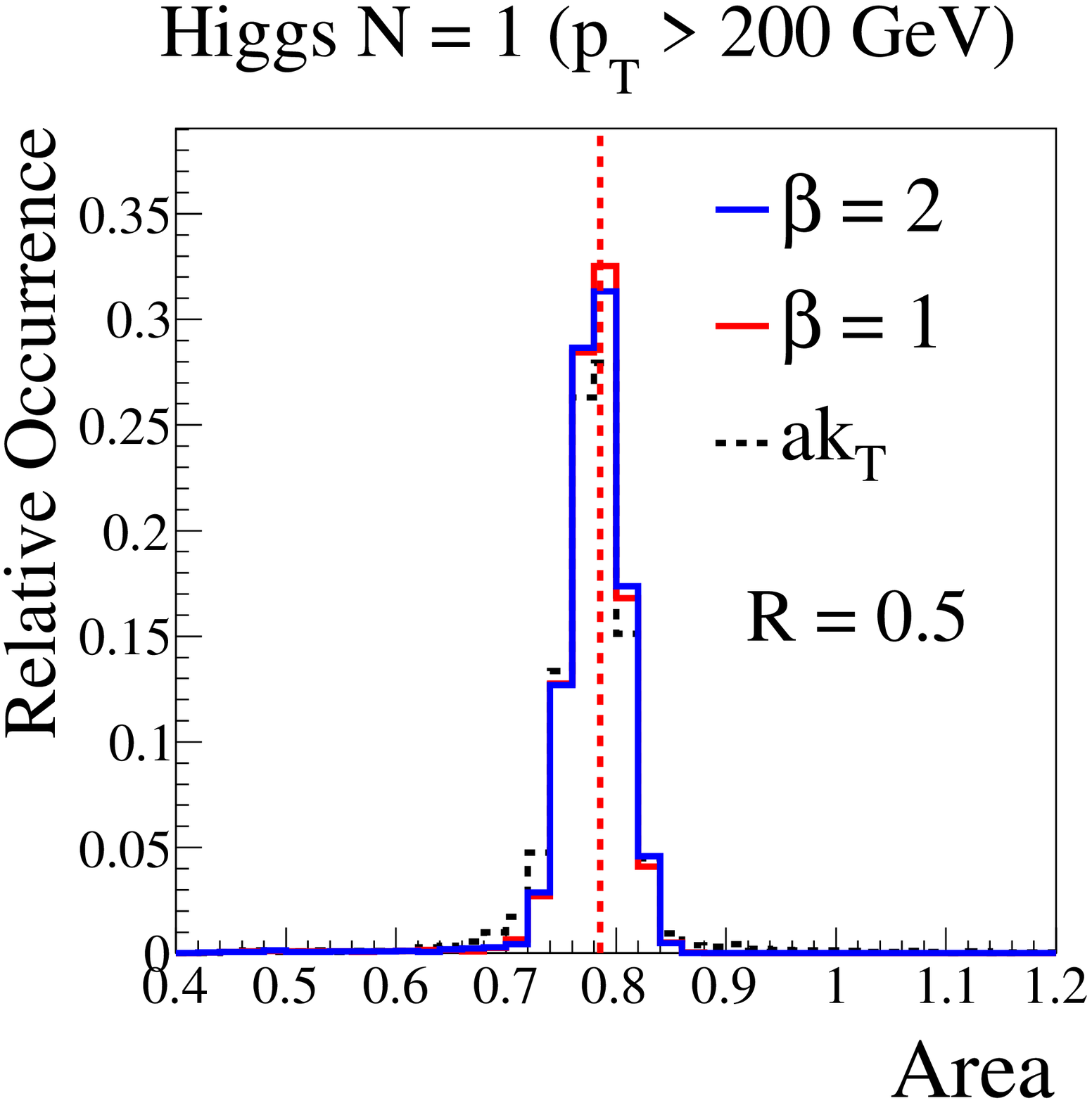}
\label{fig:higgs_1jet_area_200}
}
\subfloat[]{
\includegraphics[width=0.32\columnwidth]{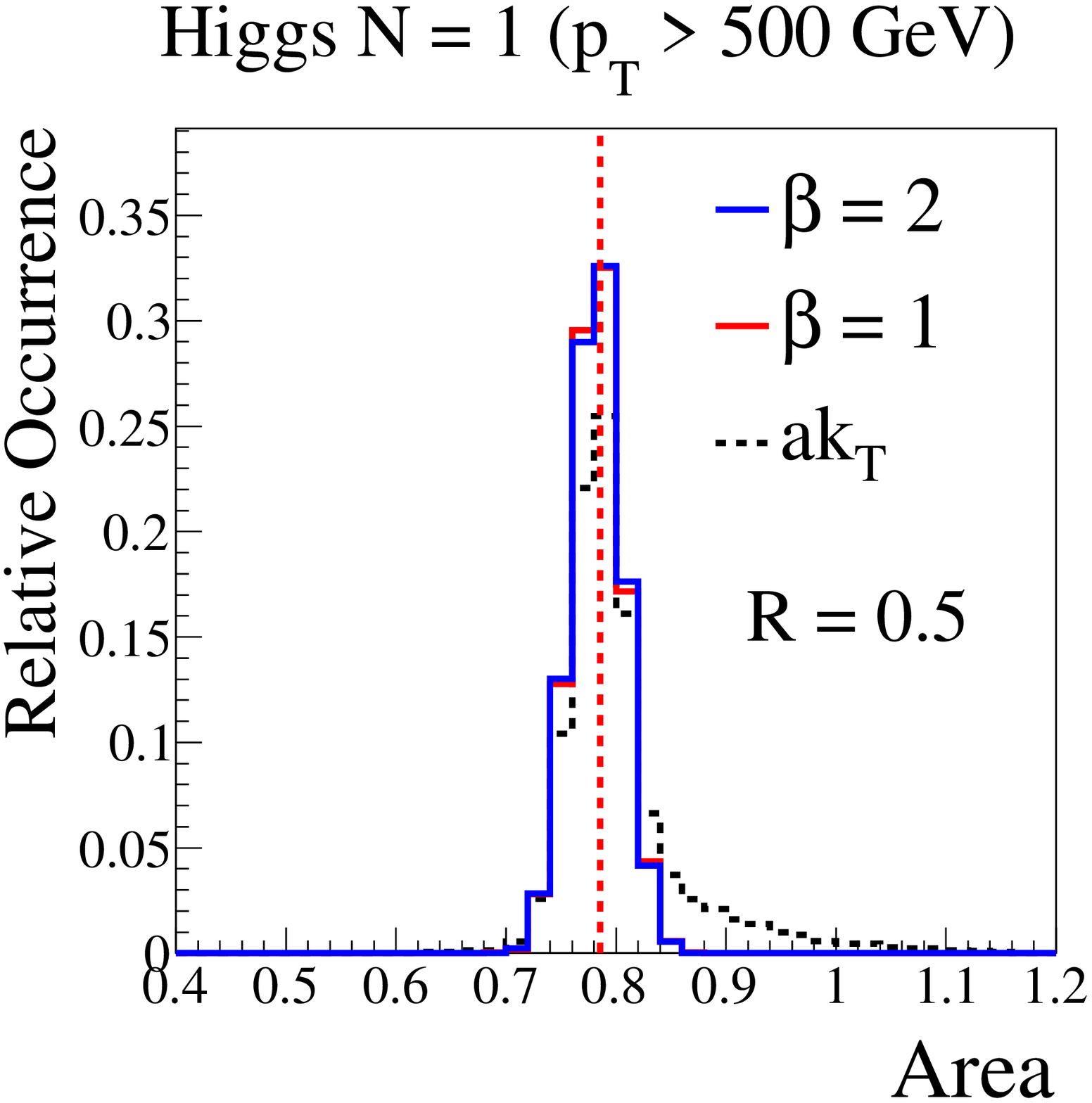}
\label{fig:higgs_1jet_area_500}
}
\subfloat[]{
\includegraphics[width=0.32\columnwidth]{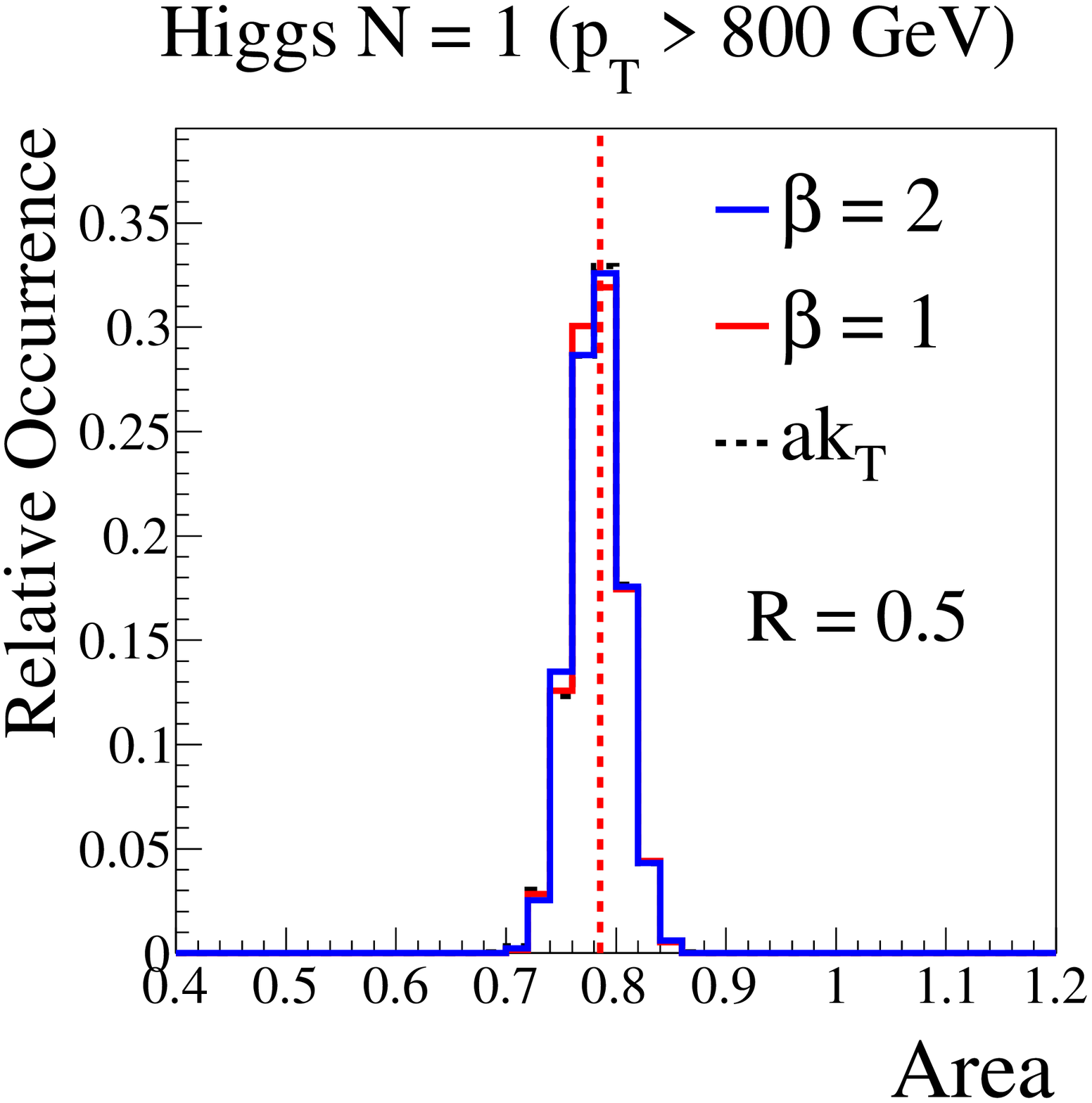}
\label{fig:higgs_1jet_area_800}
}
\caption{Top row: Comparing boosted Higgs reconstruction between XCone with $N = 1$ and the hardest anti-$k_{T}$ jet, using $\Rnaught = 0.5$.  As the Higgs $p_T$ increases from (a) 200 GeV to (b) 500 GeV to (c) 800 GeV, both methods capture the merged Higgs decay products, yielding a growing mass peak at $m_H = 125~\mathrm{GeV}$. Bottom row:  Comparing Higgs jet area using XCone $N = 1$ and the hardest anti-$k_{T}$ jet.  All distributions show the large expected peak at $A = \pi (0.5)^2$. }
\label{fig:higgs_1jet_study}
\end{figure}

Since the pioneering work in \Ref{Butterworth:2008iy}, the boosted Higgs channel has often been been analyzed by finding one fat jet with a large radius parameter, and then using substructure techniques to analyze its properties (see e.g.~\cite{Kribs:2009yh}).  Here, we compare $N = 1$ XCone to anti-$k_T$ to show that they give similar behavior in the boosted regime, though we stick to a relatively small $\Rnaught = 0.5$.\footnote{The original BDRS paper used the Cambridge/Aachen algorithm \cite{Dokshitzer:1997in,Wobisch:1998wt,Wobisch:2000dk} to identify this fat jet.  To  avoid a proliferation of curves, we only compare XCone to anti-$k_T$ in our analysis.  Regardless of the fat jet starting point, one can still recluster with Cambridge/Aachen to apply the BDRS mass drop criteria.}

\begin{figure}
\centering
\includegraphics[width = .6\columnwidth]{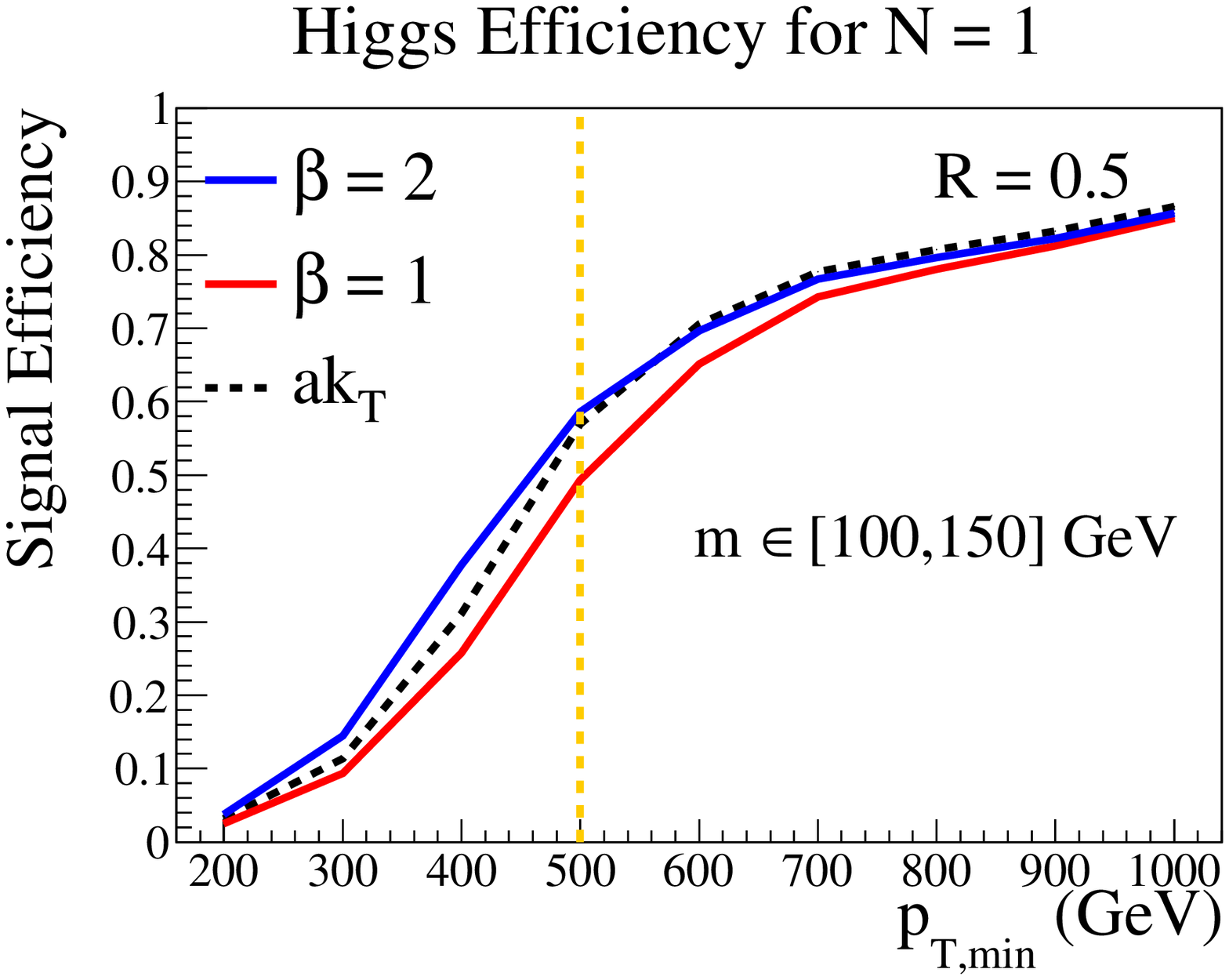}
\caption{Efficiency for $N = 1$ boosted Higgs reconstruction as a function of Higgs $p_T$, with the mass window $m_{j} \in [100,150]~\GeV$.  The efficiency grows when the $p_T$ is above the merging scale $2 m_H / \Rnaught$. XCone $\beta = 2$ outperforms $\beta =1$ in the transition region since the former centers the jet along the Higgs momentum.}
\label{fig:higgs_1jet_eff}
\end{figure}

Unlike in the dijet resonance study, the difference between $\beta = 1$ and $\beta = 2$ is more noticeable for quasi-boosted Higgs bosons.  As described in \Refs{Thaler:2011gf,Larkoski:2014uqa}, $\beta = 1$ minimization aligns the jet axis with the hardest cluster within a jet, whereas $\beta = 2$ minimization places the jet axis approximately in the direction of the jet momentum.  Roughly speaking, $\beta = 1$ finds the ``median'' jet axis direction whereas $\beta = 2$ finds the ``mean'' jet axis direction.  For the boosted Higgs case, the $\beta = 1$ jet is more likely to point in the direction of one of the two $b$-jets, while the $\beta = 2$ jet is more likely to lie in between the two $b$-jets and track the Higgs momentum direction.   Anti-$k_{T}$ (with standard $E$-scheme recombination \cite{Blazey:2000qt}) acts like $\beta = 2$ since it also aligns the jet axis with the jet momentum.

In the top row of \Fig{fig:higgs_1jet_study}, we show the single jet invariant mass as the minimum Higgs $p_T$ (at generator level) is adjusted from 200 GeV to 500 GeV to 800 GeV.  With increasing $p_T$, more of the Higgs decay products are contained inside a single jet and the peak at $m_j = 125~\GeV$ grows.  By eye, the anti-$k_T$ and $\beta = 2$ distributions are quite similar, whereas the $\beta = 1$ case has a somewhat worse performance since the jet axis is misaligned from the Higgs boson momentum.  Though not shown in the plot, the default conical geometric XCone measure \cite{Stewart:2015waa} yields a slightly better Higgs peak than the original conical measure \cite{Thaler:2011gf}.  In the bottom row of \Fig{fig:higgs_1jet_study}, we show the distribution of active jet areas, where both algorithms have a peak at $\pi \Rnaught^2$, though anti-$k_T$ has a slight high-side tail when $p_T \simeq p_T^{\rm merge}$.

We quantify the Higgs reconstruction efficiency in \Fig{fig:higgs_1jet_eff}, which shows the fraction of jets in the Higgs mass window $m_{j} \in [100,150]~\GeV$.  At very high $p_T$ values, the algorithms have very similar performance, but $\beta = 2$ does better in the vicinity of $p_T^{\rm merge}$.  This is because the $\beta = 2$ jet axis is more likely to lie in between the two $b$-jets, so the jet is more likely to capture the full Higgs decay products.  As expected, anti-$k_T$ and $\beta = 2$ are very similar.

\begin{figure}
\centering
\subfloat[]{
\includegraphics[width=0.32\columnwidth]{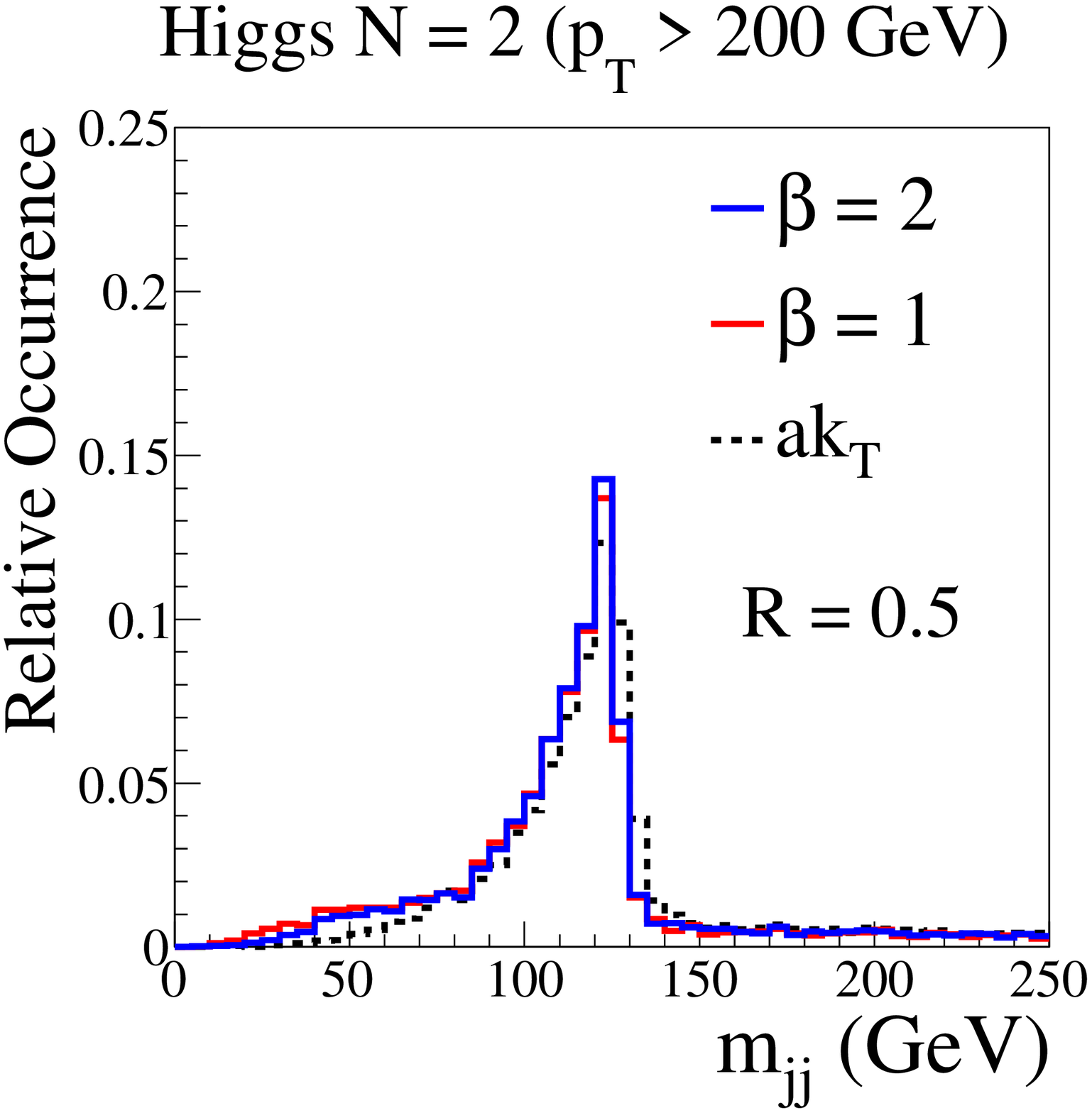}
\label{fig:higgs_2jet_mass_200}
}
\subfloat[]{
\includegraphics[width=0.32\columnwidth]{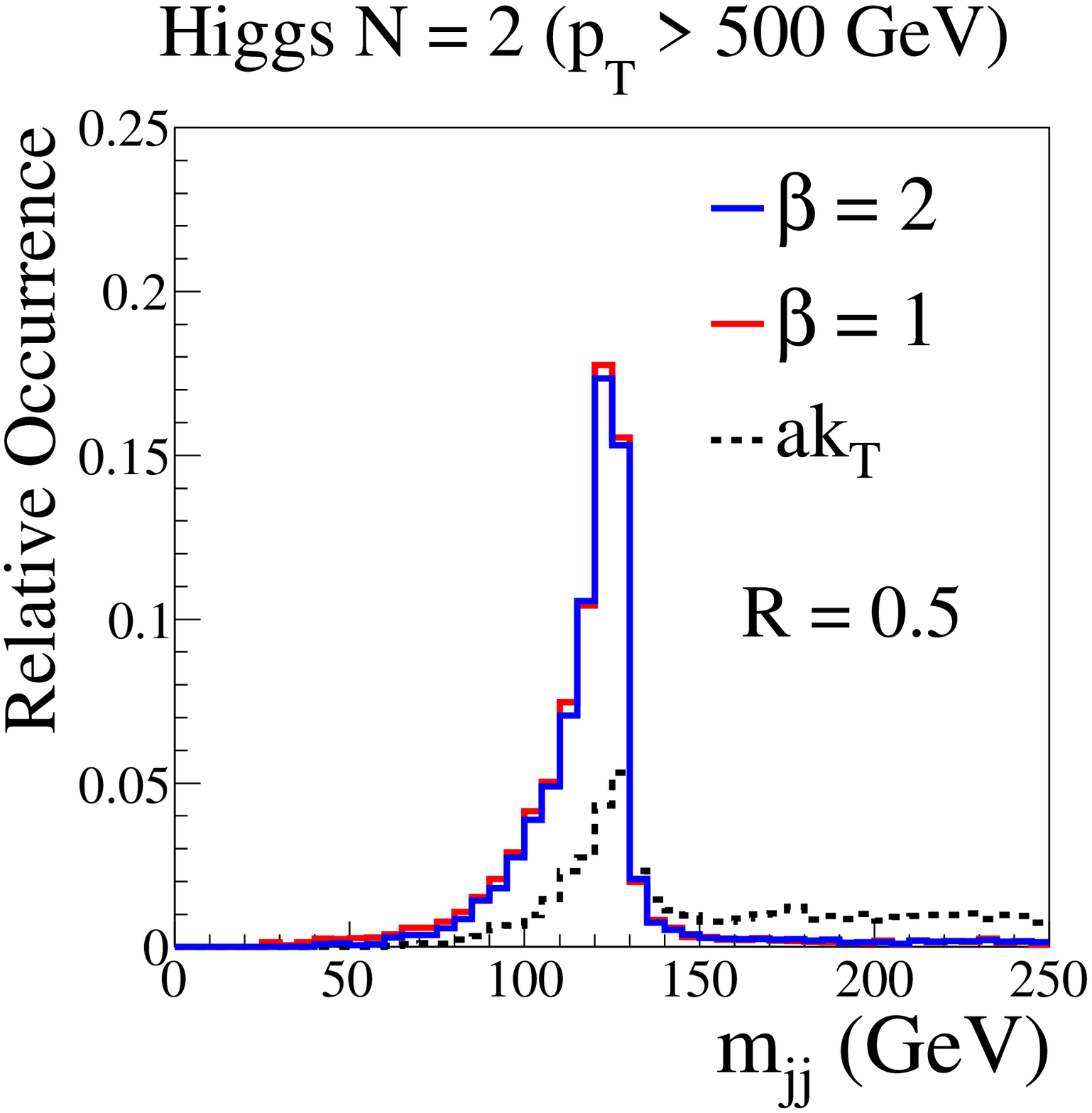}
\label{fig:higgs_2jet_mass_500}
}
\subfloat[]{
\includegraphics[width=0.32\columnwidth]{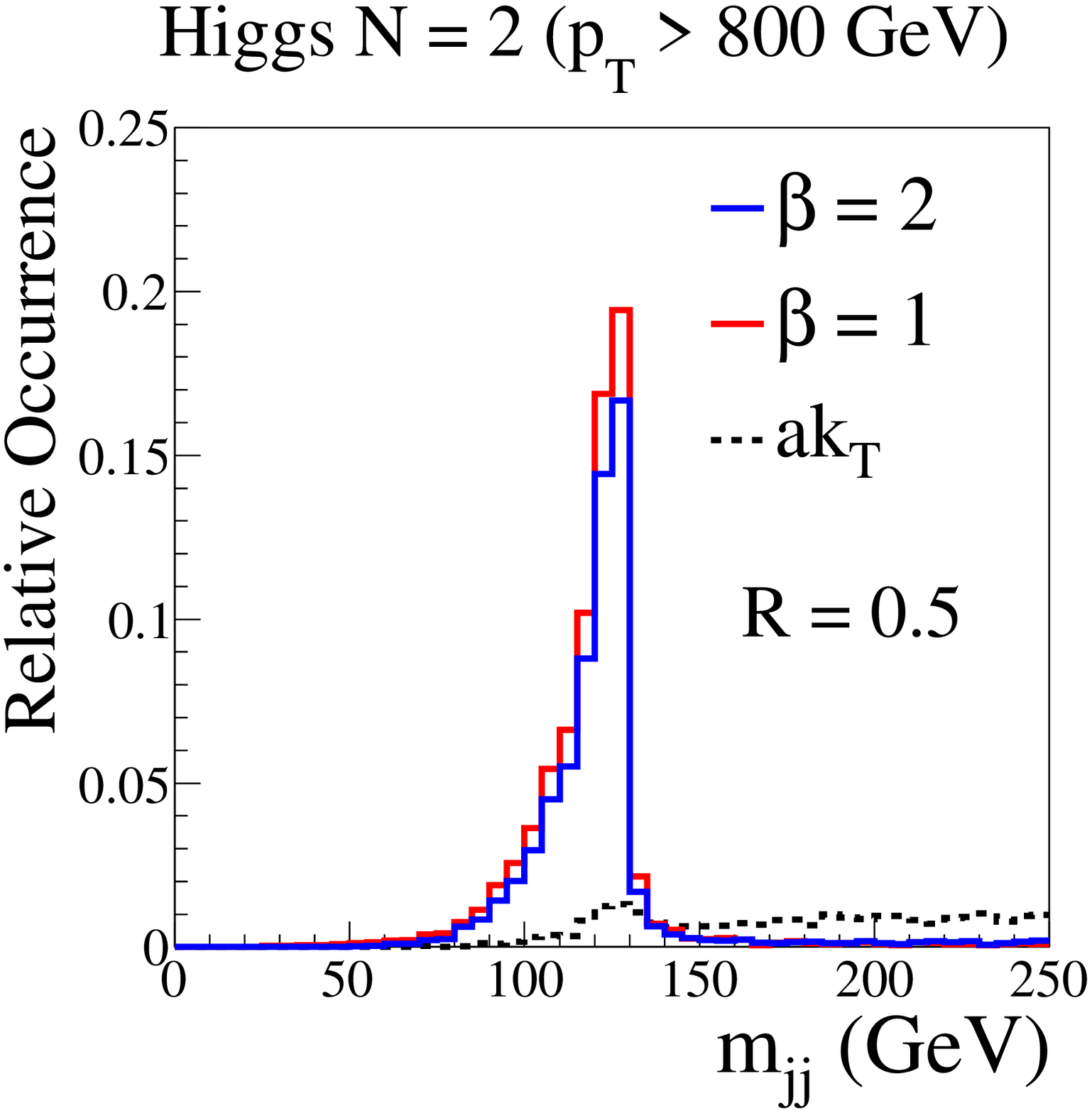}
\label{fig:higgs_2jet_mass_800}
}

\subfloat[]{
\includegraphics[width=0.32\columnwidth]{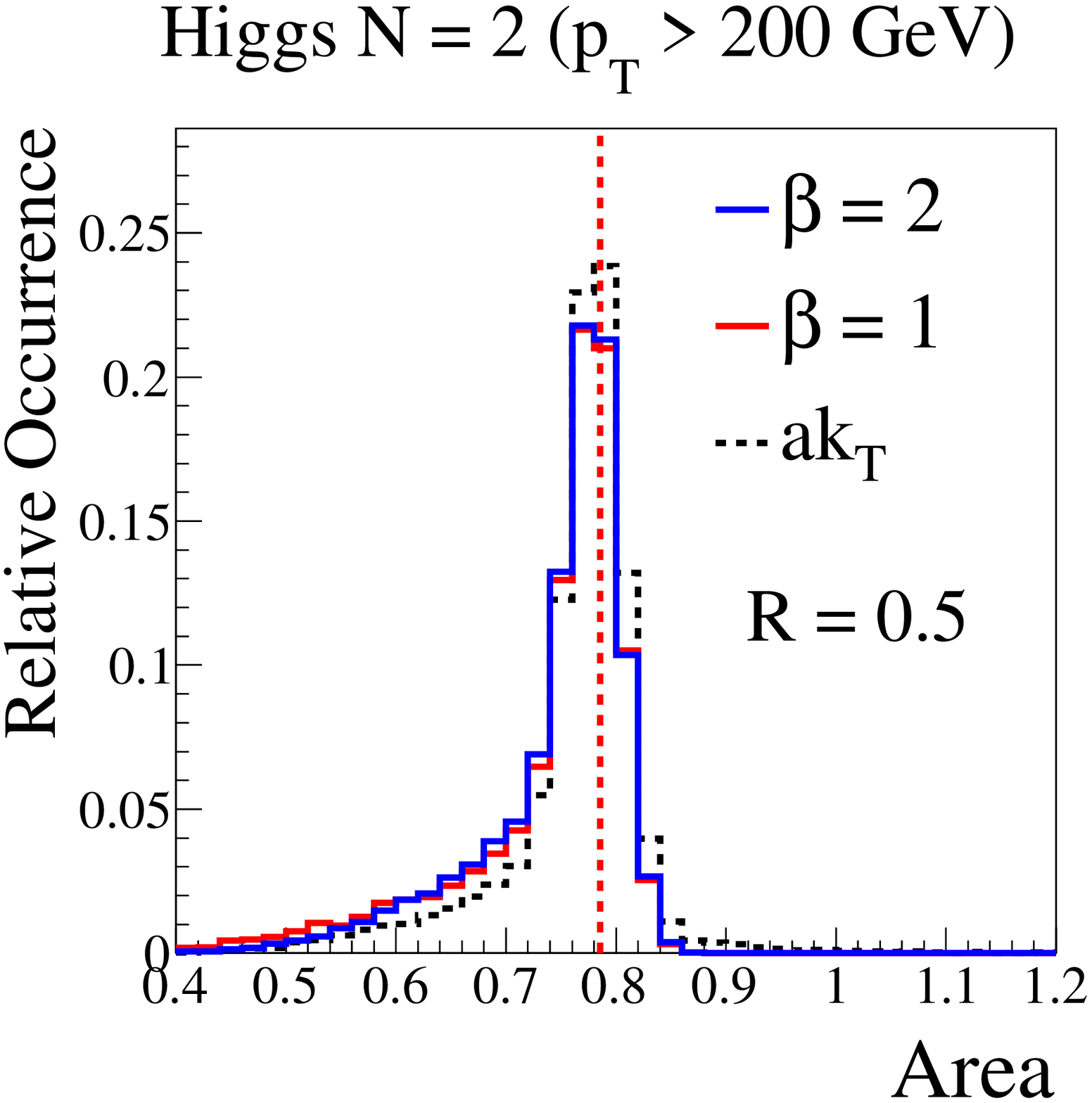}
\label{fig:higgs_2jet_area_200}
}
\subfloat[]{
\includegraphics[width=0.32\columnwidth]{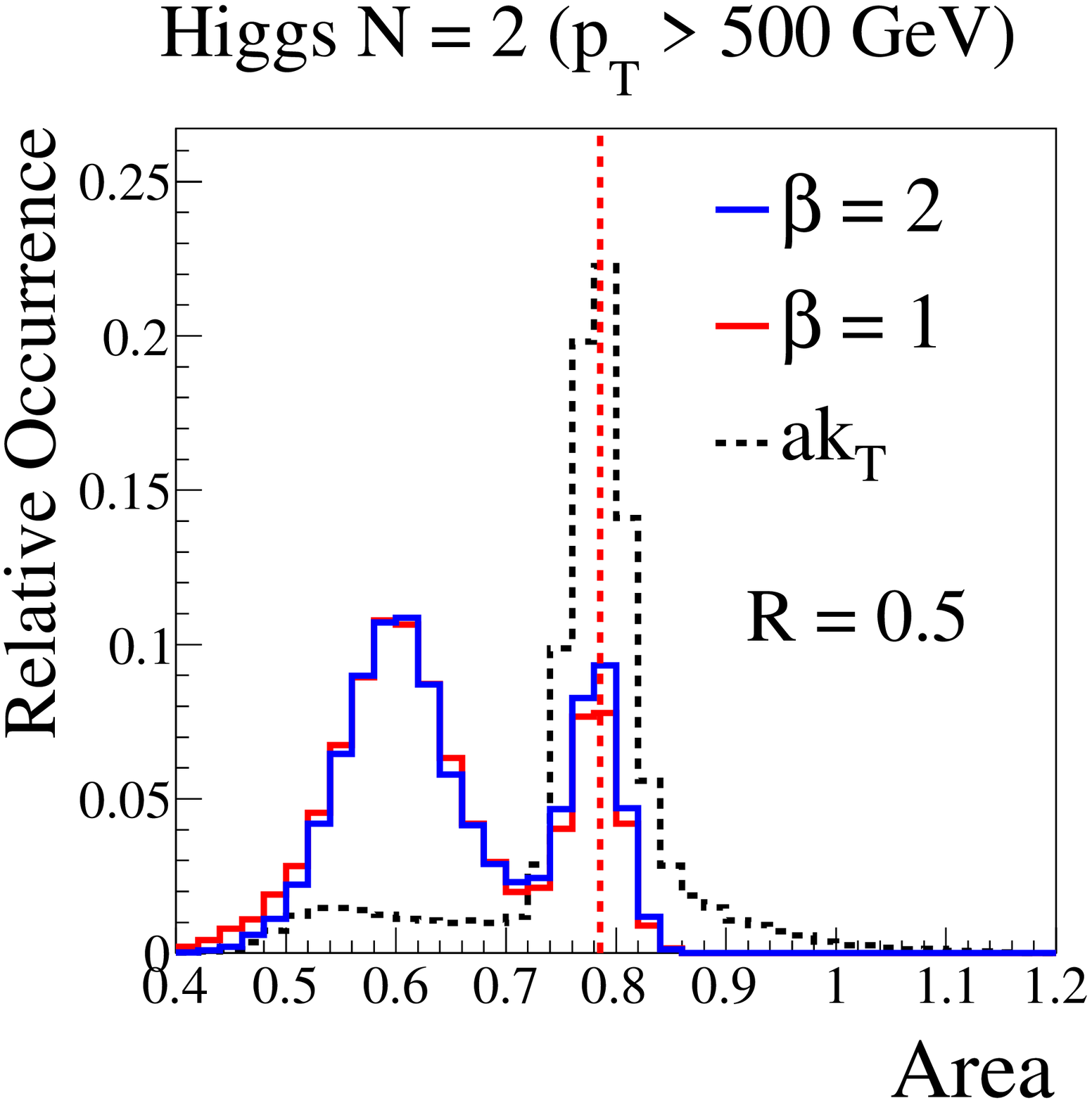}
\label{fig:higgs_2jet_area_500}
}
\subfloat[]{
\includegraphics[width=0.32\columnwidth]{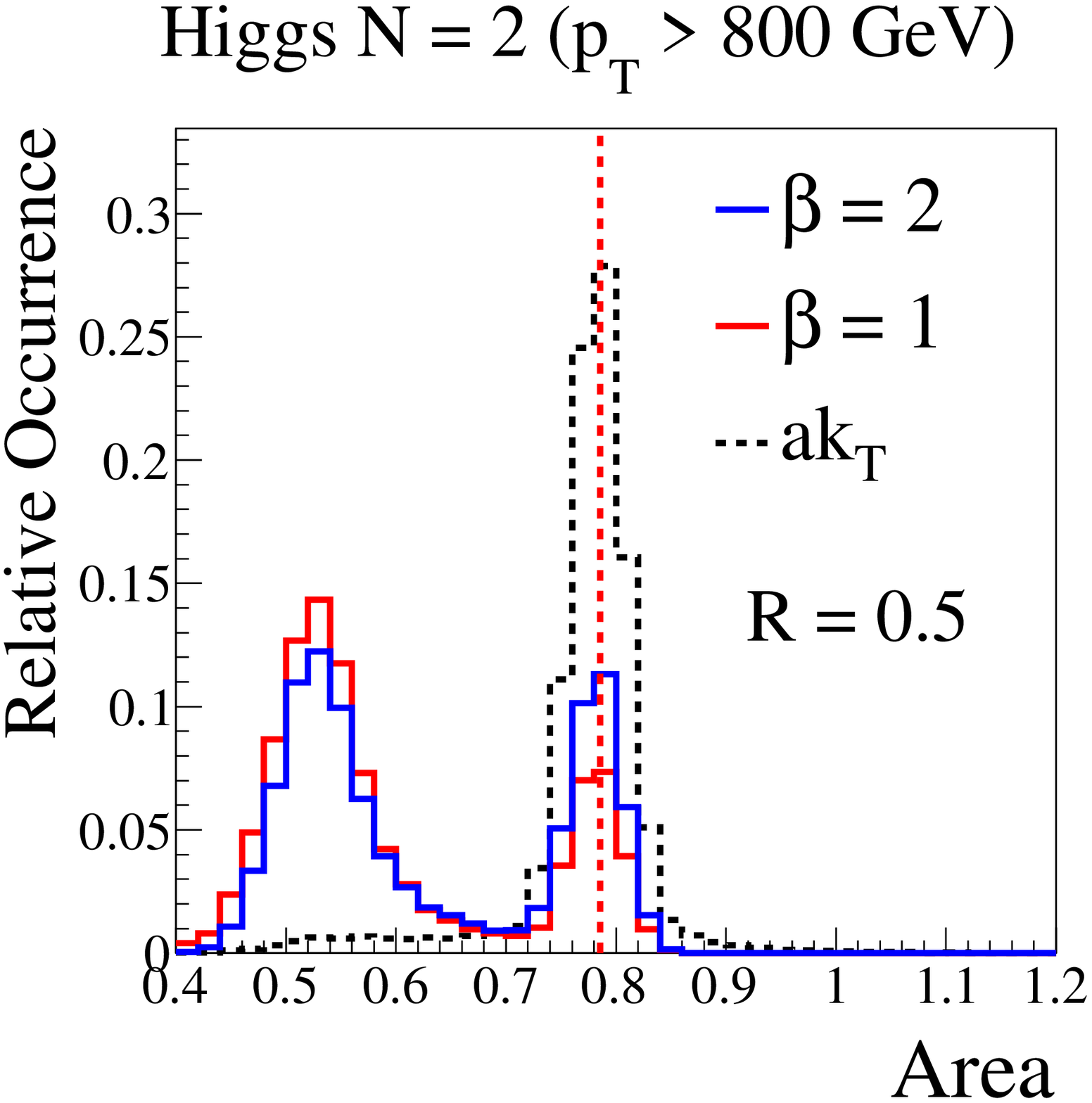}
\label{fig:higgs_2jet_area_800}
}
\caption{Same as \Fig{fig:higgs_1jet_study} but for $N = 2$.  As the Higgs $p_T$ increases from (a) 200 GeV to (b) 500 GeV to (c) 800 GeV, anti-$k_T$ suffers from jet merging, whereas XCone yields a dijet Higgs peak across the $p_{T}$ spectrum.   At low $p_{T}$, both distributions have the expected area peak at $A = \pi (0.5)^2$.  As $p_{T}$ increases, the XCone area falls to roughly half its original value, indicative of overlapping jets. See \Fig{fig:higgs_2jet_kt_study} for a comparison to exclusive $k_T$.}
\label{fig:higgs_2jet_study}
\end{figure}

\begin{figure}
\subfloat[]{
\includegraphics[width = .5\columnwidth]{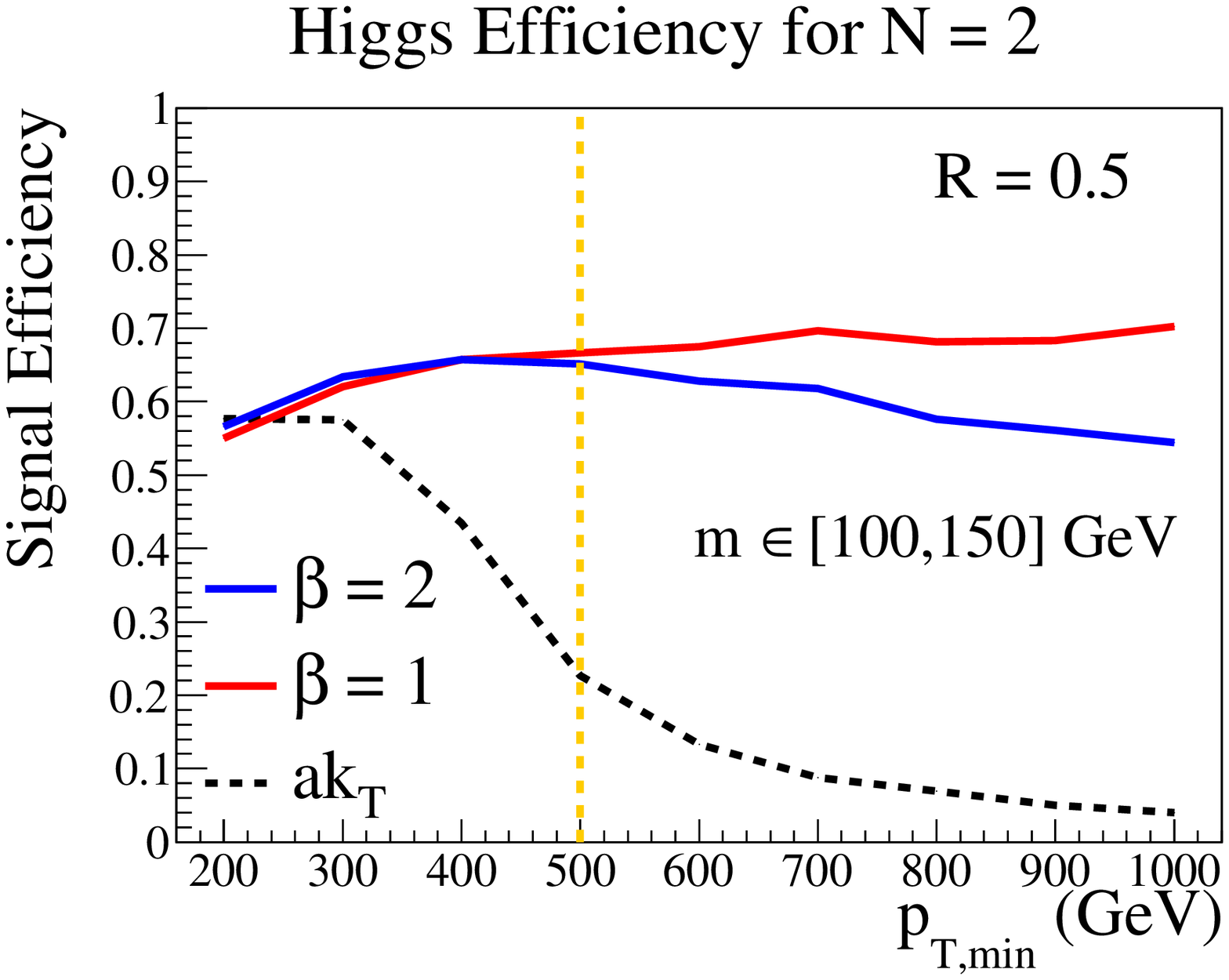}
\label{fig:higgs_2jet_eff}
}
\subfloat[]{
\includegraphics[width = .5\columnwidth]{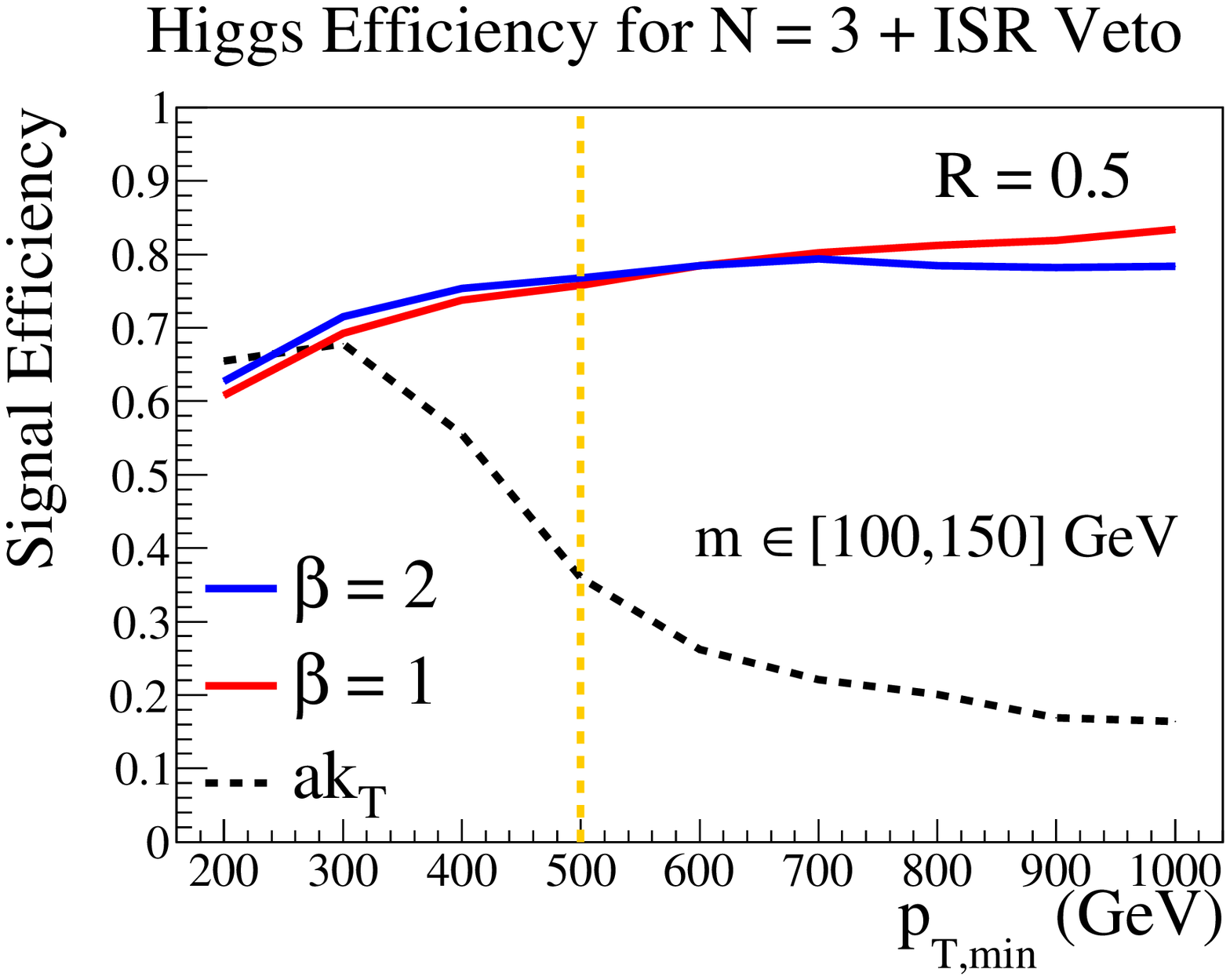}
\label{fig:higgs_2jet_eff_improved}
}
\caption{(a) Efficiency for $N = 2$ resolved dijet Higgs reconstruction as a function of Higgs $p_T$, with the mass window $m_{j} \in [100,150]~\GeV$.  We see a degradation in the efficiency of the anti-$k_{T}$ spectrum at higher $p_{T}$ due to jet merging, while XCone produces constant efficiency across the spectrum at around $65\%$.  Here, $\beta = 1$ outperforms $\beta = 2$, since the former is less susceptible to wide-angle jet contamination. (b) Same as \Fig{fig:higgs_2jet_eff}, but now allowing the Higgs to be reconstructed with either $N = 2$ or $N=3$, using the minimum pairwise mass to veto ISR.  Now, XCone $\beta = 1$ and $\beta = 2$ have comparable performance at around $75\%$. See \Fig{fig:higgs_2jet_kt_eff_both} for a comparison to exclusive $k_T$.}
\label{fig:higgs_2jet_eff_both}
\end{figure}

\subsection{N = 2 for Resolved Analysis}
\label{sec:higgs2}

For this Higgs production scenario, the real power of XCone comes from using $N = 2$.  In the unboosted regime, the standard analysis strategy is to find two $b$-jets, reconstruct their invariant mass, and look for a peak at the known Higgs mass \cite{Chatrchyan:2013zna,Aad:2014xzb}.  In the boosted regime with jet merging, though, algorithms like anti-$k_T$ are likely to find one fat Higgs jet and one ISR jet elsewhere in the event, so a dijet reconstruction strategy is no longer effective.  By contrast, since XCone is an exclusive cone algorithm, it will always identify two jets regardless of the Higgs $p_T$.   To find boosted Higgs bosons with XCone, we can simply run with $N = 2$ and perform a standard resolved jet analysis.

In the top row of \Fig{fig:higgs_2jet_study}, we show the reconstructed dijet invariant mass comparing  XCone with anti-$k_{T}$.  For low Higgs $p_T$, all algorithms find the Higgs peak with roughly the same line shape.\footnote{The low mass tail in each of the plots can be explained by neutrinos from $B$ meson decays within the $b$-jet.}  As the Higgs $p_T$ increases, the anti-$k_T$ distributions move to higher dijet masses because of a merged Higgs jet being paired with an ISR jet, whereas XCone maintains good performance regardless of $p_T$.  In the bottom row of \Fig{fig:higgs_2jet_study}, we show the jet area distributions.  Anti-$k_T$ jets peak at $\pi \Rnaught^2$ regardless of the Higgs $p_T$, whereas XCone jets transition from $\pi \Rnaught^2$ at low $p_T$ to roughly half that at high $p_T$, indicative of the desired split jet regions.

We show the Higgs reconstruction efficiency in \Fig{fig:higgs_2jet_eff} as a function of the Higgs $p_T$.  Anti-$k_{T}$ jets start to merge around $p_{T} = 300~\GeV$ and the Higgs efficiency drops significantly.  XCone has nearly flat efficiency as a function of Higgs $p_T$, even as the $p_T$ crosses beyond the $p_T^{\rm merge}$ scale.  At higher $p_T$ values, the $\beta = 2$ jets see a performance degradation, since the $\beta = 2$ jets are more influenced by ISR at wide angles.  The $\beta = 1$ jets are able to maintain their performance since the jet axes tend to always align with the momentum of the Higgs decay products.  Overall, the XCone reconstruction efficiency is around 65\% for $\Rnaught = 0.5$.

\subsection{N = 3 for ISR Vetoing}
\label{sec:Higgs3}

To improve the XCone performance, we have to account for ISR, which is the leading cause of misreconstruction.  In the presence of hard ISR, XCone can identify an ISR jet instead of finding one of the two $b$ jets.  To address this issue, we can explicitly identify the ISR jet using $N = 3$ and find the best reconstruction among the $N = 2$ and $N = 3$ options.\footnote{An alternative approach is to use a modified beam measure (such as the $\gamma = 2$ option discussed in \Ref{Stewart:2015waa}) to preferentially select central jets.}

We first run XCone with $N = 2$ and check whether the dijet mass is in the $m_{jj} \in [100,150]~\GeV$ window.  If not, we run XCone with $N = 3$ and apply the Higgs mass test on the pair of jets with the smallest invariant mass, as these are kinematically the most likely candidates to be the Higgs decay products.  Allowing two pathways for Higgs reconstruction gives improved signal efficiency, and in \Fig{fig:higgs_2jet_eff_improved}, we see that both $\beta = 1$ and $\beta = 2$ now have efficiencies around 75\%.  Applying the same 2- and 3-jet technique to anti-$k_T$ does improve the signal efficiency somewhat, though XCone still has better performance for $p_T \gtrsim 300~\GeV$.

We conclude that XCone is highly efficient in reconstructing Higgs bosons across a range of kinematics, from the resolved to quasi-boosted to boosted regimes.  Comparing \Fig{fig:higgs_1jet_eff} with \Fig{fig:higgs_2jet_eff_both}, we see that $N = 1$ does yield better signal efficiency at very high boosts, whereas $N = 2,3$ yields uniform (and still quite good) performance.  Ultimately, one may want to combine all three methods, although this would require a full understanding of background processes (see discussion in \App{app:boostedbackground}).  In a full analysis, one would also want to exploit $b$-tagging to better identify the Higgs candidate and mitigate non-$b$ backgrounds.  This is especially important when using the combined $N = 2,3$ method, where background events have two pathways to land in the signal window.  But the main take away from this study is that XCone allows traditional resolved analyses to be extended into the boosted regime, providing a $p_{T}$-independent method for Higgs reconstruction.  

\section{Boosted Top Quarks and High-Multiplicity Final States}
\label{sec:tops}

Given the success of XCone in reconstructing boosted Higgs bosons, we now test whether XCone can handle the increasingly complex final states possible at LHC collision energies.  An important process at the LHC is pair production of top quarks with fully hadronic decays:
\be
pp \rightarrow t\bar{t}, \qquad t \to W b \to q \bar{q}' b.
\ee
At low $m_{t\bar{t}}$, the final state consists of six resolved jets.  At high $m_{t\bar{t}}$, the jets are arranged into two fat jets with three-prong substructure, and a variety of substructure techniques have been developed to tag these boosted tops (see, e.g.~\cite{Brooijmans:1077731,Thaler:2008ju,Kaplan:2008ie,Almeida:2008yp}).  Here, we show that XCone with $N = 6$ can identify each of the six individual (sub)jets, regardless of the $m_{t\bar{t}}$ value, allowing the same analysis strategy to be effective in both the resolved and boosted regimes (see \App{app:exclusive_top} for similar behavior from exclusive $k_T$).  We also show a more efficient $N = 2 \times 3$ method where the event is first partitioned into hemispheres using $N=2$ and then separated into subjets by applying $N = 3$ in each hemisphere.  

\begin{figure}
\centering
\subfloat[]{
\includegraphics[width = 0.45\columnwidth]{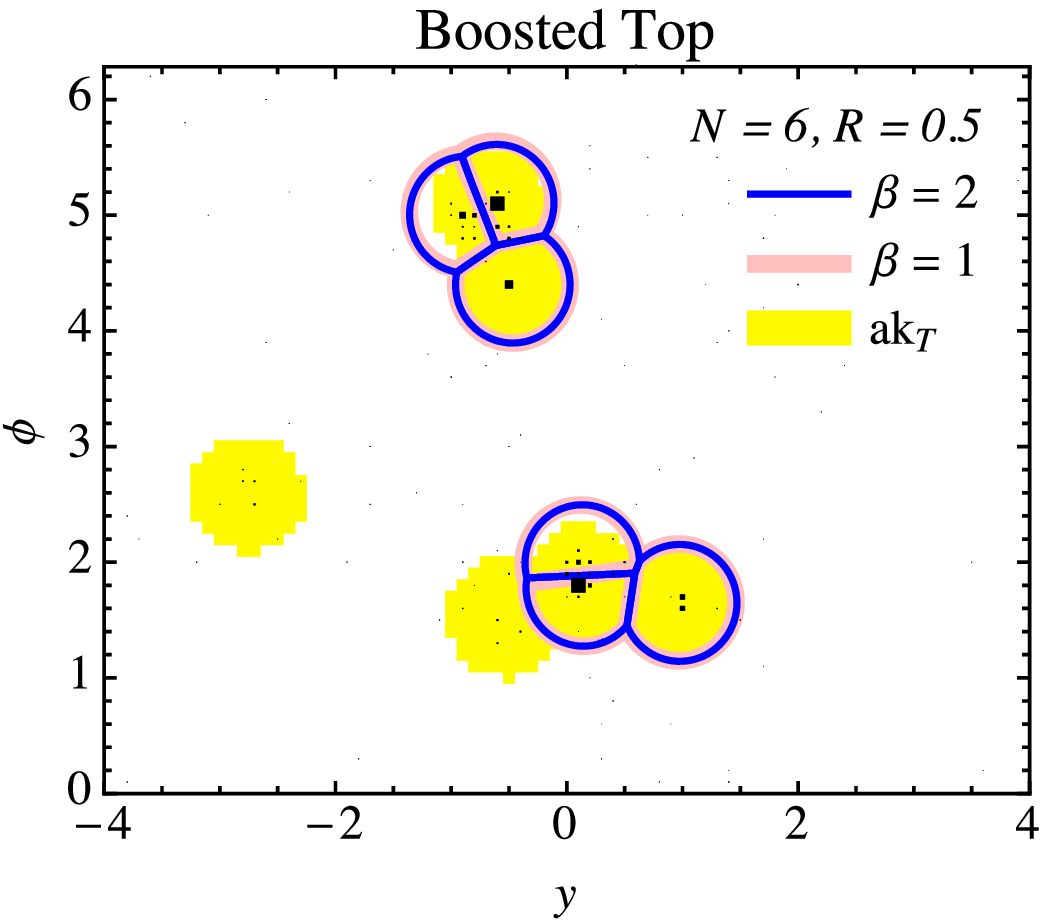}
\label{fig:top_display_compare:a}}
$\qquad$
\subfloat[]{
\includegraphics[width = 0.45\columnwidth]{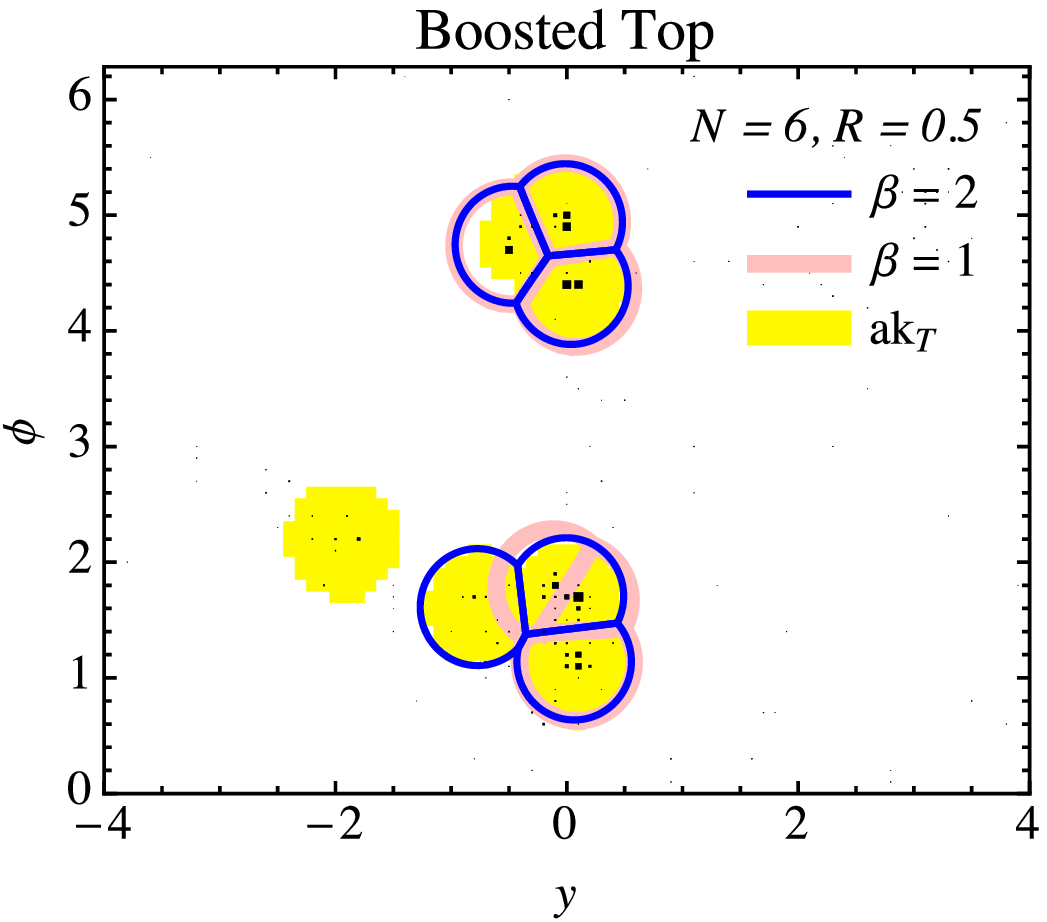}
\label{fig:top_display_compare:b}}
\caption{Two reconstructed events from the BOOST 2010 top sample, using XCone with $N = 6$ and $\Rnaught = 0.5$.  As is often the case, $N=6$ and $N = 2 \times 3$ give identical results for these events.  Compared to anti-$k_T$, XCone directly identifies three prong substructure through the initial jet finding.  While $\beta = 2$ and $\beta = 1$ often give similar jet regions, they can differ more substantially, as shown on the right.}
\label{fig:top_display_compare}
\end{figure}

\begin{figure}
\centering
\subfloat[]{
\includegraphics[width = 0.40\columnwidth]{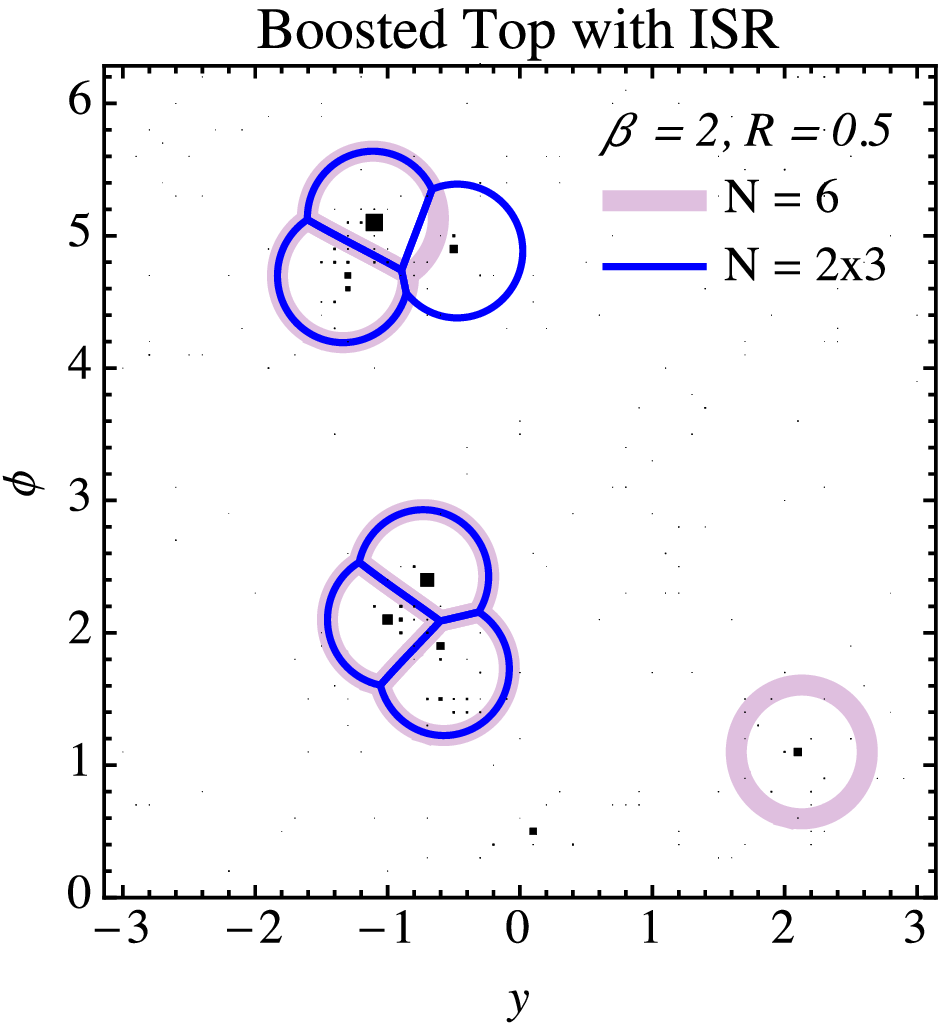}
\label{fig:top_display:a}}
$\qquad$
\subfloat[]{
\includegraphics[width = 0.40\columnwidth]{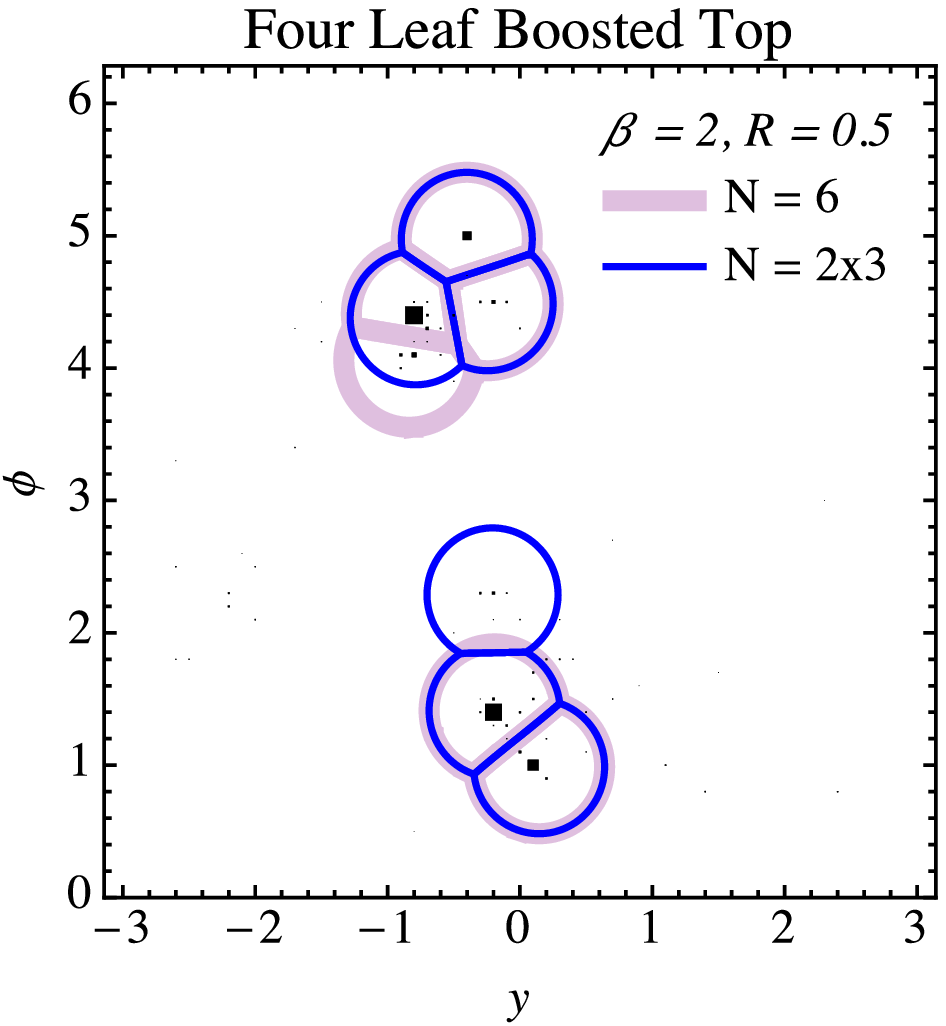}
\label{fig:top_display:b}}
\caption{Pathological $N=6$ reconstructions that are improved by using $N = 2 \times 3$.  For these hand selected events, $N = 2 \times 3$ correctly identifies the boosted tops, while $N=6$ identifies (a) an additional hard jet from ISR and (b) a fourth leaf in one of the boosted tops.  Not shown are examples where the $N = 2 \times 3$ method finds an ISR jet, which can sometimes be resolved using $N = 2 \times 4$ or $N = 7$.}
\label{fig:top_display}
\end{figure}

Our study is based on the BOOST 2010 events samples \cite{Abdesselam:2010pt}, which were generated for the 7 TeV LHC.\footnote{While the BOOST 2011 report \cite{Altheimer:2012mn} included updated benchmark samples, those event files were lost due to ``spring cleaning'' at the host servers.  Instead of generating fresh boosted top samples at 14 TeV collision energies, we have decided to use the BOOST 2010 event samples to enable easier comparisons and verifications of our results.}  For the boosted top signal, we use the Herwig $t\bar{t} \rightarrow \mathrm{hadrons}$ samples where the generator-level top $p_T$ ranges from $200$--$800$ GeV in bins of $100$ GeV.  We also apply XCone to the Herwig dijet background sample in the same $p_{T}$ bins.  As in the boosted Higgs study, we take $\Rnaught = 0.5$.  When comparing to traditional fat jet studies, we use anti-$k_T$ jets with $\Rnaught = 1.0$ as recommended in the BOOST 2010 report \cite{Abdesselam:2010pt}.  For brevity, we do not include a straight $N=2$ fat jet study for XCone, since the results are similar to those found in \Sec{sec:dijet}.  Example event displays from XCone are shown in \Figs{fig:top_display_compare}{fig:top_display}.

At the outset, we want to emphasize that XCone is able to handle partially overlapping jets, as expected in the quasi-boosted regime.  In the highly boosted limit, however, the subjets are fully overlapping, so substructure methods based on fat jets are typically more effective at signal/background separation.  While it is possible to combine XCone with jet shapes like $N$-subjettiness \cite{Thaler:2010tr,Thaler:2011gf} for improved performance in the highly boosted limit, we find in preliminary studies that there is no real advantage to using $N=6$ over a more traditional fat jet analysis with $N=2$.  The key advantage of XCone is that it yields relatively uniform performance over a broad $p_T$ range, and while specialized techniques can achieve better performance at extreme kinematics, XCone allows resolved techniques to be applied even when jets are overlapping.

\subsection{N = 6 Baseline Analysis}
\label{sec:Top6}

\begin{figure}
\centering
\subfloat[]{
\includegraphics[width = 0.45\columnwidth]{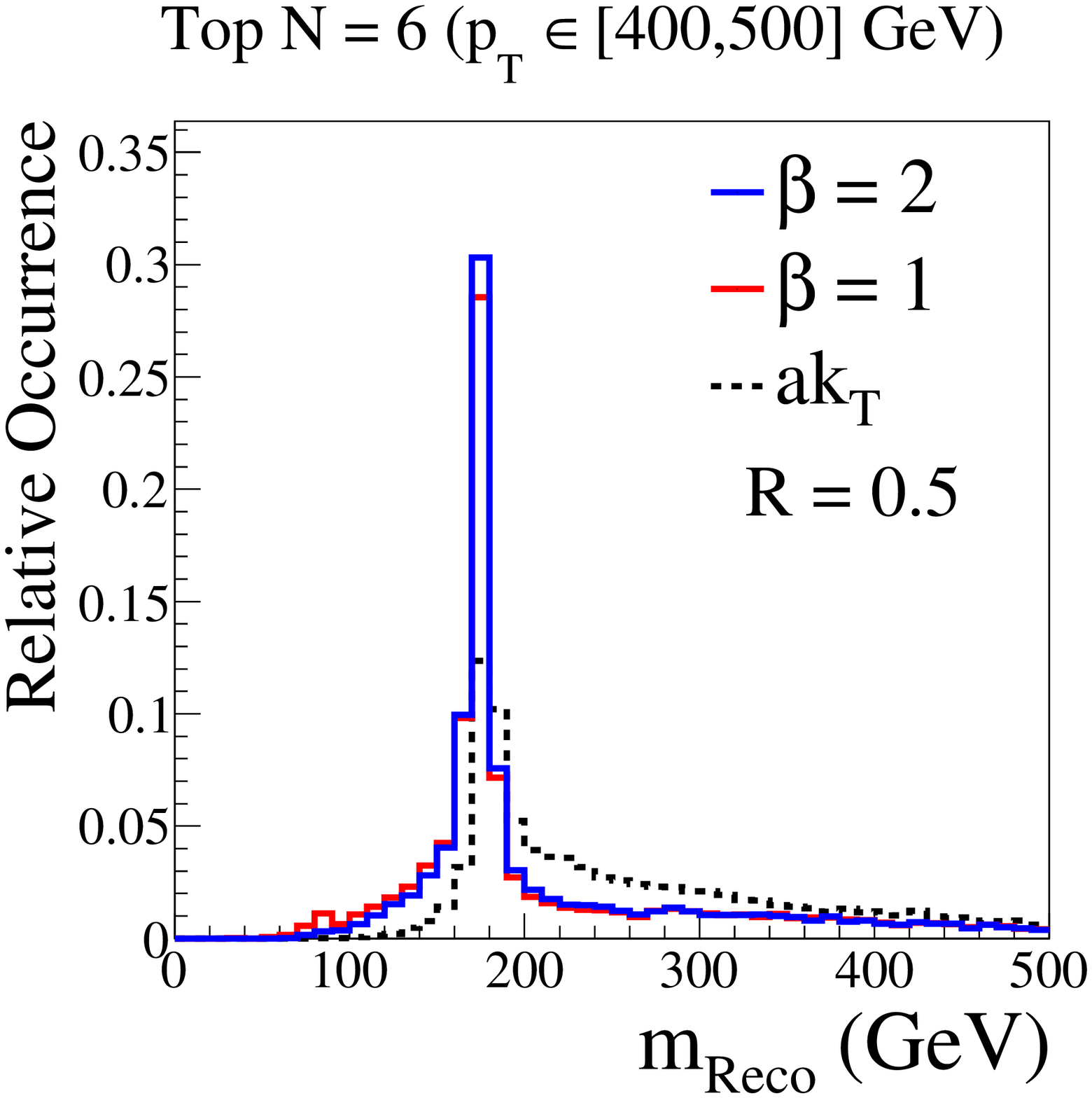}
\label{fig:top_ttbar_rawmass_400}}
$\quad$
\subfloat[]{
\includegraphics[width = 0.45\columnwidth]{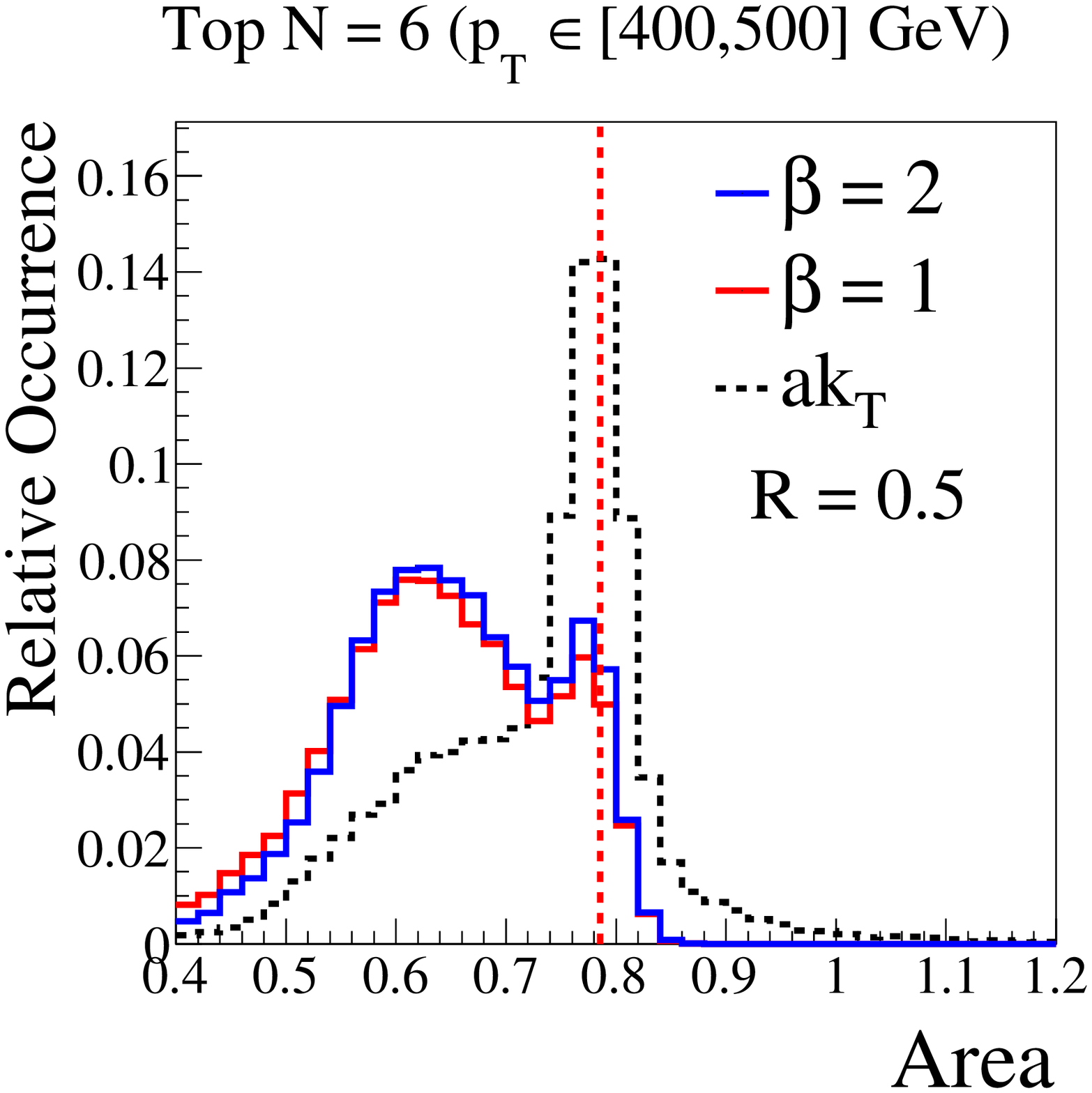}
\label{fig:top_ttbar_area_400}}

\subfloat[]{
\includegraphics[width = 0.45\columnwidth]{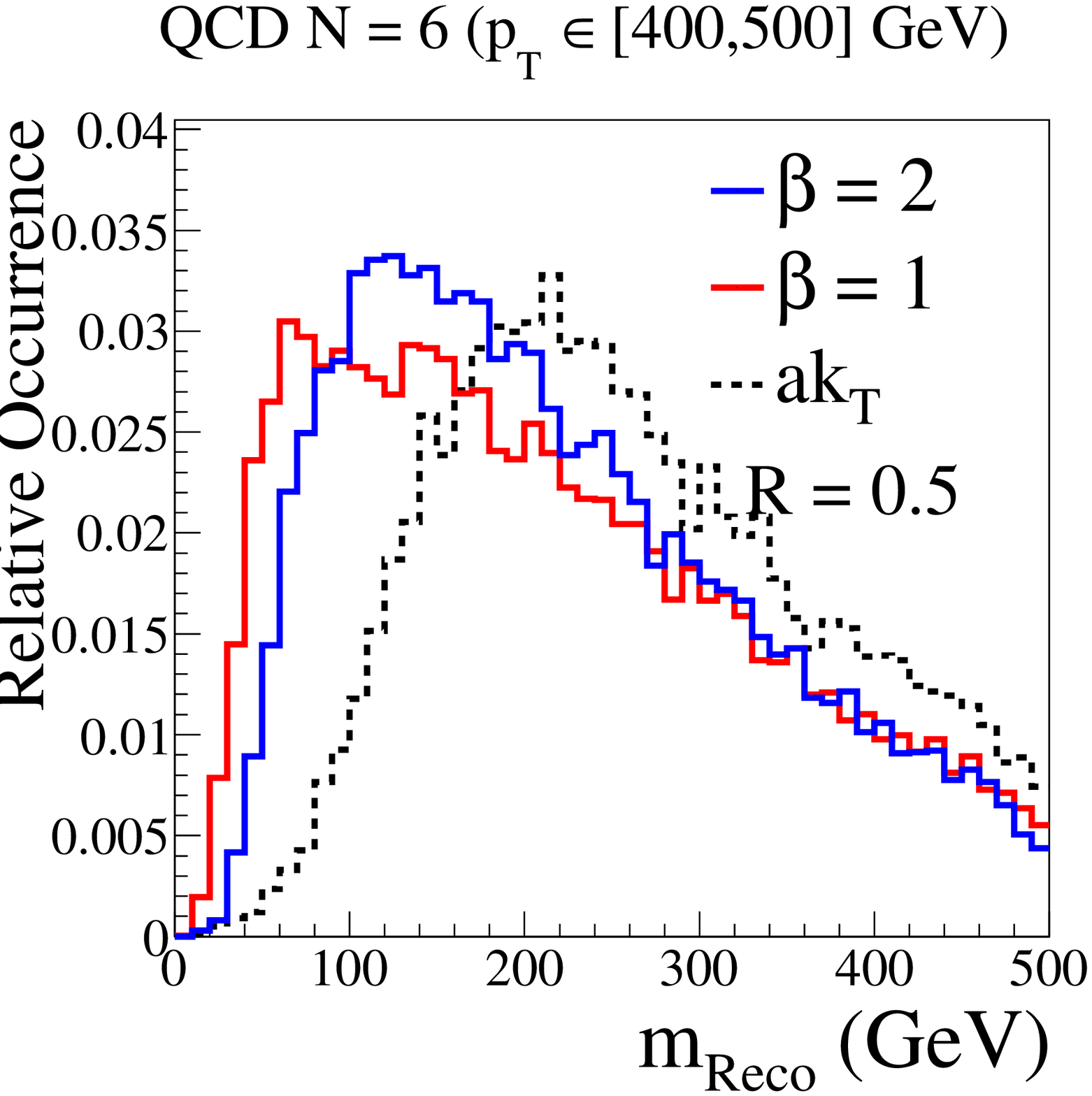}
\label{fig:top_qcd_rawmass_400}}
$\quad$
\subfloat[]{
\includegraphics[width = 0.45\columnwidth]{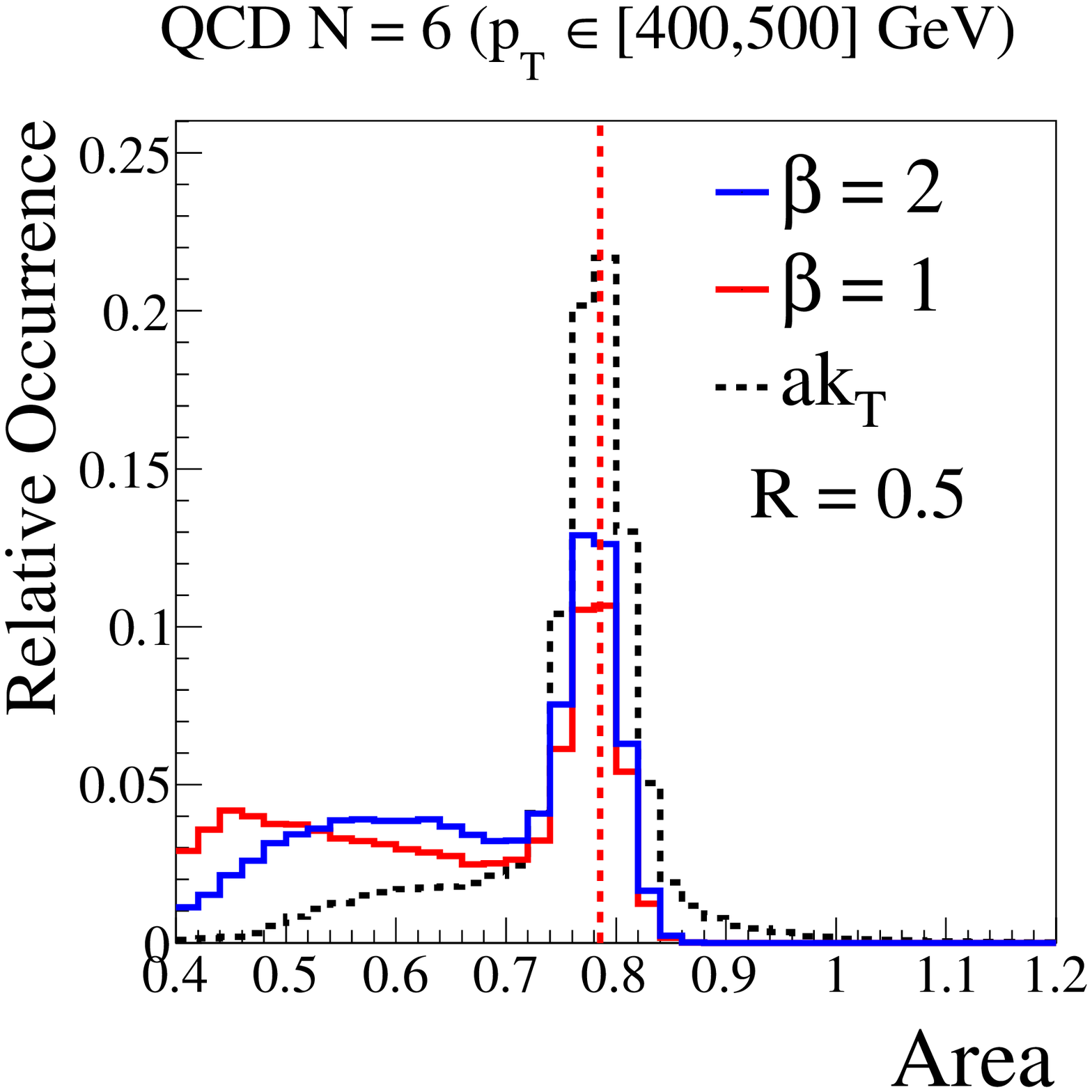}
\label{fig:top_qcd_area_400}}
\caption{Top row:  Comparing resolved three-jet top reconstruction between XCone with $N = 6$ and the six hardest anti-$k_T$ jets, in the $p_T \in [400,500]~\GeV$ bin.  Here, top candidates are identified by minimizing the sum of the three-jet masses.  (a)  Candidate top mass distributions, showing that XCone does not have as pronounced of a high mass tail due to ISR.  (b)  Area of all six jets, showing a peak at $(2/3) \pi \Rnaught^2$  for XCone expected of clover jet configuration compared to a peak at  $\pi \Rnaught^2$ for anti-$k_T$ expected of separated jets.   Bottom row:  Same for the QCD background.  See \Fig{fig:top_kt_6jet_study} for a comparison to exclusive $k_T$.}
\label{fig:top_6jet_study}
\end{figure}

\begin{figure}
\centering
\subfloat[]{
\includegraphics[width = 0.32\columnwidth]{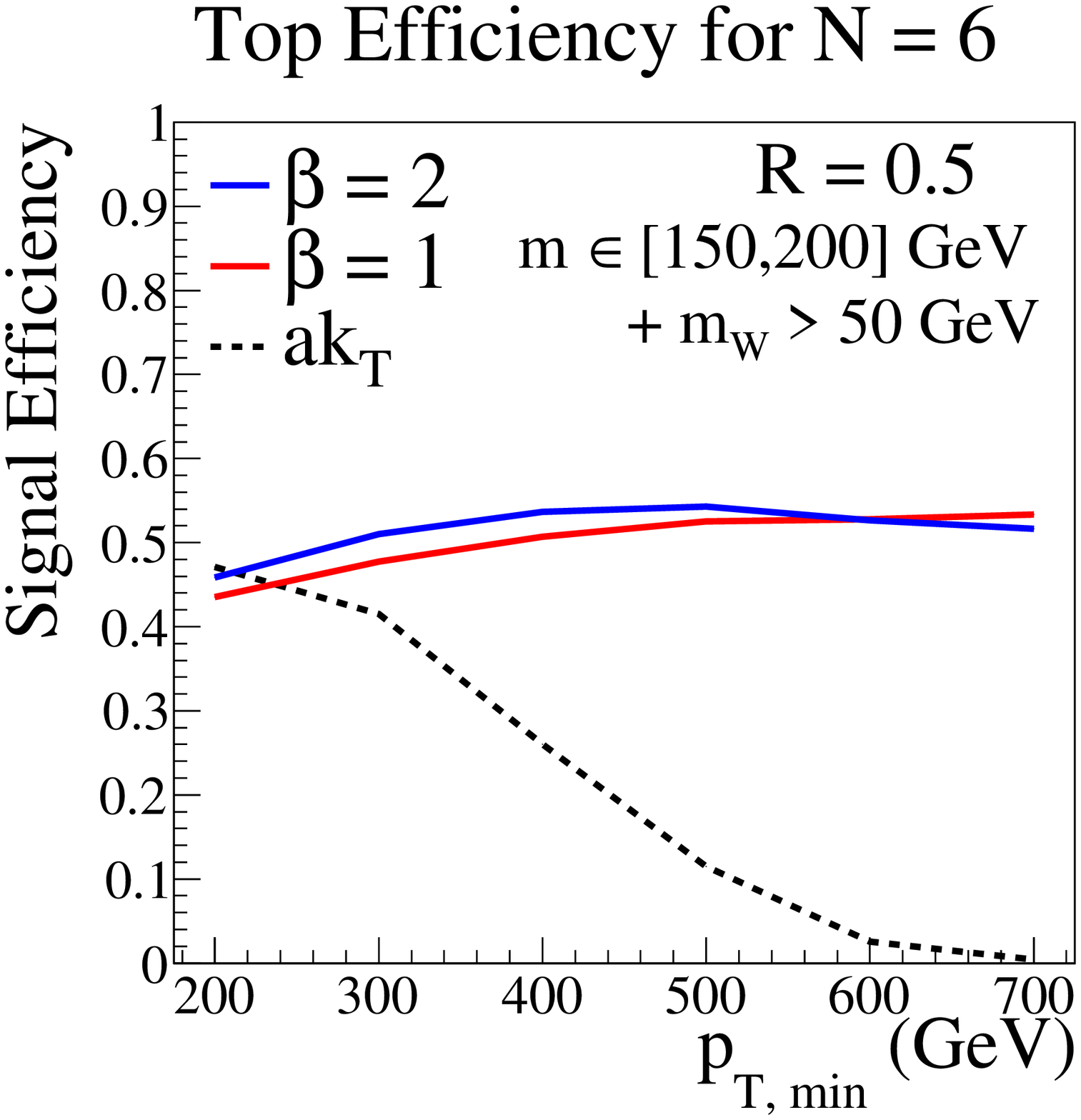}
\label{fig:top_ttbar_raweff}}
\subfloat[]{
\includegraphics[width = 0.32\columnwidth]{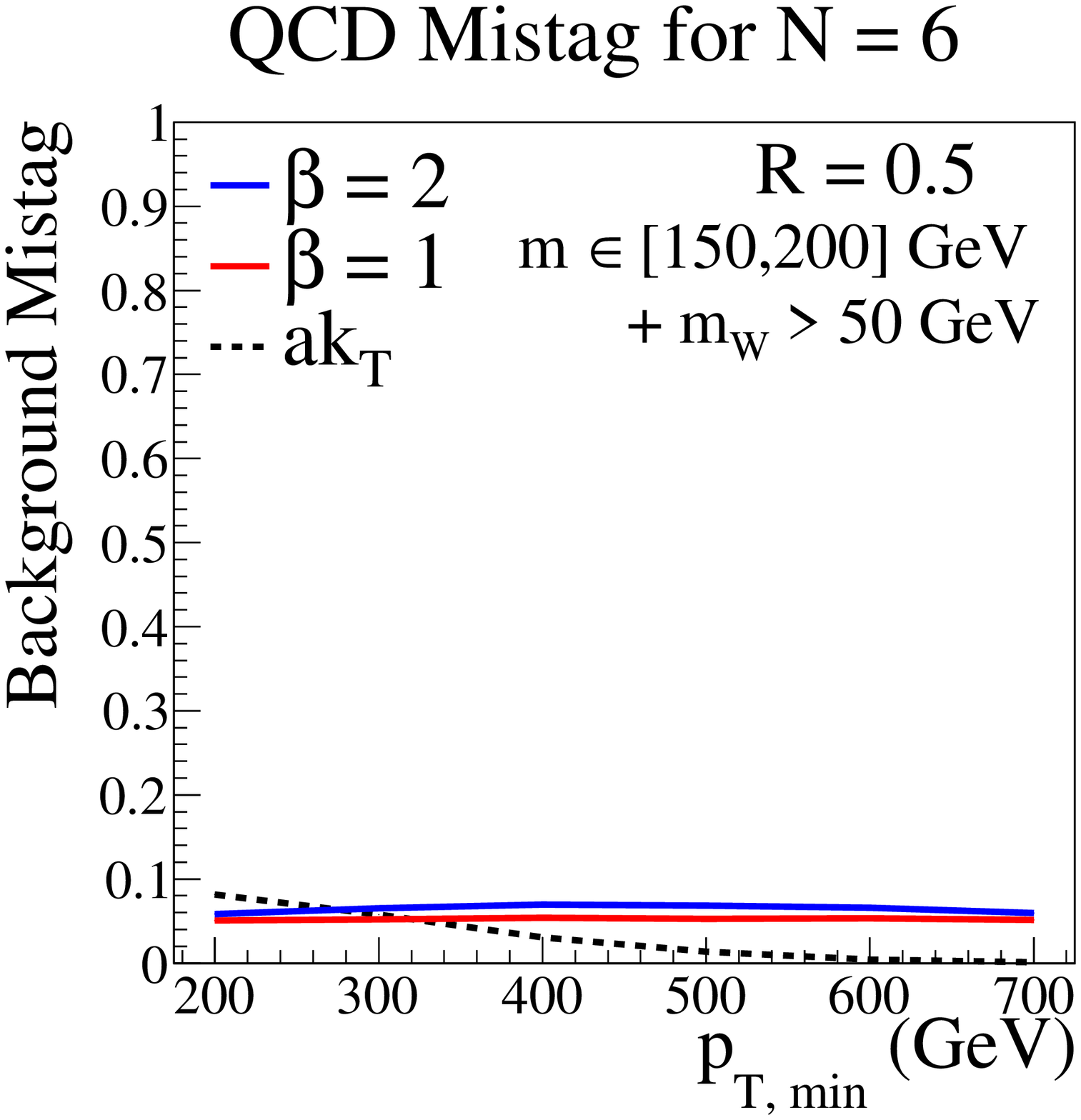}
\label{fig:top_qcd_raweff}}
\subfloat[]{
\includegraphics[width = 0.32\columnwidth]{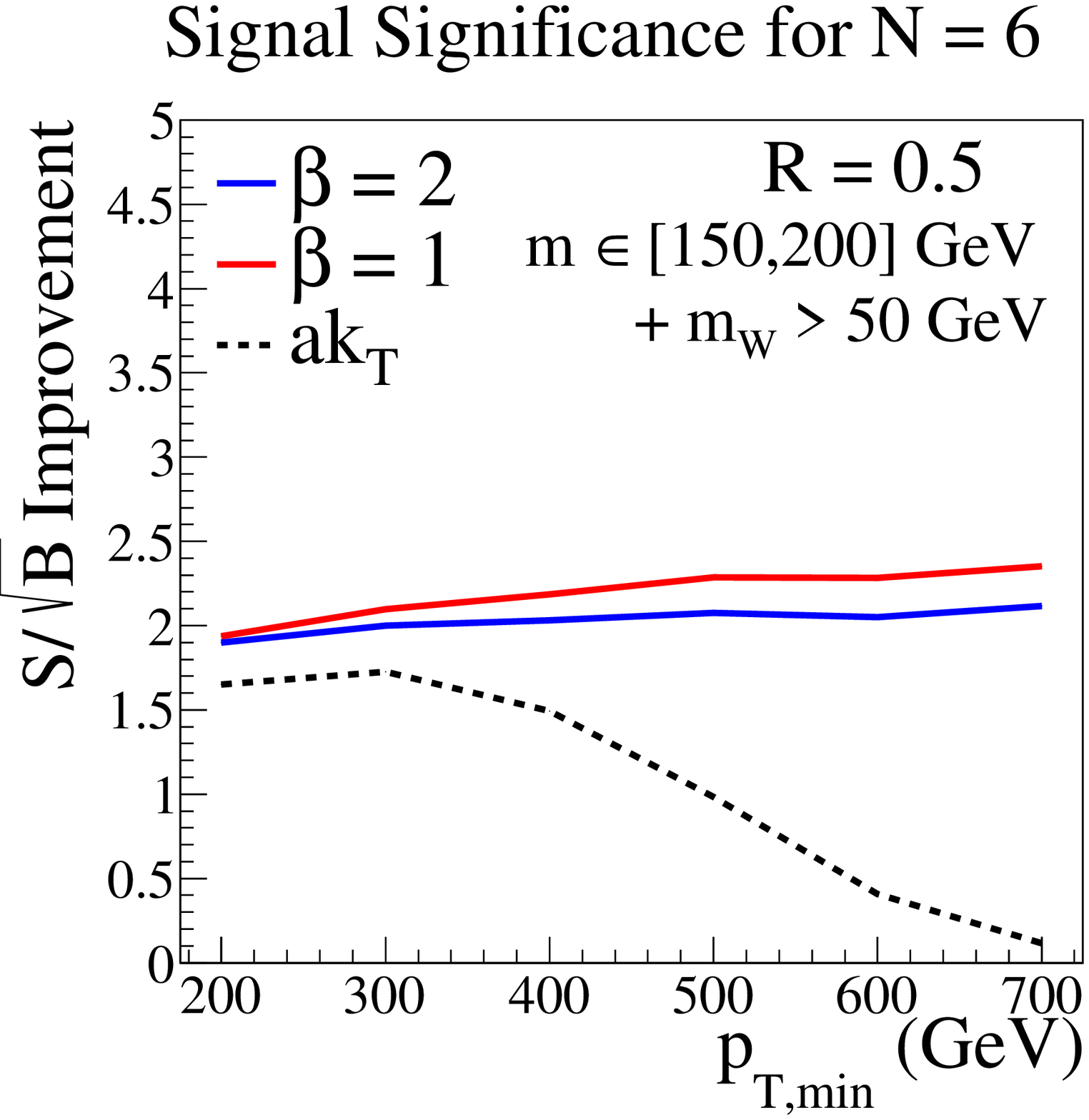}
\label{fig:top_signalsig_Wmasscut}}
\caption{Comparing boosted top and XCone performance with $N = 6$ as a function of top $p_T$ for (a) signal efficiency, (b) background mistag, and (c) signal significance gain.  By avoiding jet mergers, XCone has improved performance compared to anti-$k_{T}$ across the entire $p_{T}$ range.  See \Fig{fig:top_kt_efficiency_study} for a comparison to exclusive $k_T$.}
\label{fig:top_efficiency_study}
\end{figure}

The most straightforward application of XCone for hadronic tops is using $N = 6$ to resolve six jets.  Like in \Sec{sec:Higgs3}, one can try to improve the performance by explicitly identifying ISR jets using $N=7$, but we find that the $N = 2 \times 3$ method shown below is more effective at dealing with the combinatorial complexity of this final state.  For these studies, we have not incorporated $b$-tagging information, though it would be straightforward to use XCone in conjunction with recent subjet $b$-tagging methods \cite{CMS:2013vea,ATL-PHYS-PUB-2014-013}.

After running XCone with $N = 6$, we want to partition the jets into two top candidates in a way that is $p_{T}$-independent.  We do this by finding all ${}_6C_3 = 20$ ways of partitioning the jets into two three-jet clusters, and then finding the configuration that minimizes the scalar sum of the three-jet masses.  Much like in the boosted Higgs case, we expect that minimizing the mass is most likely to yield the correct top candidates.  For an apples-to-apples comparison, we apply the same analysis strategy on the six hardest anti-$k_T$ jets.  In a more sophisticated analysis, one could use a $\chi^2$-minimization approach to find the best top candidates, also incorporating $W$-mass and $b$-tagging information.

In \Fig{fig:top_ttbar_rawmass_400}, we show the reconstructed top jet mass in the $p_{T} \in [400,500]$ GeV bin, comparing XCone and anti-$k_T$ with $N = 6$.  The XCone distributions show a better resolved top peak, which is more symmetric around the top mass and has a substantially reduced high-mass tail.  In the area distributions in \Fig{fig:top_ttbar_area_400}, we see that XCone jets are peaked at roughly $(2/3) \pi \Rnaught^2$, where the factor of $2/3$ is expected since the jets are arranged in a clover configuration around the boosted top direction.  The anti-$k_{T}$ jets are peaked around $\pi \Rnaught^2$, as expected since anti-$k_T$ jets do not typically overlap.  Because of subjet mergers, the six found anti-$k_T$ jets are not all associated with the top quarks, leading to large invariant mass values from ISR jets.

The equivalent dijet background distributions are shown in \Figs{fig:top_qcd_rawmass_400}{fig:top_qcd_area_400}.  In the absence of genuine three-prong substructure, the XCone jets tend to be scattered throughout the event, leading to large reconstructed invariant masses and $\pi \Rnaught^2$ areas.  The effect is even stronger in the anti-$k_T$ jets, since they avoid overlaps.  There is a noticeable difference between the $\beta = 1$ and $\beta = 2$ QCD dijet distributions, indicating that the mean versus median effects described in \Sec{sec:Higgs1} are significant for quark and gluon jets without well-defined substructure (a feature exploited in \Ref{Curtin:2012rm}). 

To define the top signal region, we take a top mass window of $m_{jjj} \in [150, 200]~\GeV$.\footnote{Compared to the typical mass ranges used in the BOOST 2010 report  \cite{Abdesselam:2010pt}, this range is smaller and more symmetric.  Because the peak from XCone is more narrow and symmetric around the top mass, we can use this tighter mass window without much loss in signal efficiency.}  We also apply a $W$-tagging cut as described in the CMS analysis \cite{CMS:2014fya}, by analyzing each pairwise combination of the three subjects and requiring a minimum pairwise invariant mass cut of $m_{jj,\min} > 50~\GeV$.  In \Figs{fig:top_ttbar_raweff}{fig:top_qcd_raweff} we show the efficiency and mistag rates for the top and dijet samples as a function of (generator-level) $p_T$.  XCone has a signal efficiency of around 60\% across the entire $p_{T}$ range, showing the desired scale invariance.   While anti-$k_T$ starts with the same 60\% efficiency in the resolved regime, the efficiency drops considerably with $p_T$ due to jet merging, analogous to what was found in \Sec{sec:higgs2}.  Both methods have around a 10\% background mistag rate, which is relatively stable as a function of $p_T$. 

The improvement in signal significance $(S/\sqrt{B})$ is shown in \Fig{fig:top_signalsig_Wmasscut}, where we see that the performance remains relatively flat across the entire $p_{T}$ range, with performance comparable to or better than anti-$k_{T}$.  Note that we have not included $b$-tagging information nor additional jet shape information, so background rejection factors can be much larger in practice.  From this study, we see that a simple application of XCone allows for a $p_T$-independent analysis strategy even for complicated final states.

\subsection{N = 2 $\times$ 3 Improved Analysis}
\label{sec:Top23}

To further improve on the performance of XCone, we can take into account the event topology.  Even with a moderate boost, the top decay products tend to arrange themselves into two hemispheres, a feature that is exploited, for example, in the HEPTogTagger \cite{Plehn:2010st} (see also \cite{Plehn:2009rk}).  Thus, we can use XCone in multiple stages, first dividing the event into separate top candidate regions with $N = 2$ and $\Rnaught \to \infty$, and then finding jets in each of those regions using $N = 3$ and $\Rnaught = 0.5$.\footnote{For a theoretical analysis of the top mass using a related hemisphere approach, see refs. \cite{Fleming:2007xt,Fleming:2007qr}.}

There are two advantages of this $N = 2 \times 3$ approach over the $N = 6$ approach.  First, it reduces combinatorial confusion and increases computational efficiency.  Second, it ensures that each top candidate has the potential to involve three jets.  Even without ISR (as in \Fig{fig:top_display:a}), the $N = 6$ method can yield one four-leaf top clover and one two-leaf top clover (as in \Fig{fig:top_display:b}), something that is avoided with $N = 2 \times 3$.  While it is possible to get even higher signal efficiencies by applying $N = 2 \times 4$ and vetoing ISR jets (analogous to \Sec{sec:Higgs3}), such an approach tends to also increase the background mistag rate, so we will not show $N = 2 \times 4$ results here.

We can compare XCone to traditional top reconstruction methods in two different ways.  For a traditional boosted strategy (``Bst''), we can run anti-$k_T$ with $\Rnaught = 1.0$ to find two fat jets and then run exclusive $k_T$ with $N = 3$ and $\Rnaught = 0.5$ on the fat jet constituents to identify subjets.  For a traditional resolved strategy (``Res''), we can run exclusive $k_T$ with $N = 2$ and $\Rnaught \to \infty$ to find hemisphere regions, and then run anti-$k_T$ with $\Rnaught = 0.5$ to find the three hardest jets in each hemisphere.  As we will see, XCone $N = 2 \times 3$ effectively interpolates between these behaviors as a function of $p_T$, reproducing (and sometimes surpassing) the best performance of the traditional strategies in their respective domains.  To highlight the advantages of XCone, we will be working in the regime of high signal acceptance; in the regime of high background rejection, it is well known that anti-$k_T$-based boosted strategies are highly effective when combined with jet substructure methods.

\begin{figure}
\centering
\subfloat[]{
\includegraphics[width = 0.45\columnwidth]{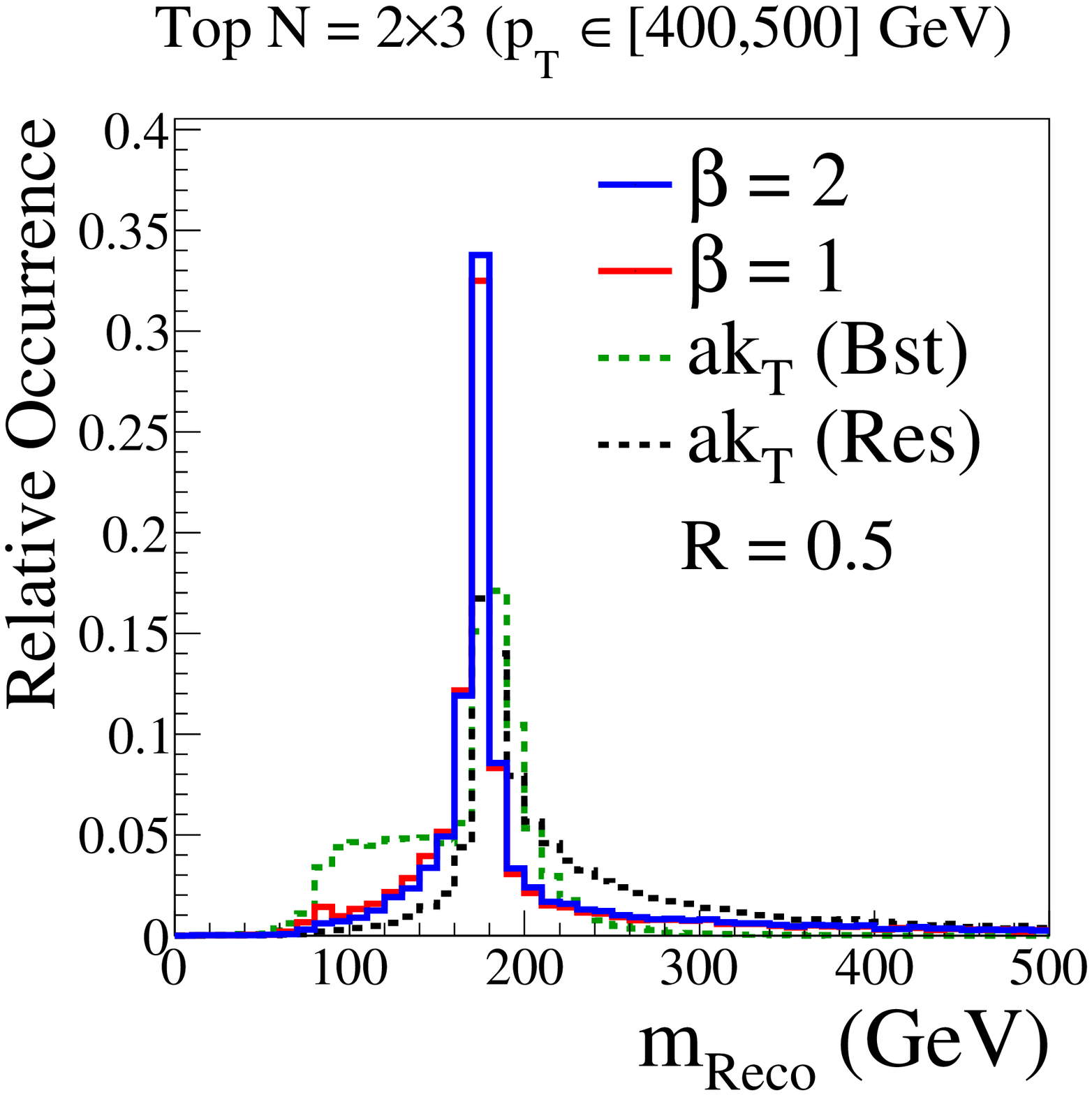}
\label{fig:top_ttbar_mass_2groups}}
$\quad$
\subfloat[]{
\includegraphics[width = 0.45\columnwidth]{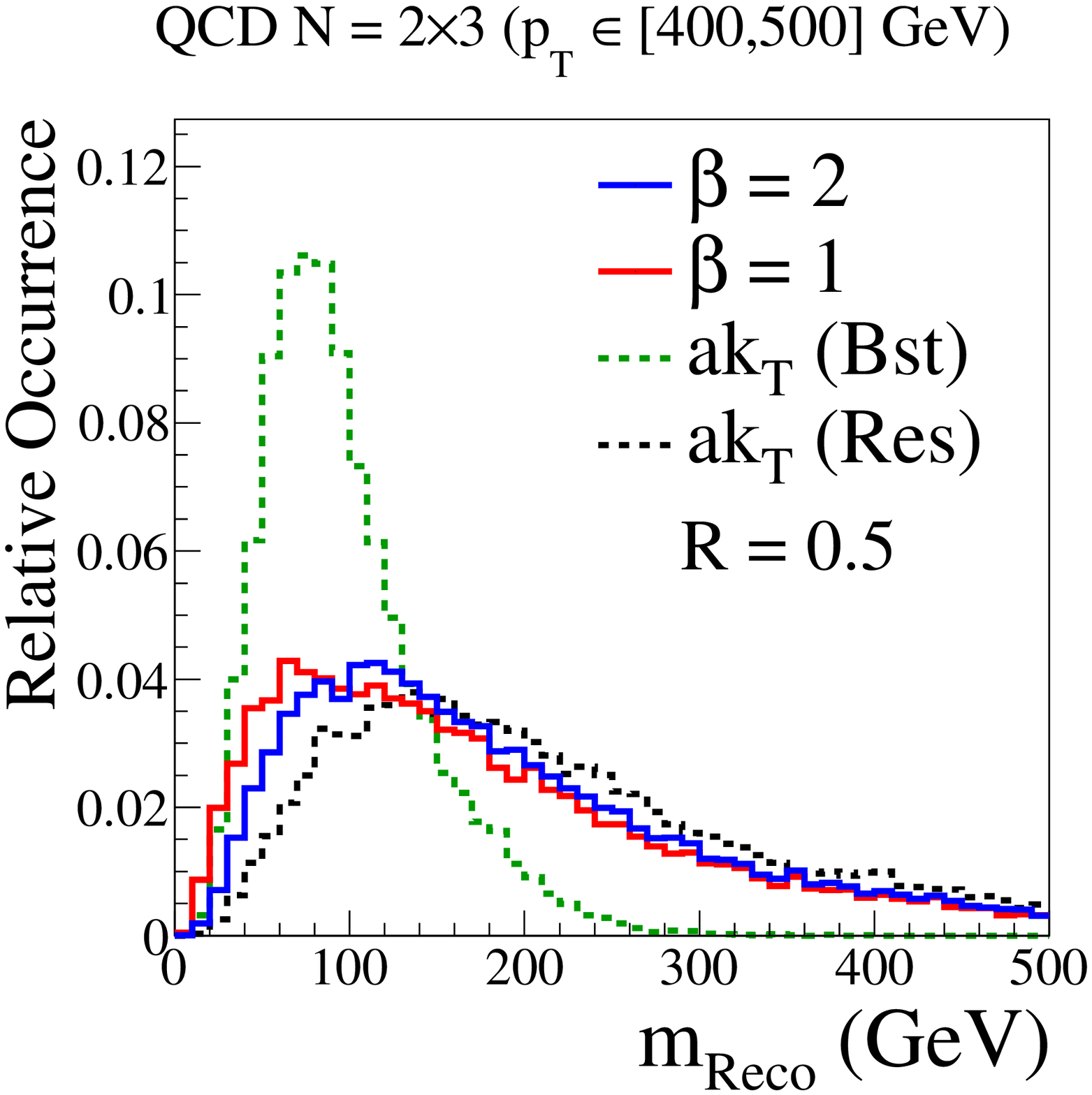}
\label{fig:top_qcd_mass_2groups}}
\caption{Reconstructed mass distributions with the $N = 2\times3$ strategy, for (a) top signal events and (b) QCD background events. Here we compare the XCone $N = 2\times3$ method to two traditional methods: a boosted strategy (``Bst'') where two anti-$k_T$ $\Rnaught = 1.0$ jets have three exclusive $k_T$ subjets, and a resolved strategy (``Res'') where two exclusive $k_T$ hemispheres have three anti-$k_T$ $\Rnaught = 0.5$ jets. }
\end{figure}

\begin{figure}
\centering
\subfloat[]{
\includegraphics[width = 0.32\columnwidth]{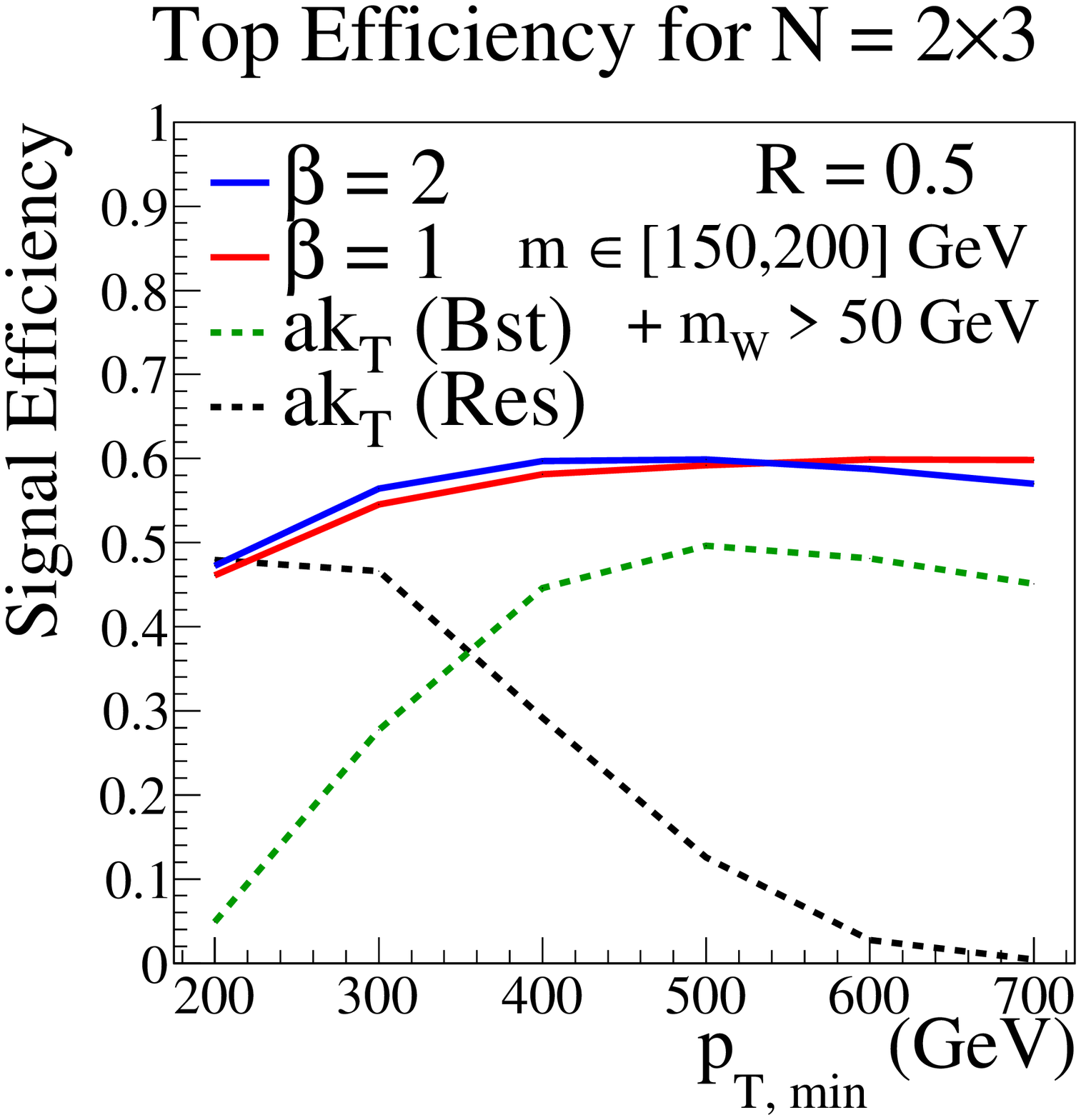}
\label{fig:top_ttbar_eff_2groups_Wmasscut}}
\subfloat[]{
\includegraphics[width = 0.32\columnwidth]{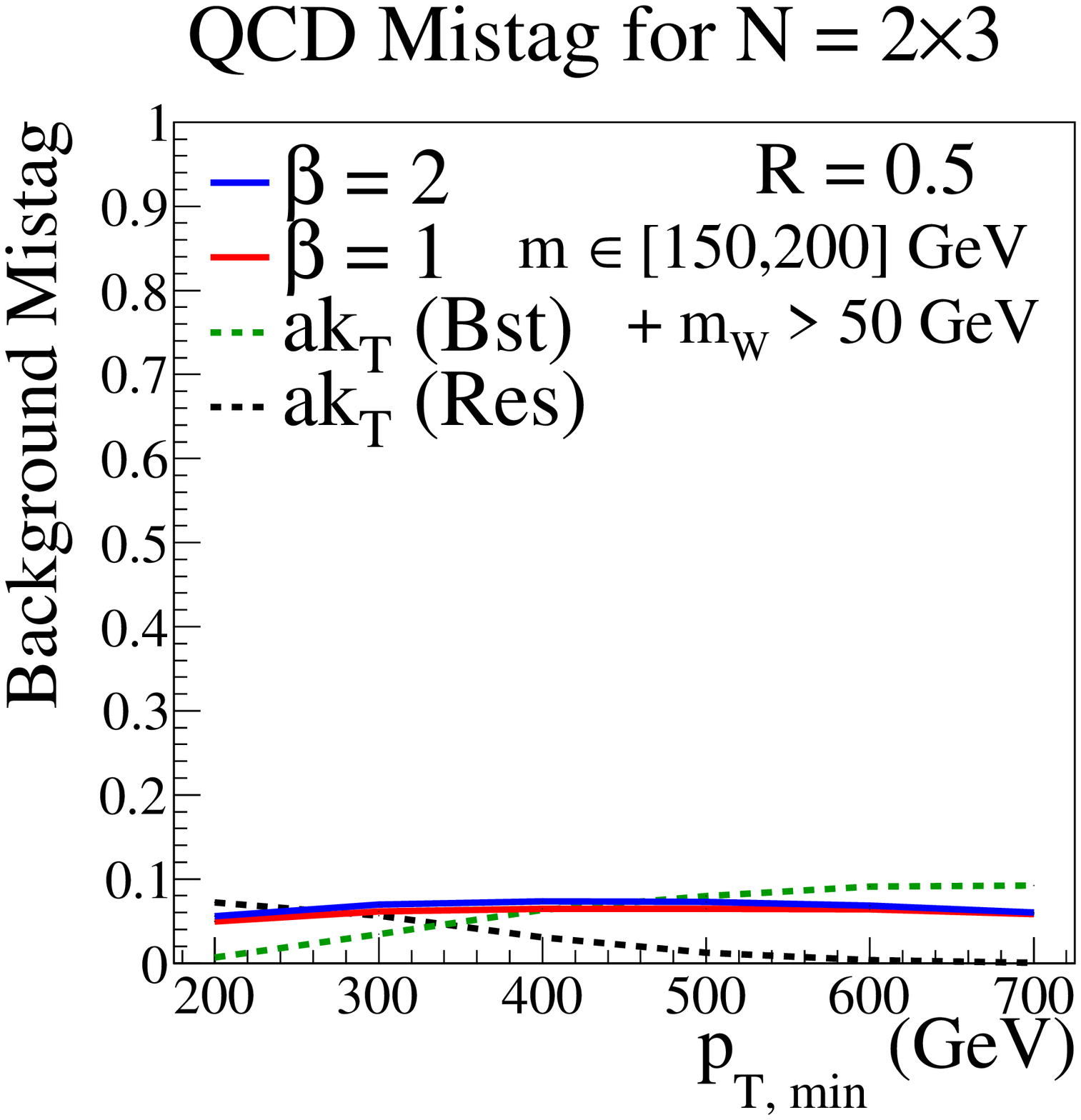}
\label{fig:top_qcd_eff_2groups_Wmasscut}}
\subfloat[]{
\includegraphics[width = 0.32\columnwidth]{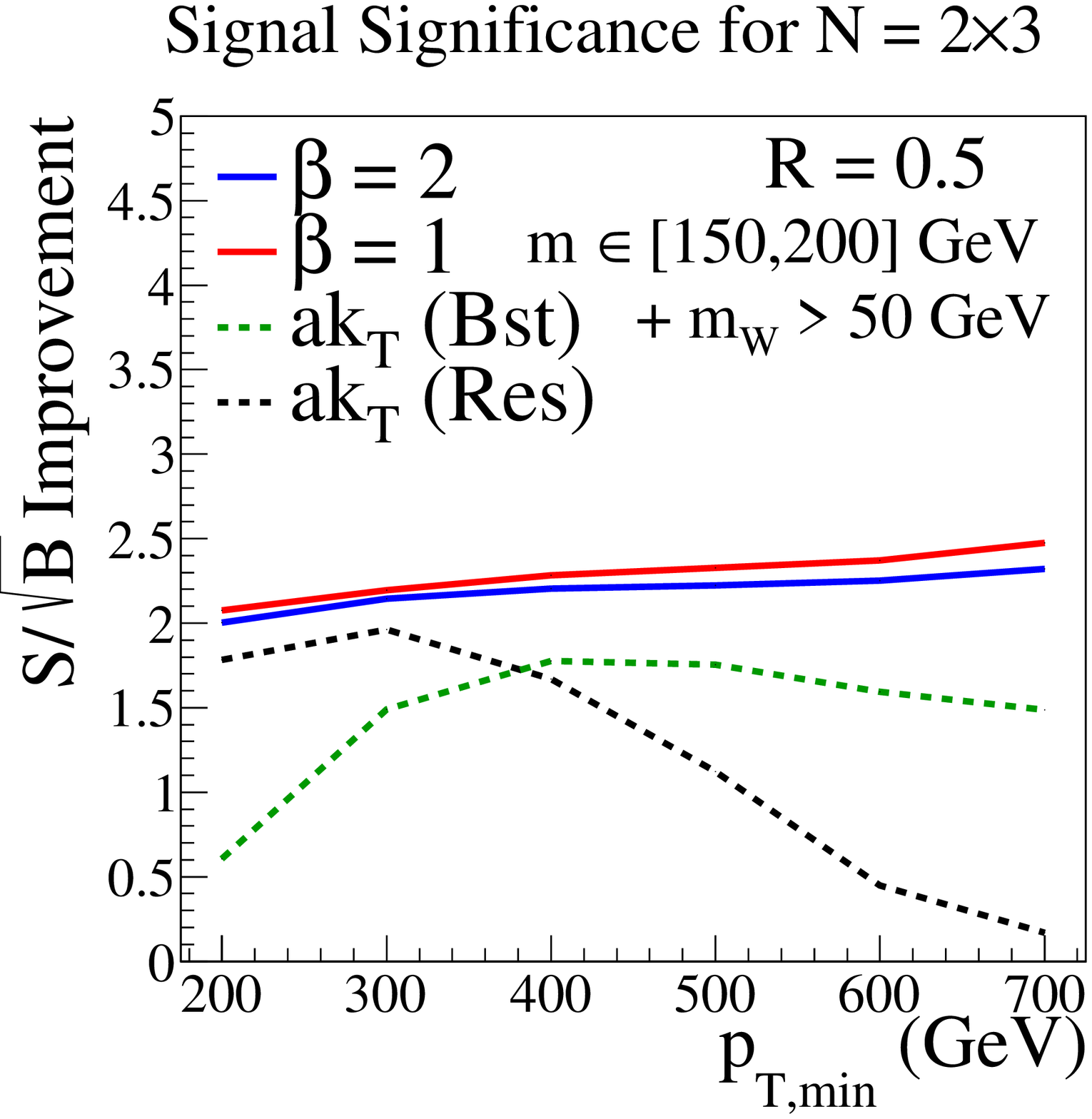}
\label{fig:top_signalsig_2groups_Wmasscut}}
\caption{Comparing boosted top and XCone performance with $N = 2\times3$ as a function of top $p_T$ for (a) signal efficiency, (b) background mistag, and (c) signal significance gain.  XCone interpolates between the traditional resolved strategy at low $p_T$ and the traditional boosted strategy at high $p_T$, with improved performance in the boosted regime due to the addition of soft subjets (at moderate $p_T$) and grooming away of jet contamination (at higher $p_T$).
}
\end{figure}

The resulting top mass distributions are shown in \Fig{fig:top_ttbar_mass_2groups}, again in the $p_{T} \in [400,500]$ GeV bin.  The top mass distribution for XCone $N = 2\times3$ jets is similar to the $N = 6$ case, continuing to maintain a peaked, symmetric shape around the top mass.  Crucially, $N = 2\times3$ reduces the high-mass tail since there are more correctly reconstructed top quarks.  The traditional boosted strategy results in a $W$ shelf caused by the $\Rnaught = 1.0$ jet radius not containing the full top decay products, while the traditional resolved strategy has a high-side mass tail from the inclusion of ISR jets.  XCone avoids both of these pitfalls, giving an excellent overall reconstruction.  

Similarly, as shown in \Fig{fig:top_qcd_mass_2groups}, the background mass distribution for XCone falls in between the traditional boosted and resolved strategies. Like in the $N = 6$ case, both $N = 2 \times 3$ XCone jets and traditional resolved jets tend to be scattered throughout the hemisphere, leading to large invariant masses. However, the additional hemisphere constraint from using $N = 2\times3$ helps to control this effect, giving smaller masses on average for $N = 2 \times 3$ than $N = 6$.

The signal efficiency and background mistag rates are given in \Figs{fig:top_ttbar_eff_2groups_Wmasscut}{fig:top_qcd_eff_2groups_Wmasscut}, again for the $m_{jjj} \in [150, 200]~\GeV$ top mass window and $m_{jj,\min} > 50~\GeV$ $W$ mass cut.  For the signal efficiency, it is clear the XCone interpolates between the good traditional resolved performance at low $p_T$ and the good traditional boosted performance at high $p_T$, yielding approximately a 60\% reconstruction efficiency throughout the $p_T$ range.  For the background mistag rate, XCone holds steady at 10\%, where again, further improvements are possible using $b$-tagging or substructure information.  

The improvement in signal significance $(S/\sqrt{B})$ is shown in \Fig{fig:top_signalsig_2groups_Wmasscut}, where across the whole $p_{T}$ range, the performance of XCone matches or surpasses that of the traditional boosted and resolved strategies.  In the boosted regime, the improvement of XCone over traditional methods is due to two different effects already alluded to above.  At moderate $p_T$, XCone can identify soft subjets that fall outside of the normal anti-$k_{T}$ cluster radius (see also \Ref{Freytsis:2014hpa}), increasing the jet mass into the top window.  At high $p_T$, $\Rnaught = 0.5$ acts similarly to the subjet radius in filtering \cite{Butterworth:2008iy} or trimming \cite{Krohn:2009th}, removing jet contamination and decreasing the jet mass into the top window.   We conclude that XCone gives a powerful way to extend conventional resolved analysis strategies into the boosted regime, especially if the goal is obtaining high signal efficiency.

\section{Conclusions}
\label{sec:conclude}

In this paper, we presented three case studies to show how XCone can be used in a range of LHC analyses, producing comparable or better results than conventional methods.  Remarkably, a single benchmark cone size of $\Rnaught = 0.5$ was able to successfully reconstruct dijet resonances, boosted Higgs bosons, and boosted top quarks.   We are particularly encouraged by XCone's ability to smoothly interpolate between the resolved and boosted regimes, and we anticipate that further improvement will be possible by combining XCone jet finding with additional jet substructure discriminants.

The focus of this paper has been on the physics applications of XCone, but it is also important to note the relative computational efficiency of XCone compared to other algorithms.  In the case of boosted tops with $N = 6$, we find that the anti-$k_{T}$ algorithm runs at an average speed of $0.3$ ms per event on a typical laptop, while XCone takes around $7.3$ ms per event, which is roughly $25$ times slower, though still relatively fast.  If computational speed is a priority, one can use XCone with the seed axes directly as the jet axes (i.e.\ no minimization step), which is dubbed ``PseudoXCone'' in the \textsc{Nsubjettiness} contrib.  This takes around $0.7$ ms per event, only $2.3$ times slower than anti-$k_{T}$.  As shown in the companion paper, $95\%$ of the seed axes are located at or near the global $N$-jettiness minimum \cite{Stewart:2015waa}, so the effectiveness of PseudoXCone is comparable to that of XCone.  As an example, there are only percent-level differences in the boosted top signal significance by using PseudoXCone instead of XCone.

Beyond the examples shown in this paper, XCone should also work well in even more complex final states, including multi-jet searches for physics beyond the standard model.  A key challenge for any exclusive approach is how to best veto ISR jets, and it may be that multiple $N$ values will be needed reach optimal performance.  One intriguing possibility for new physics searches is consider different $N$-jettiness beam measures.  In particular, the conical geometric measure in \Ref{Stewart:2015waa} has a $\gamma = 2$ option (compared to the recommended $\gamma = 1$ default), which preferentially select jets in the central region.  This would help not only to veto ISR jets but also to control QCD backgrounds.

In an experimental context, a key practical question is how to calibrate the energy scale of XCone jets.  For widely separated jets, we have seen that XCone jets are essentially the same as anti-$k_T$ jets, so existing calibration strategies should be straightforward to adapt to XCone.  The case of overlapping jets is more complicated in XCone, since jet regions can have varying shapes and sizes.  As already mentioned, though, XCone jet regions are fully determined by the location of the jet axes, independent of the details of the jet constituents, so one could develop a calibration strategy that accounts for, say, the active jet area \cite{Cacciari:2007fd,Cacciari:2008gn} when determining the jet energy scale.  Indeed, such a strategy is already used for anti-$k_T$ jets in the context of pileup mitigation, though further studies are needed to understand systematics in the XCone case.

Ultimately, the choice of jet algorithm should depend on the physics process of interest.  For situations with widely separated jets, exclusive cone algorithms like XCone and inclusive cone algorithms like anti-$k_T$ yield rather similar performance, so there is no obvious reason to prefer one algorithm over the other.  XCone does have the appealing features of yielding well-defined rigid jet geometries and fitting nicely into $N$-jettiness factorization theorems (see \cite{Stewart:2015waa}), though this has to be balanced against the ubiquity, simplicity, and computational efficiency of anti-$k_T$.  For this reason, the advantages of XCone are most prominent in the boosted regime of overlapping (sub)jets, where standard anti-$k_T$ reconstruction is simply not applicable while XCone can dynamically split jet regions.

We have emphasized how XCone allows analysis strategies developed in the more familiar resolved regime to be extrapolated into the boosted regime.  One could also adopt the reverse strategy of taking analysis strategies based on boosted jets and extrapolating them into the resolved regime.  The HEPTogTagger \cite{Plehn:2010st} is an example of this reverse strategy, since it uses fat jets to achieve good signal efficiency at low $p_T$.  The variable $R$ jet algorithm \cite{Krohn:2009zg} can also be used to implement a reverse strategy, since it can match the jet radius of a fat jet to the $p_T$ of the boosted object of interest.  A priori, it is not clear whether resolved-to-boosted or boosted-to-resolved approaches will be more performant, so it is important to develop both types of strategies, as well as hybrid strategies like our $N=2\times3$ method.\footnote{A similar dichotomy is present for jet grooming methods, where one can take an outside-in approach of first finding a large radius jet and then partitioning it into subjets as in trimming \cite{Krohn:2009th}, or take an inside-out approach of first finding small radius jets and then combining them into larger objects as in jets-from-jets \cite{Nachman:2014kla}.} 

Given the value of exclusive jet algorithms for lepton colliders but the advantages of conical jets for hadron colliders, we expect that exclusive cone jet algorithms will find useful applications beyond the ones studied in this paper.  Because XCone effectively separates jet axis finding from jet region finding, one could even imagine a generalized exclusive cone strategy that dynamically chooses different jet radii $\Rnaught$ depending on the final state, making it possible to treat even more extreme kinematics.  We look forward to seeing how XCone and its generalizations perform in future analyses, as the LHC continues to pursue physics in and beyond the standard model over a wide kinematic range.

\acknowledgments{
We thank Matteo Cacciari, Gavin Salam, Gregory Soyez, Iain Stewart, Frank Tackmann, Chris Vermilion, Wouter Waalewijn, and Ken Van Tilburg for helpful conversations.  This work was supported by the U.S. Department of Energy (DOE) under cooperative research agreement DE-SC00012567.  J.T.\ is also supported by the DOE Early Career research program DE-SC0006389 and by a Sloan Research Fellowship from the Alfred P.\ Sloan Foundation.  T.W.\ is also supported by the MIT Undergraduate Research Opportunities Program (UROP) through the Paul E. Gray Endowed Fund.}

\appendix

\section{Comparison to Exclusive $k_T$}
\label{app:exclusive}

The focus of this paper has been on comparing XCone to anti-$k_{T}$, which is essentially comparing an exclusive cone algorithm to an inclusive cone algorithm.  It is also instructive to compare XCone to other exclusive non-cone algorithms, which we do in this appendix.   Given the popular use of sequential recombination algorithms, the exclusive variants of $k_{T}$ \cite{Catani:1993hr,Ellis:1993tq} and Cambridge/Aachen (C/A) \cite{Dokshitzer:1997in,Wobisch:1998wt,Wobisch:2000dk} clustering provide the most useful comparisons.

Because C/A clusters according to angles alone, it turns out to yield rather poor performance when dealing with events with multiple angular scales.  This can also be understood from the study in \Ref{Stewart:2015waa}, which shows that C/A does not provide good seed axes for $N$-jettiness minimization.  For this reason, we only compare to the exclusive $k_{T}$ algorithm, as it provides the most similar performance to XCone.  Indeed, the $\beta = 1$ minimization procedure in XCone starts from seeds derived from $k_{T}$ clustering.\footnote{For $\beta = 1$ minimization, XCone uses the winner-take-all recombination scheme \cite{Bertolini:2013iqa,Larkoski:2014uqa,Salambroadening}.  For the exclusive $k_T$ studies here, we stick with standard $E$-scheme recombination.  For $\beta = 2$ minimization, XCone uses a generalized $k_T$ measure halfway between $k_T$ and C/A.}  

As explained in \Sec{sec:dijet}, the distinction between exclusive cone algorithms and inclusive cone algorithms is largely irrelevant in the case of well-separated jets.  By contrast, the distinction between cone jets and $k_T$-style jets is rather important for well-separated jets, especially at large jet radius.  Famously, $k_{T}$-style jets have distinctly non-conical boundaries which are determined by the configuration of soft radiation within the jet, yielding a broad spectrum of jet areas \cite{Cacciari:2007fd,Cacciari:2008gn}.  For this reason, it is not really fair to compare XCone to exclusive $k_T$ with well-separated jets, since XCone simply inherits the advantages of anti-$k_T$ and other cone-like algorithms.

\begin{figure}
\centering
\subfloat[]{
\includegraphics[width=0.32\columnwidth]{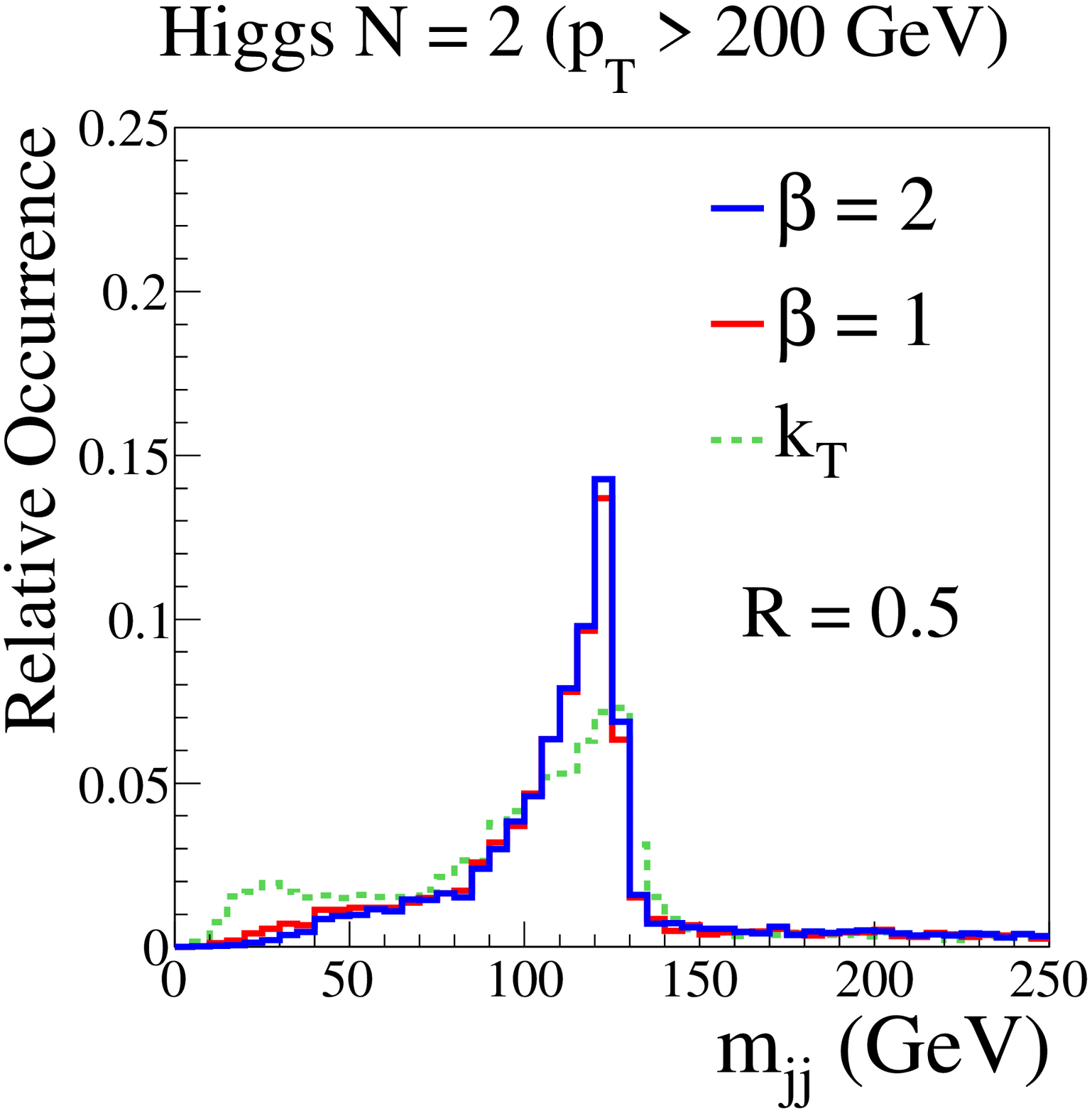}
\label{fig:higgs_2jet_kt_mass_200}
}
\subfloat[]{
\includegraphics[width=0.32\columnwidth]{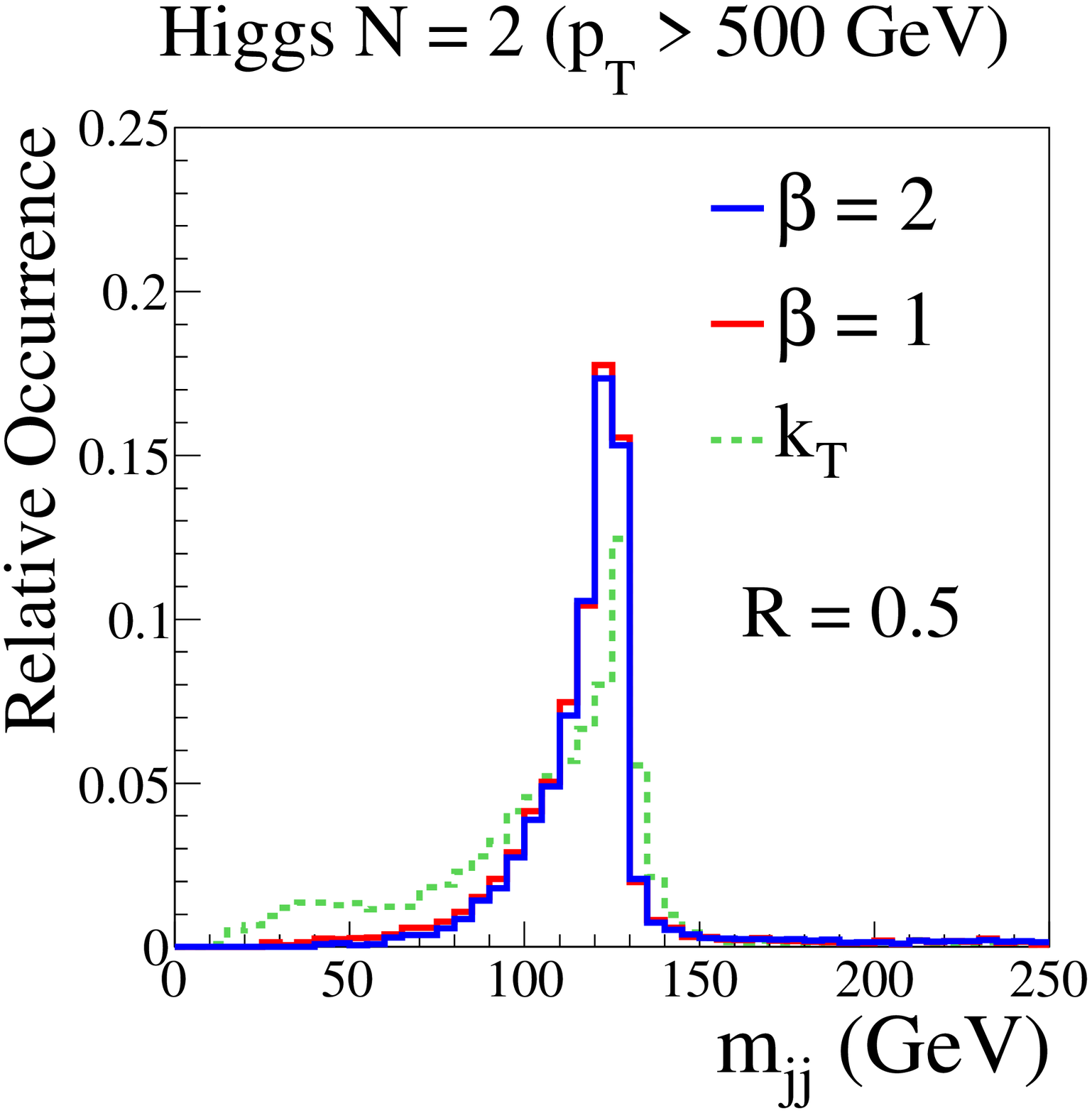}
\label{fig:higgs_2jet_kt_mass_500}
}
\subfloat[]{
\includegraphics[width=0.32\columnwidth]{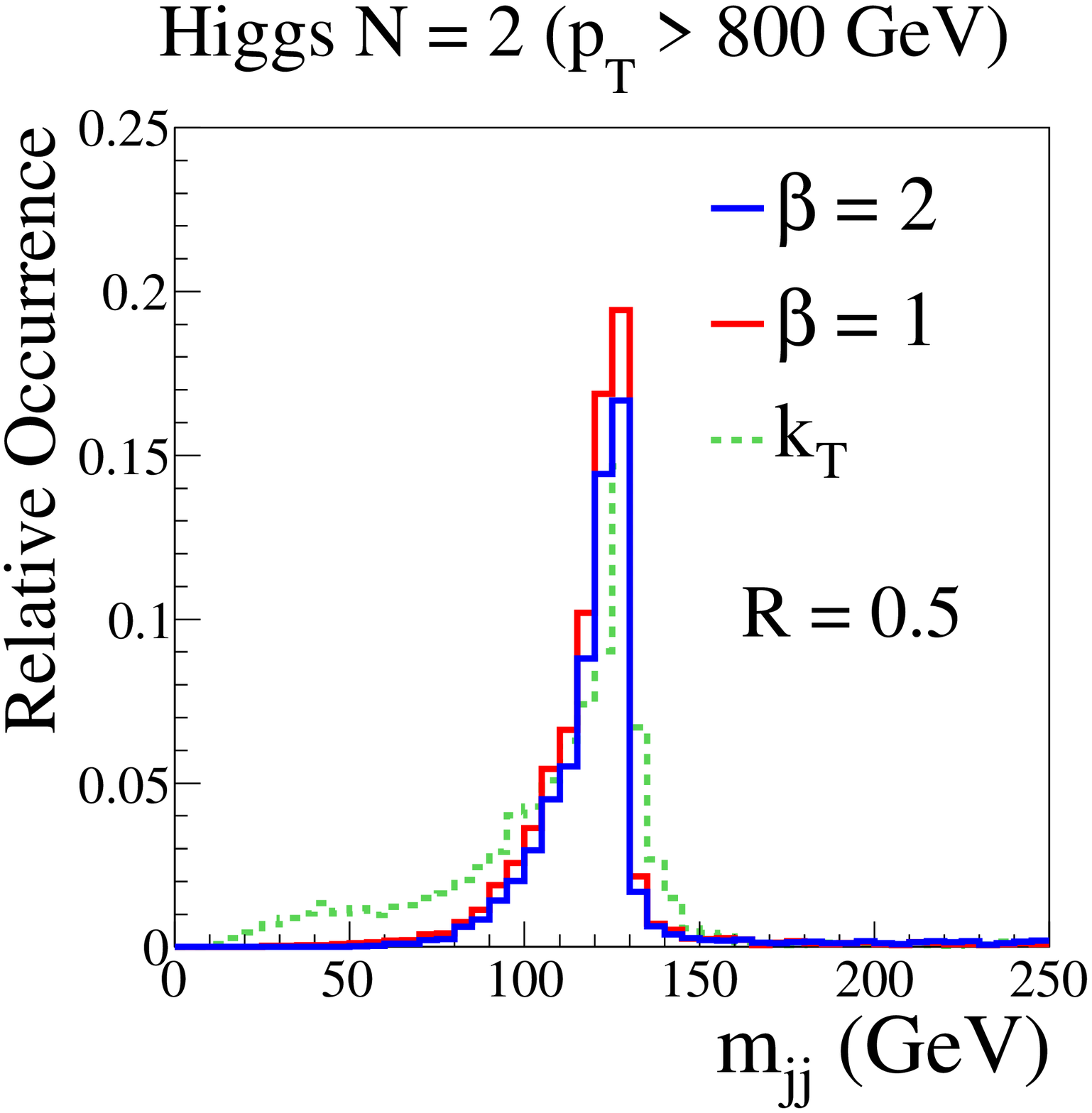}
\label{fig:higgs_2jet_kt_mass_800}
}

\subfloat[]{
\includegraphics[width=0.32\columnwidth]{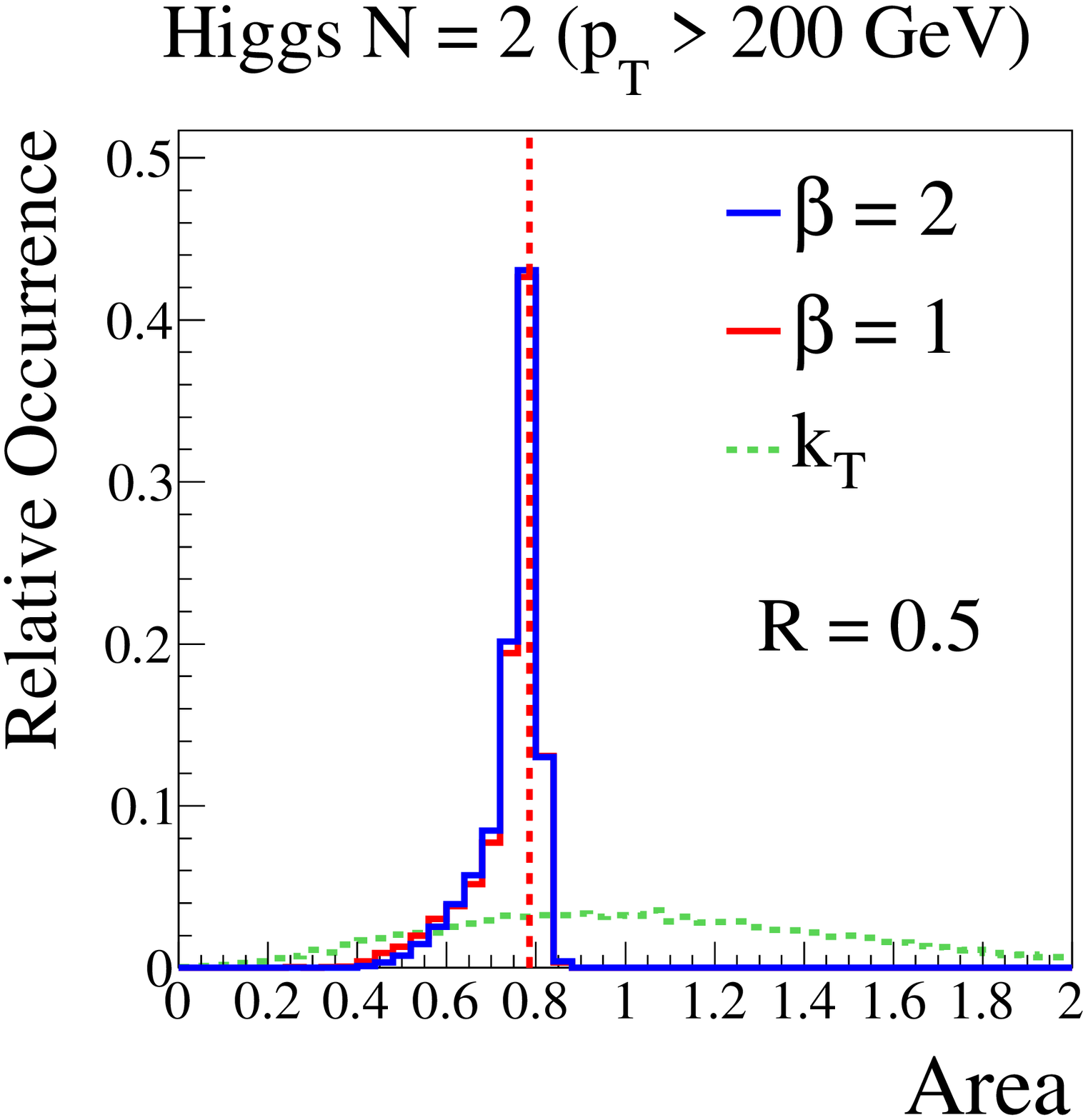}
\label{fig:higgs_2jet_kt_area_200}
}
\subfloat[]{
\includegraphics[width=0.32\columnwidth]{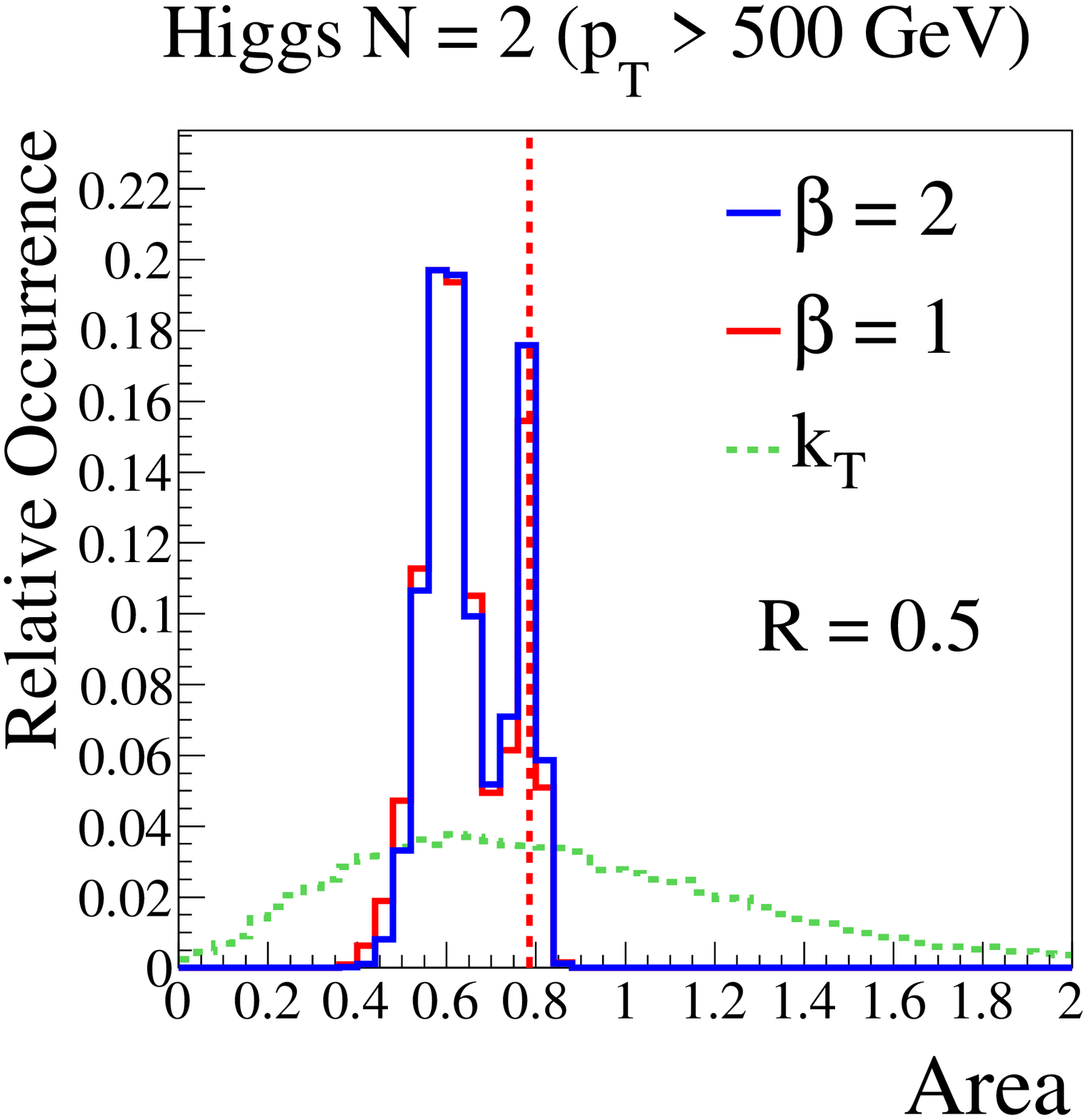}
\label{fig:higgs_2jet_kt_area_500}
}
\subfloat[]{
\includegraphics[width=0.32\columnwidth]{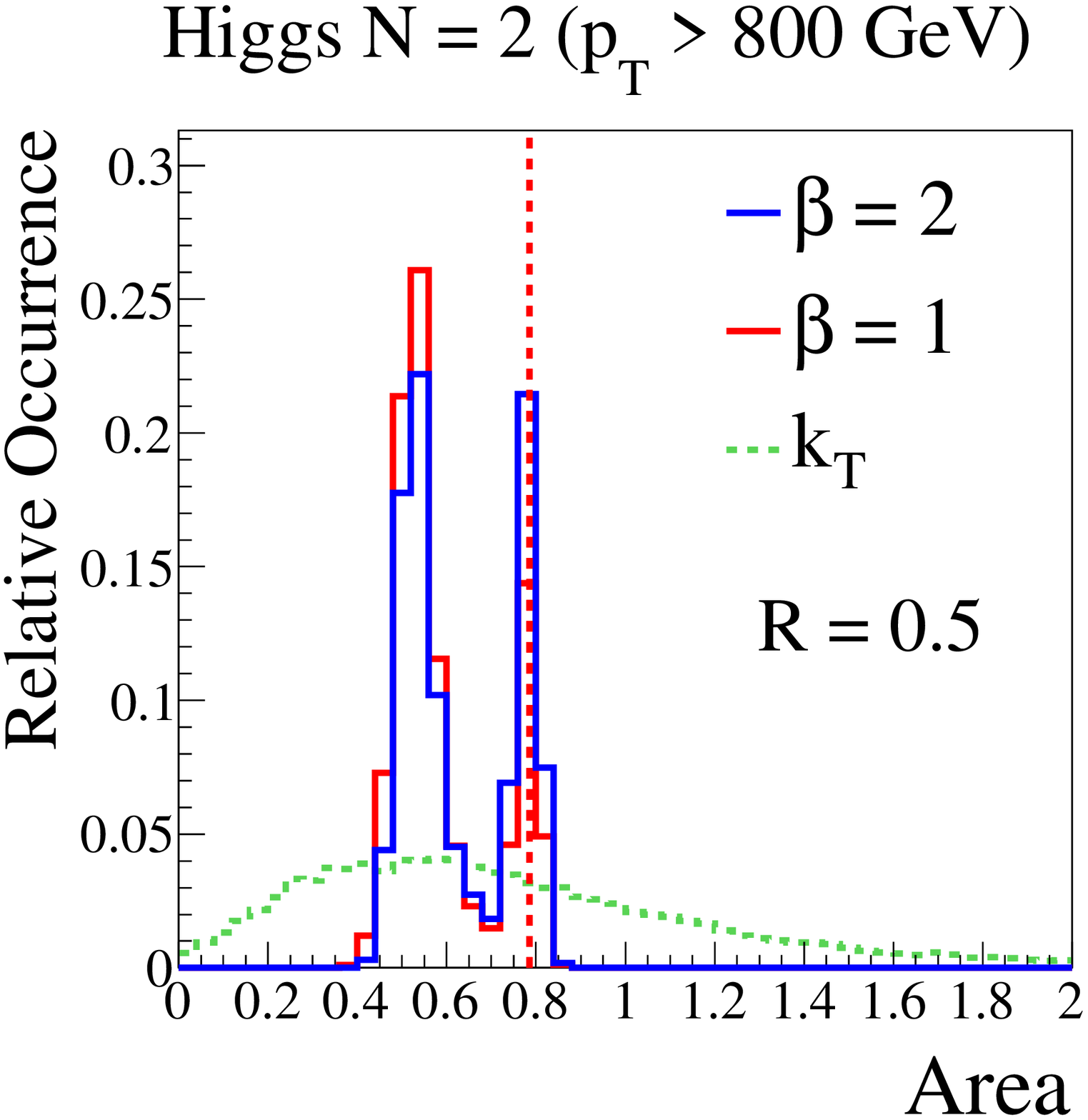}
\label{fig:higgs_2jet_kt_area_800}
}
\caption{Same as \Fig{fig:higgs_2jet_study}, but with exclusive $k_{T}$. As the Higgs $p_T$ increases from (a) 200 GeV to (b) 500 GeV to (c) 800 GeV, both XCone and exclusive $k_{T}$ yield a dijet Higgs peak. Exclusive $k_{T}$ shows a larger low-mass tail, though, indicating that it does not always capture all of the Higgs decay products.  This is due to the irregular shape of the $k_{T}$ jets, as quantified by the jet areas shown in the bottom row.}
\label{fig:higgs_2jet_kt_study}
\end{figure}

The comparison between XCone and exclusive $k_{T}$ is most instructive in the boosted regime, where area effects are less important and the distinction between inclusive and exclusive jets is more pronounced.   We restrict our attention to $N=2$ and $N=3$ boosted Higgs reconstruction (see \Secs{sec:higgs2}{sec:Higgs3}) and $N = 6$ boosted top reconstruction (see \Sec{sec:Top6}).  We do not show $N = 2\times3$ for boosted tops, since \Sec{sec:Top23} already presented a similar study with exclusive $k_{T}$.

\subsection{Boosted Higgs Bosons with N = 2}
\label{app:exclusive_higgs}

\begin{figure}
\subfloat[]{
\includegraphics[width = .5\columnwidth]{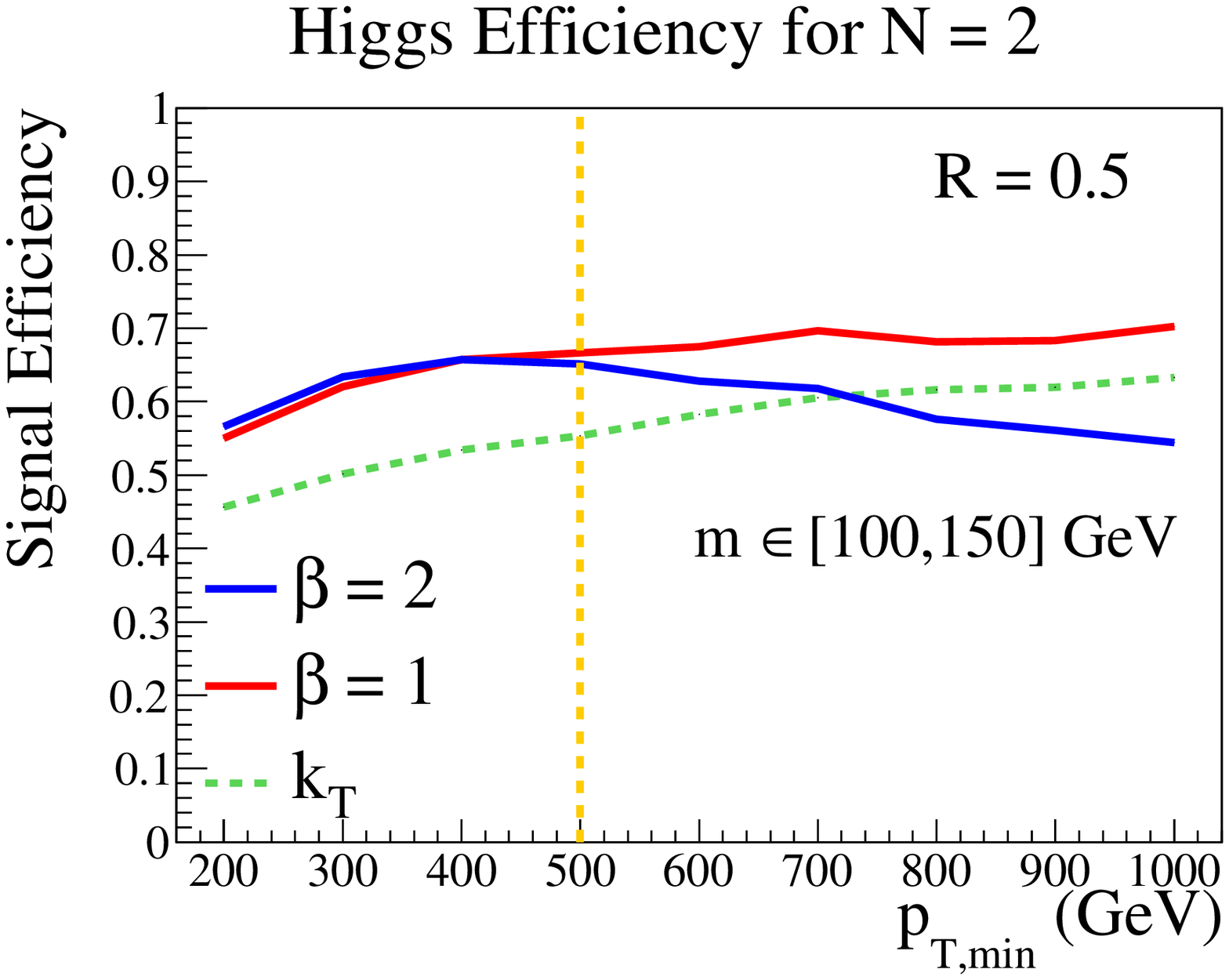}
\label{fig:higgs_2jet_kt_eff}
}
\subfloat[]{
\includegraphics[width = .5\columnwidth]{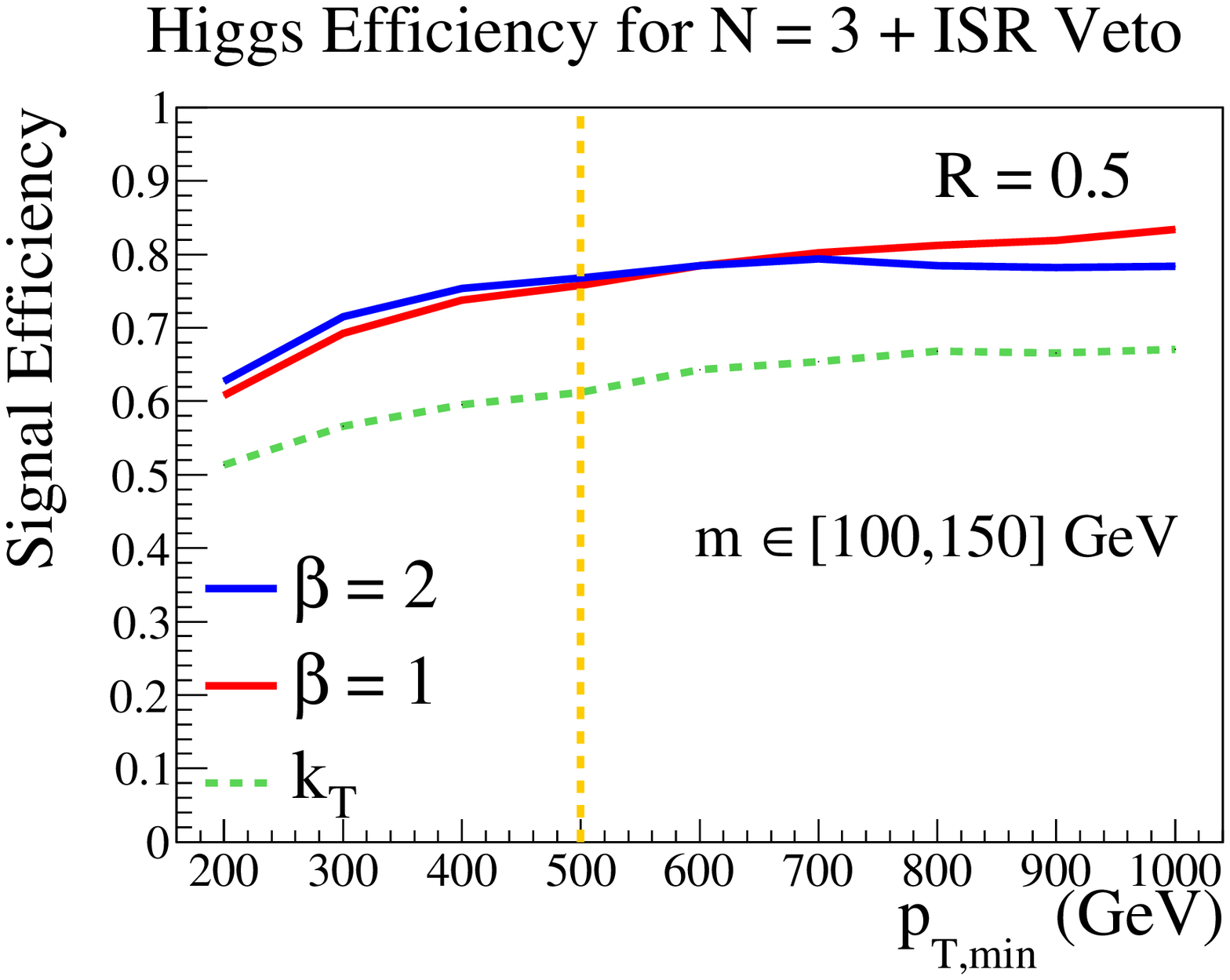}
\label{fig:higgs_2jet_eff_kt_improved}
}
\caption{Same as \Fig{fig:higgs_2jet_eff_both}, but with exclusive $k_{T}$. The efficiency of both XCone and exclusive $k_{T}$ remain roughly constant across the entire kinematic range.  XCone shows a distinct advantage over exclusive $k_{T}$, however, as its more regular shape allows it to consistently capture the Higgs decay products.}
\label{fig:higgs_2jet_kt_eff_both}
\end{figure}

We saw that XCone with $N = 2$ can reconstruct $H \to b\bar{b}$ decays in both the resolved and boosted regimes, and $N = 3$ can be used to improve performance by accounting for possible ISR in the event.  Here, we can repeat the exact same analysis strategy using the exclusive $k_T$ algorithm, with the same choice of $\Rnaught = 0.5$ and $N = 2,3$.  The XCone distributions shown here are identical to the ones shown in \Sec{sec:higgs}.

The top row of \Fig{fig:higgs_2jet_kt_study} shows that the reconstructed Higgs mass distributions are rather similar between the two algorithms, indicating that exclusive $k_{T}$ also successfully finds the Higgs decay products across a wide kinematic range.  However, exclusive $k_{T}$ has a noticeably larger low-mass tail, indicating that the algorithm does not always capture all of the Higgs decay products.  This is due to the irregularity of the resulting $k_T$ jet shapes.  As shown in the bottom row of \Fig{fig:higgs_2jet_kt_study}, the exclusive $k_T$ jet area distributions are rather broad, though they roughly peak at $\pi R^2$ as expected.  Small-area jets give rise to the low-mass tail while large-area jets are responsible for the high-mass tail.  By keeping a more uniform jet area distribution, XCone achieves better Higgs mass resolution.

The Higgs efficiency of the two algorithms is shown in \Fig{fig:higgs_2jet_kt_eff_both}, using both the $N =2$ and $N =3$ methods.  Both XCone and exclusive $k_{T}$ follow a similar pattern, displaying roughly constant efficiencies across the entire $p_{T}$ range with improved performance going into the boosted regime.  XCone yields a higher overall efficiency, though, due to the jet shape issue described above.  We conclude that the conical nature of XCone gives it an advantage over exclusive $k_{T}$ for boosted Higgs reconstruction.

\subsection{Boosted Top Quarks with N = 6}
\label{app:exclusive_top}

\begin{figure}
\centering
\subfloat[]{
\includegraphics[width = 0.45\columnwidth]{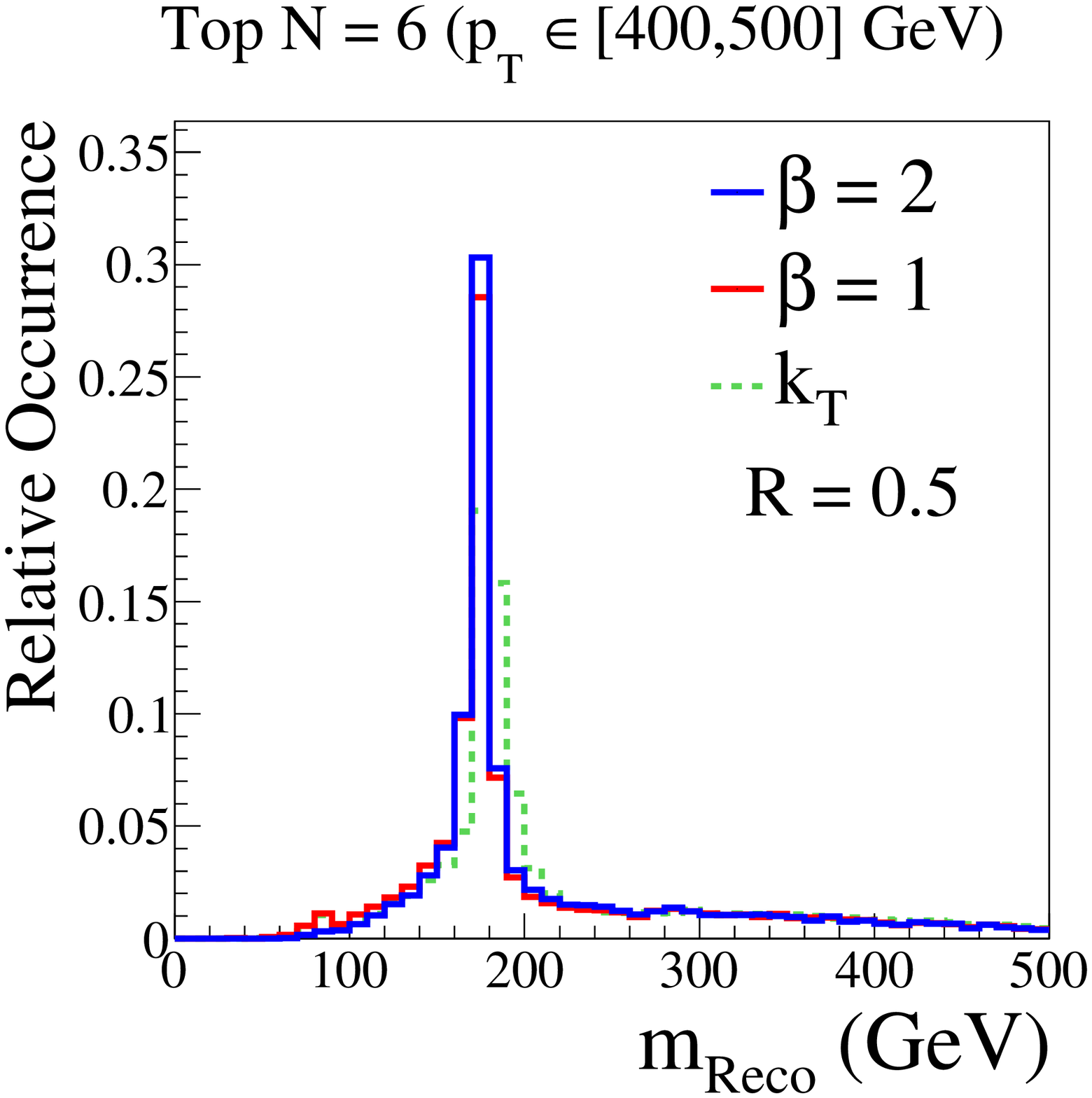}
\label{fig:top_ttbar_kt_rawmass_400}}
$\quad$
\subfloat[]{
\includegraphics[width = 0.45\columnwidth]{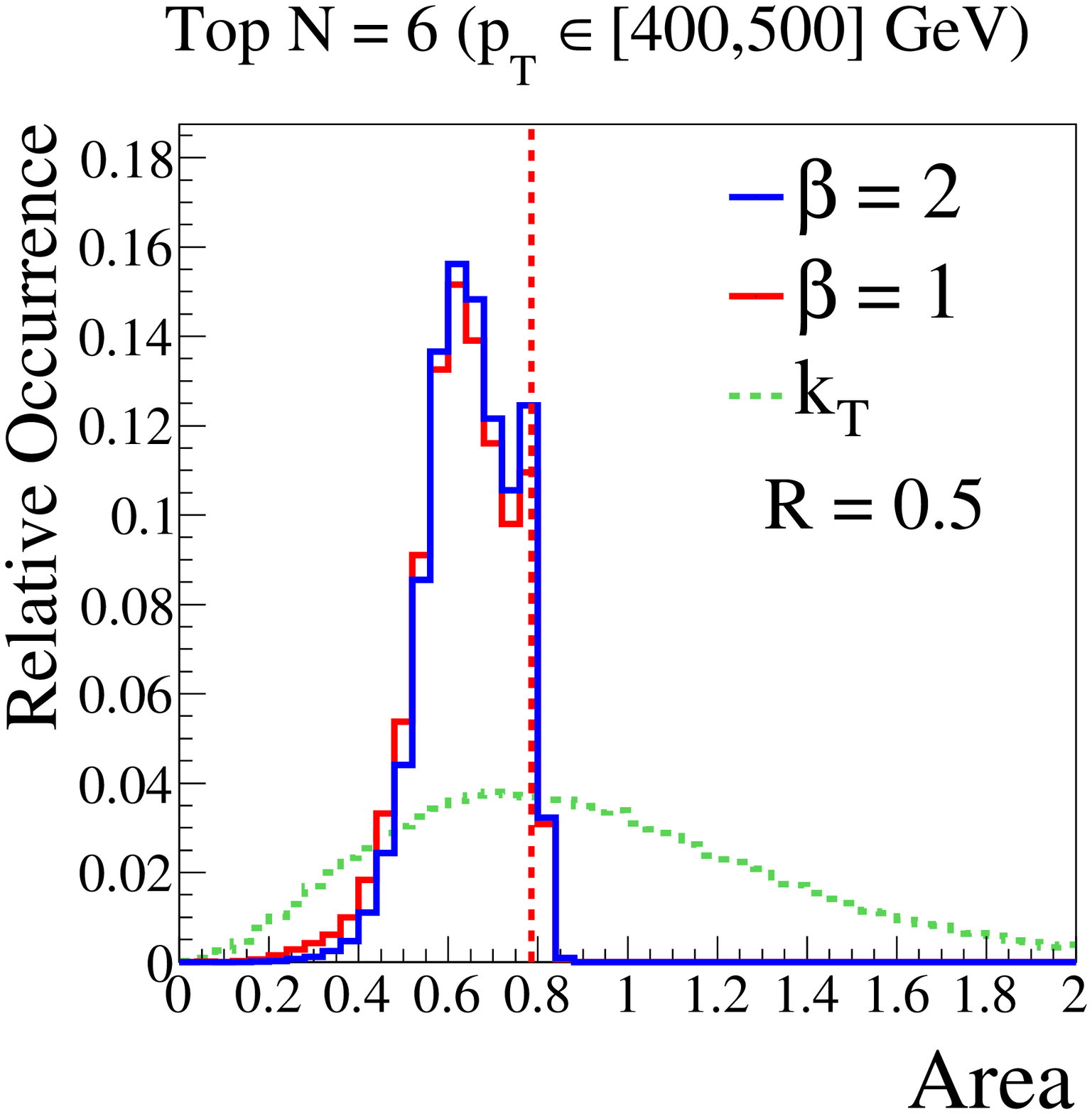}
\label{fig:top_ttbar_kt_area_400}}

\subfloat[]{
\includegraphics[width = 0.45\columnwidth]{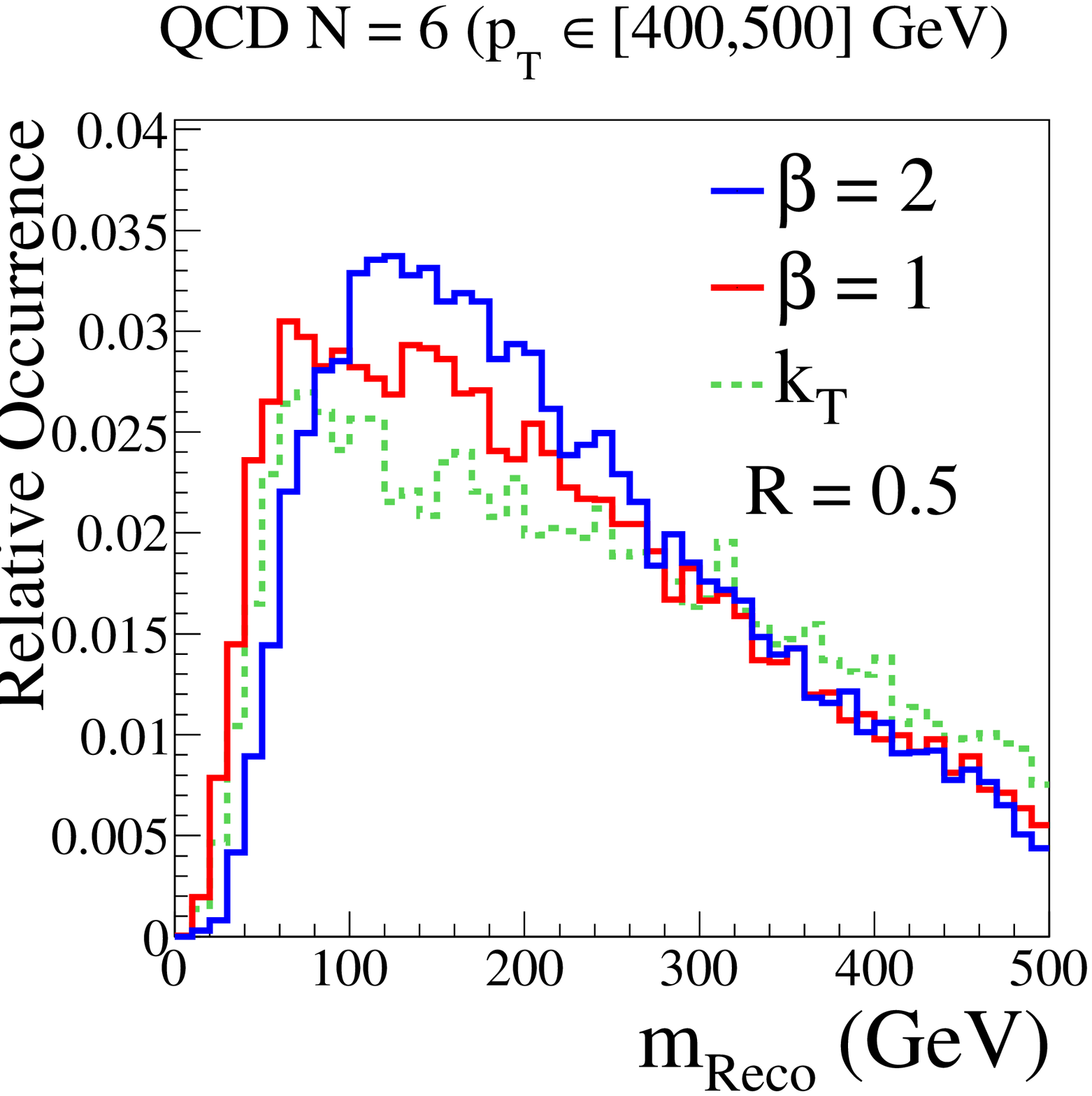}
\label{fig:top_qcd_kt_rawmass_400}}
$\quad$
\subfloat[]{
\includegraphics[width = 0.45\columnwidth]{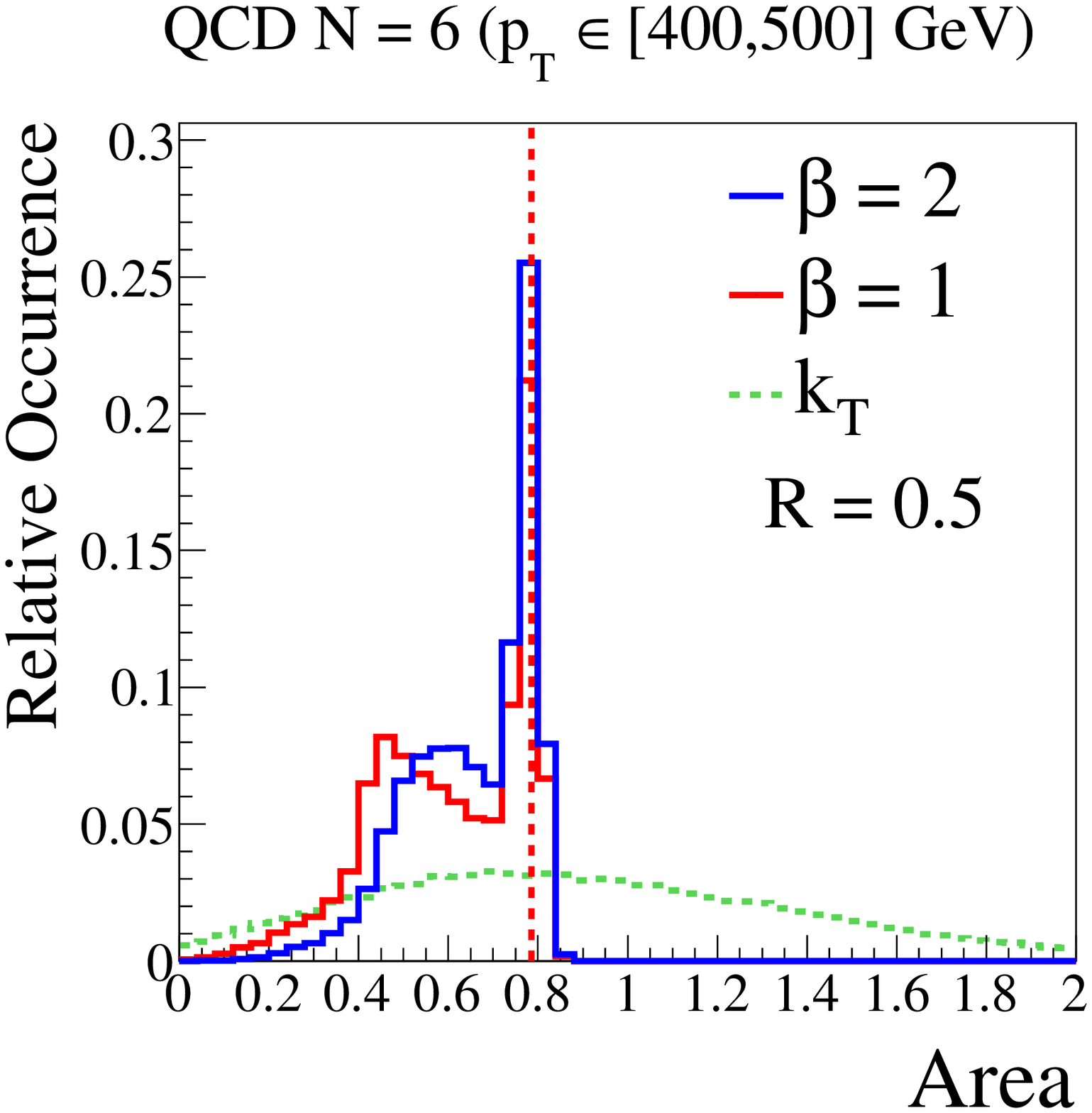}
\label{fig:top_qcd_kt_area_400}}
\caption{Same as \Fig{fig:top_6jet_study}, but with exclusive $k_{T}$. The mass distributions in (a) and (c) are very similar to XCone for both signal and background, but the area distributions in (b) and (d) show that the exclusive $k_{T}$ jets have a more inconsistent shape than the XCone jets, as expected. }
\label{fig:top_kt_6jet_study}
\end{figure}

\begin{figure}
\centering
\subfloat[]{
\includegraphics[width = 0.32\columnwidth]{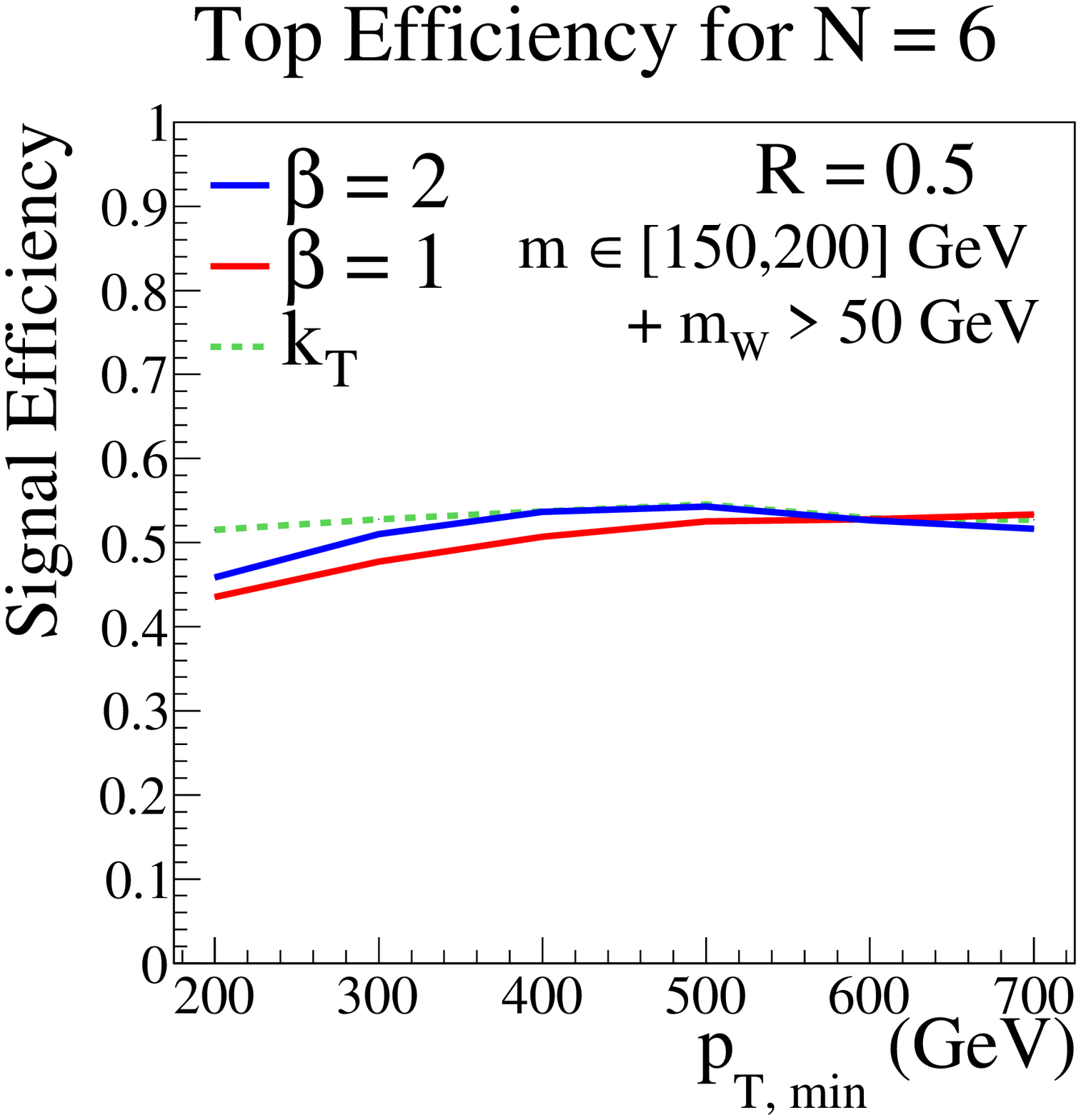}
\label{fig:top_ttbar_kt_raweff}}
\subfloat[]{
\includegraphics[width = 0.32\columnwidth]{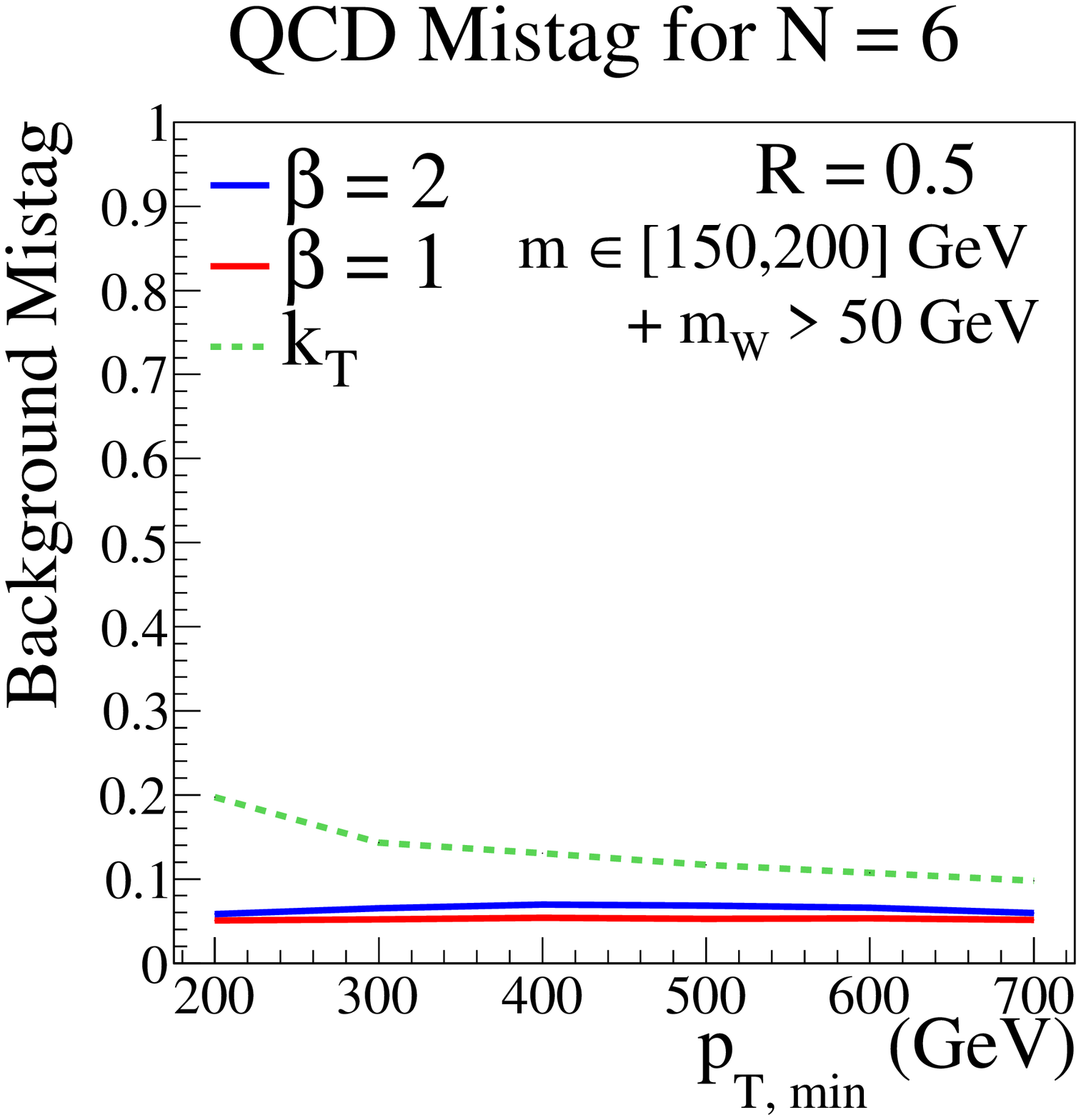}
\label{fig:top_qcd_kt_raweff}}
\subfloat[]{
\includegraphics[width = 0.32\columnwidth]{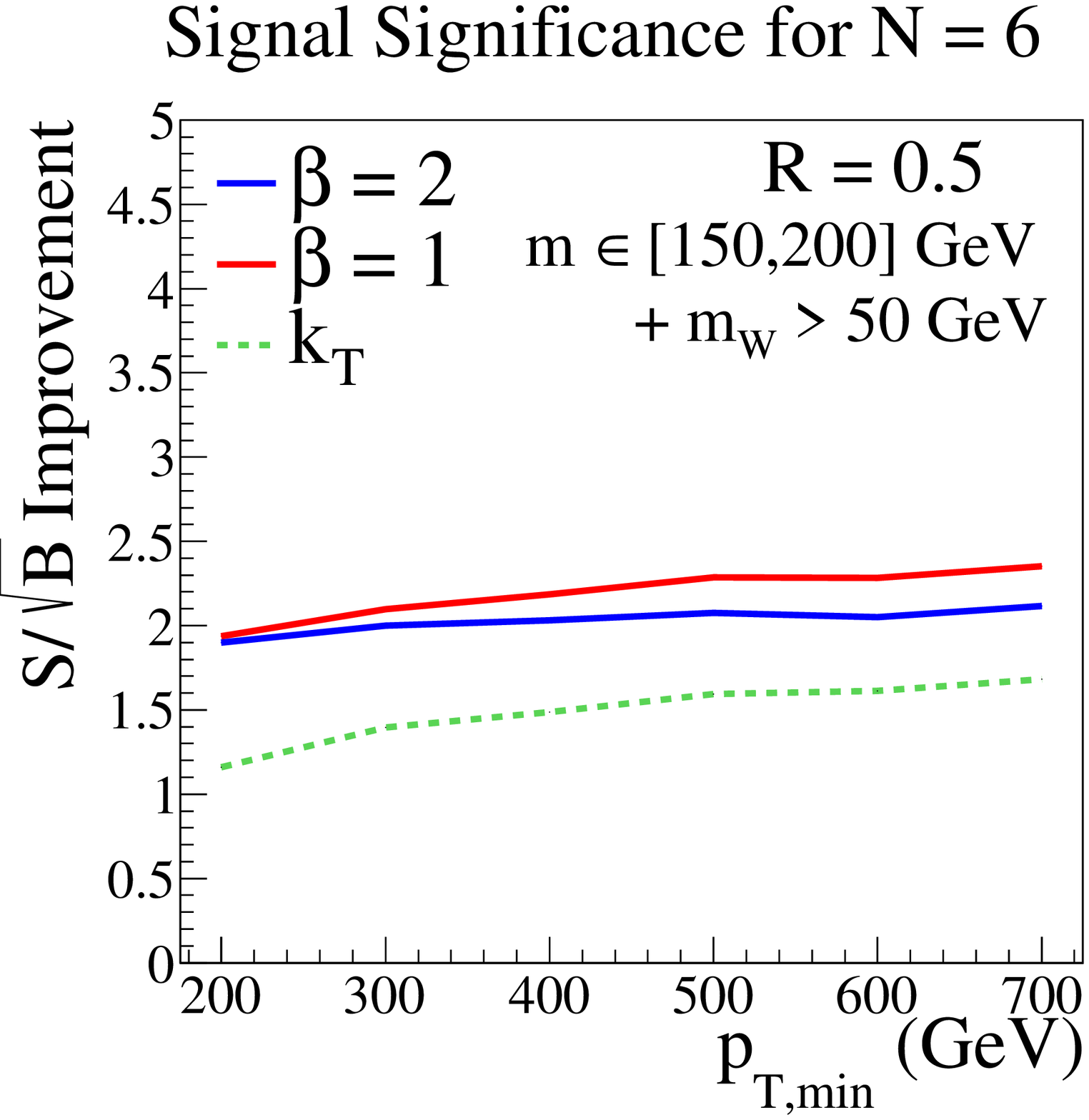}
\label{fig:top_kt_signalsig_Wmasscut}}
\caption{Same as \Fig{fig:top_efficiency_study}, but with exclusive $k_{T}$. The signal efficiency is similar between XCone and exclusive $k_{T}$, but the background mistag is higher for exclusive $k_{T}$, giving XCone the edge in terms of signal significance.}
\label{fig:top_kt_efficiency_study}
\end{figure}

The comparison between exclusive $k_{T}$ and XCone is also instructive for high multiplicity final states, such as boosted top reconstruction.  We apply the same analysis strategy as described in \Sec{sec:Top6}, comparing exclusive $k_{T}$ to XCone with $R = 0.5$ and $N = 6$. 

The mass distributions for exclusive $k_{T}$ jets are shown in the left column of \Fig{fig:top_kt_6jet_study}.  For the boosted top signal, the top mass peak is very similar to that of XCone, with a somewhat less pronounced peak and an offset to higher mass values.  This same high-mass shift is seen in the QCD background distributions.  In the right column of \Fig{fig:top_kt_6jet_study}, one sees again the broad structure of the exclusive $k_T$ jet area distributions.  Even though exclusive $k_{T}$ seems to successfully find the same top decay products as XCone, the jet shapes are rather different, and the high-mass shift is due to the large jet areas sometimes found by exclusive $k_T$.

The signal efficiency and background mistag rates  are shown in \Fig{fig:top_kt_efficiency_study}.  The signal efficiency of XCone and exclusive $k_{T}$ are very similar across the whole $p_{T}$ range, with exclusive $k_T$ even yielding better performance at low $p_T$.  With exclusive $k_{T}$, however, the $W$ mass cut of $50 \,\GeV$ is less effective at controlling backgrounds, yielding a higher background mistag rate than XCone.  This allows XCone to achieve higher signal significance than exclusive $k_{T}$. 

That said, it may be possible to achieve the same performance gains of XCone by using a hybrid cone and $k_T$ clustering scheme.\footnote{We thank Gavin Salam for discussions on this point.}  For example, one could mitigate the high-mass shift coming from large area jets by running an $N = 1$ cone algorithm on each of the $N = 6$ exclusive $k_T$ jets.  Alternatively, one could first find the $N = 6$ hardest jets from an inclusive cone algorithm and then run $N=6$ exclusive $k_T$ on the combined jet constituents.  While perhaps not as elegant as XCone, this hybrid approach would still result in a fixed number of approximately conical jets.  We leave a study of these hybrid methods for future work.

\section{Background Considerations for Boosted Higgs}
\label{app:boostedbackground}

In \Sec{sec:higgs}, we showed that XCone yields excellent signal efficiency for Higgs reconstruction, with an approximately flat response in the case of $N=2$.  While a dedicated background study is beyond the scope of this paper, in this appendix we show that XCone gives sensible results when applied to one of the main background sources:  $Z + \text{jets}$.  We have not included the effect of $b$-tagging, though we suspect that XCone would work well with subjet $b$-tagging methods \cite{CMS:2013vea,ATL-PHYS-PUB-2014-013}.  We use Pythia 8.176 \cite{Sjostrand:2006za,Sjostrand:2007gs} at the $\sqrt{s} = 14$ TeV LHC to simulate $pp \rightarrow Z + \text{jets}$, forcing the $Z$ to decay to neutrinos.  As a proxy for the Higgs $p_T$, we use the recoil $p_T$ of the $Z$ boson.  We follow the same analysis strategy as in \Sec{sec:higgs}, showing the background mistag rate for the $N = 1$, $N = 2$, and $N = 3$ + ISR veto strategies.

\begin{figure}
\centering
\subfloat[]{
\includegraphics[width=0.32\columnwidth]{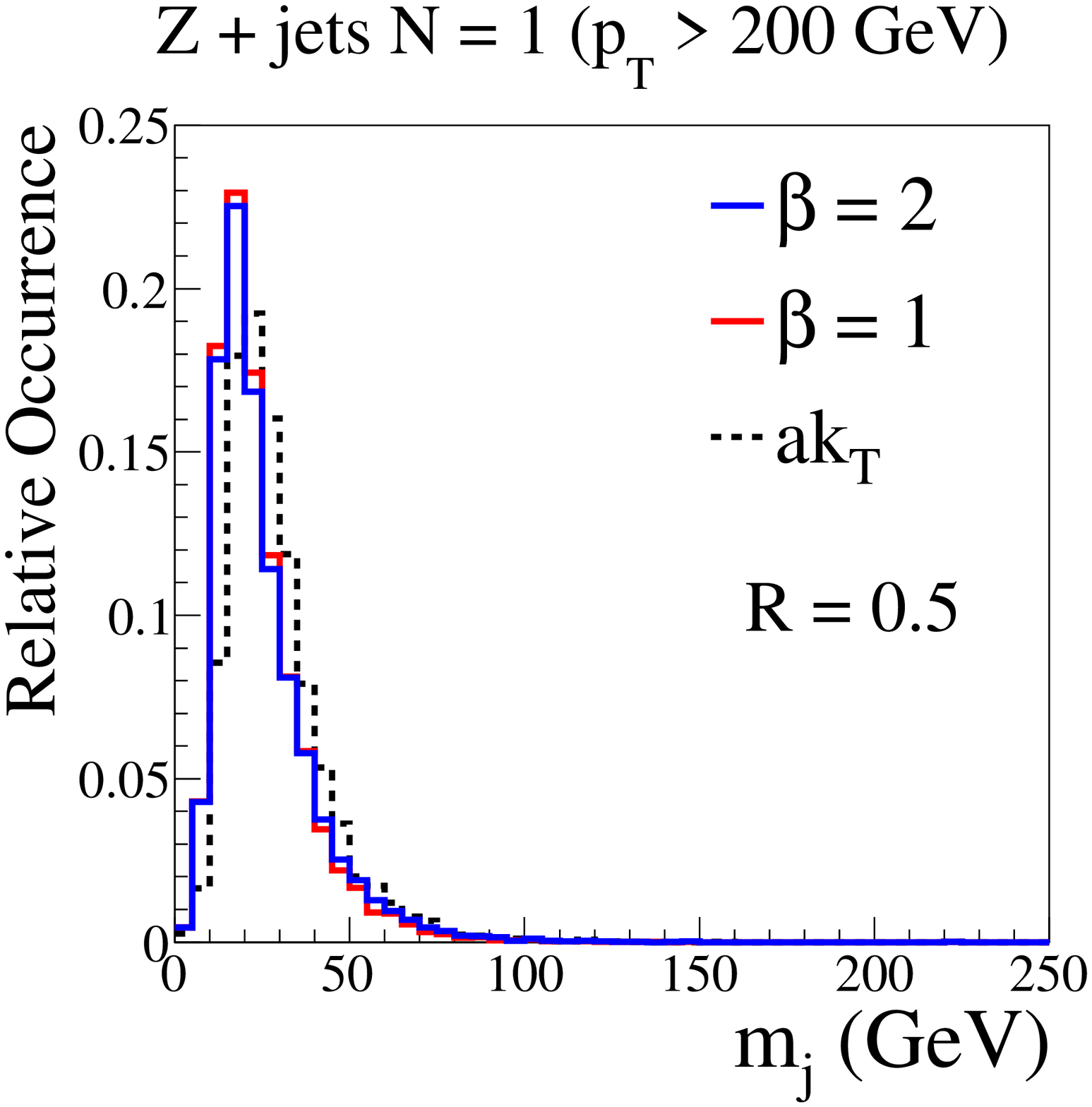}
\label{fig:back_higgs_1jet_mass_200}
}
\subfloat[]{
\includegraphics[width=0.32\columnwidth]{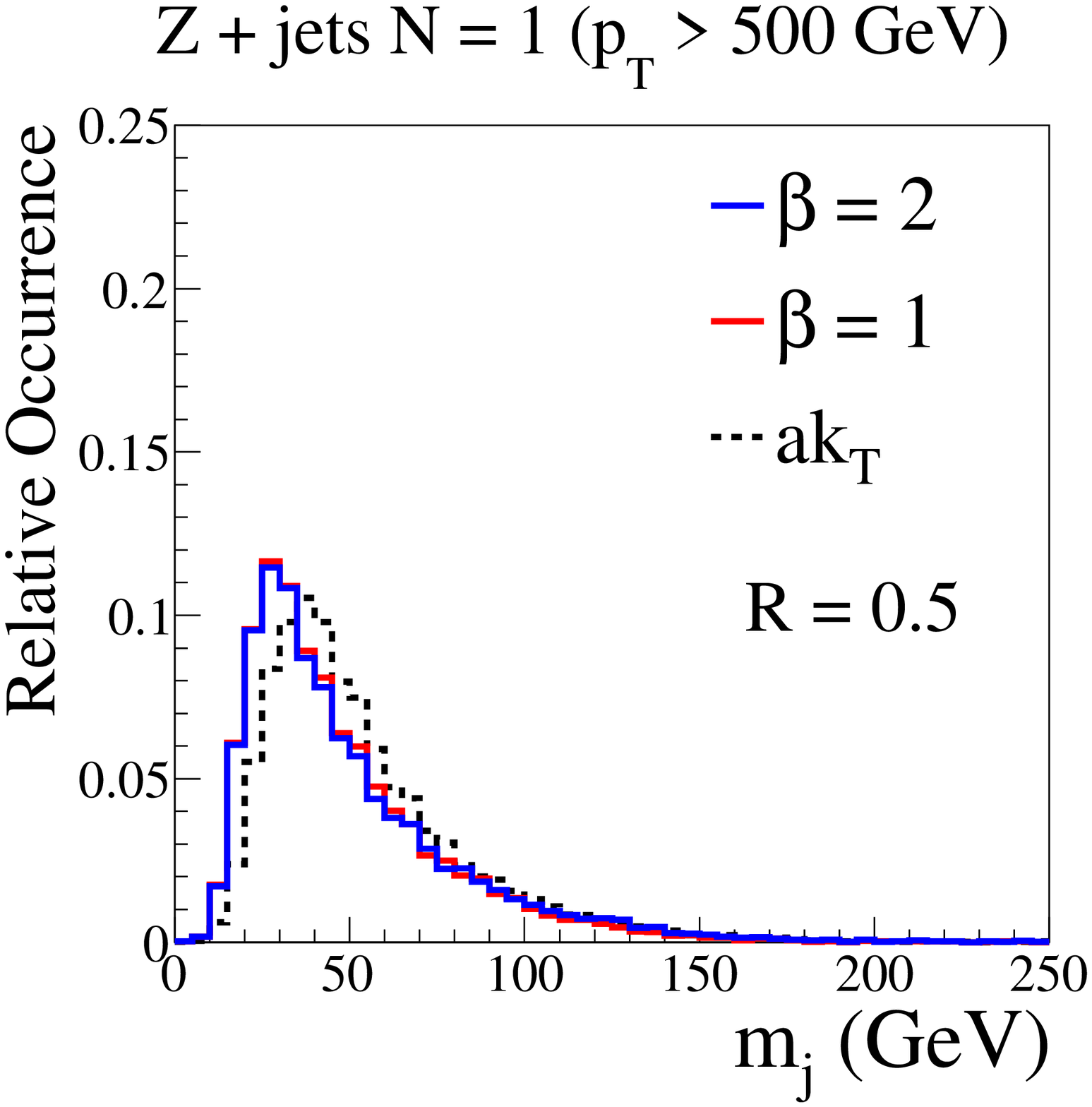}
\label{fig:back_higgs_1jet_mass_500}
}
\subfloat[]{
\includegraphics[width=0.32\columnwidth]{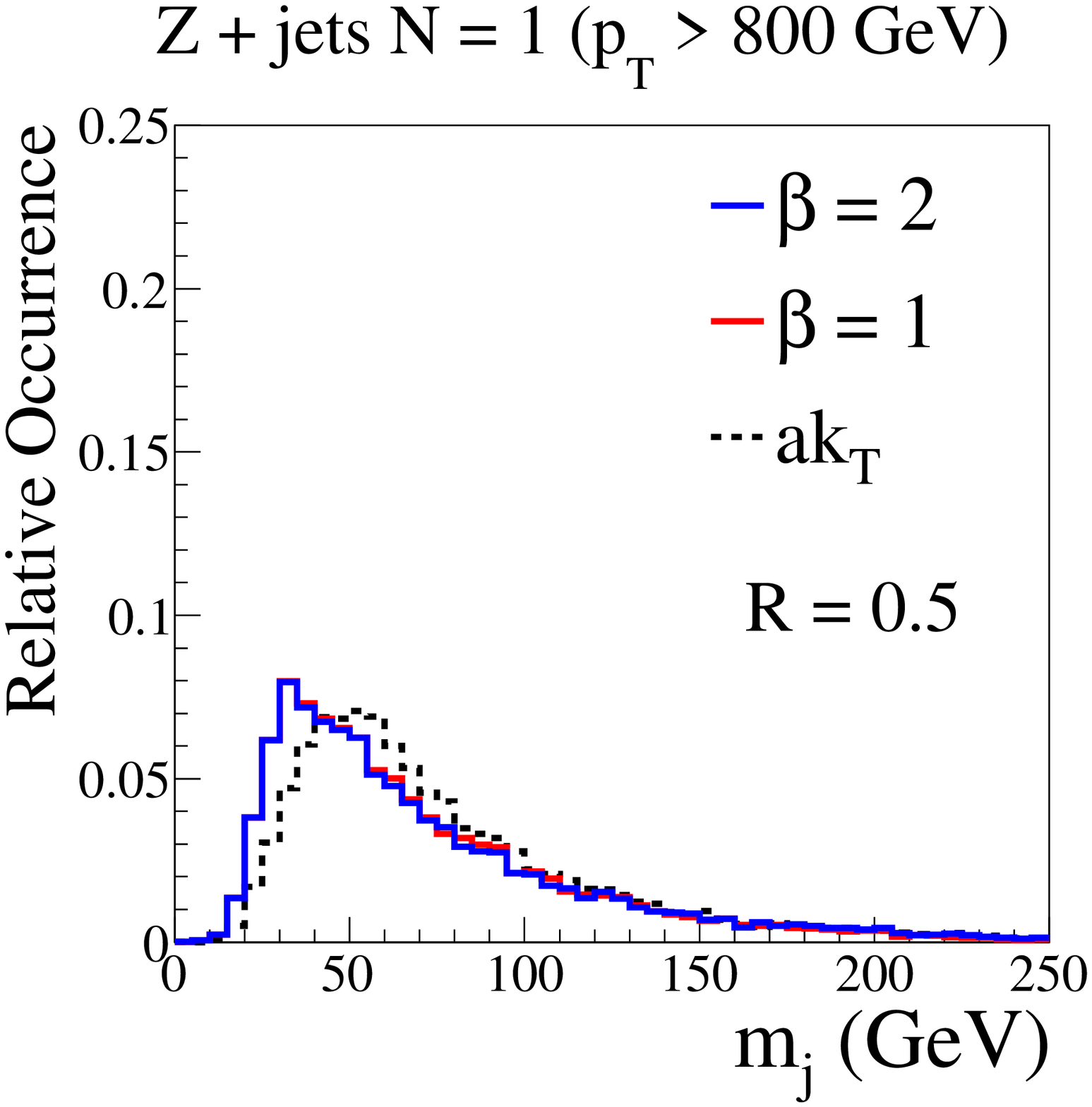}
\label{fig:back_higgs_1jet_mass_800}
}
\caption{$Z + \text{jets}$ reconstruction for $N = 1$, as a potential background to associated Higgs production.  As the recoil $Z$ $p_T$ increases from (a) 200 GeV to (b) 500 GeV to (c) 800 GeV, the single jet invariant mass increases, populating the Higgs signal region.  See \Fig{fig:higgs_1jet_study} for the corresponding signal study.}
\label{fig:back_higgs_1jet_study}
\end{figure}

\begin{figure}
\centering
\includegraphics[width = .6\columnwidth]{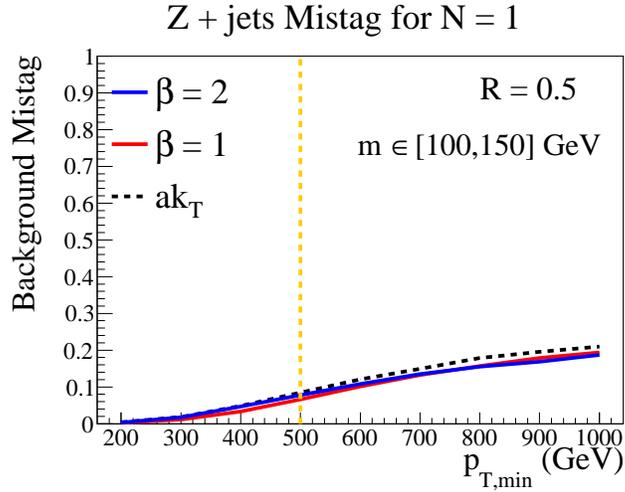}
\caption{$Z + \text{jets}$ mistag rate for $N = 1$ boosted Higgs reconstruction as a function of $Z$ recoil $p_T$, with the mass window $m_{j} \in [100,150]~\GeV$.  The mistag rate grows with $p_T$, with similar behavior seen between XCone and anti-$k_T$.   See \Fig{fig:higgs_1jet_eff} for the corresponding signal study.}
\label{fig:back_higgs_1jet_eff}
\end{figure}

As shown in \Fig{fig:back_higgs_1jet_study} for $N=1$, the reconstructed single jet mass increases for both XCone and anti-$k_T$ as a function of recoil $Z$ $p_T$.  In \Fig{fig:back_higgs_1jet_eff}, this leads to a mistag rate that increases roughly linearly with $p_T$, as expected since the invariant mass of an ordinary QCD jet rises as a function of jet $p_T$.  This is part of the reason why jet substructure techniques like mass drop \cite{Butterworth:2008iy} are needed to control the background jet mass distribution.  Just as in \Fig{fig:higgs_1jet_eff}, XCone and anti-$k_T$ exhibit very similar performance.

\begin{figure}
\centering
\subfloat[]{
\includegraphics[width=0.32\columnwidth]{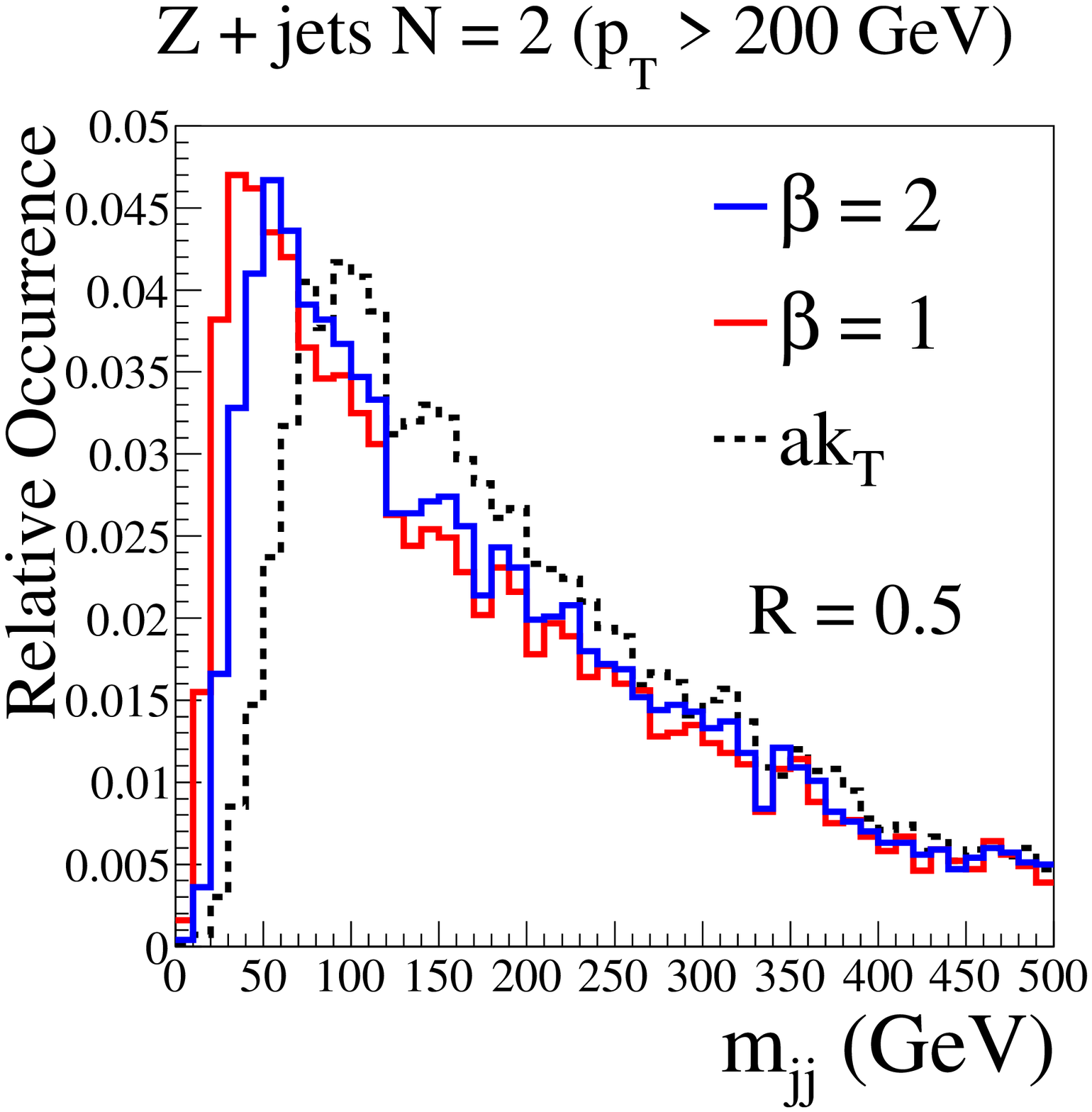}
\label{fig:back_higgs_2jets_mass_200}
}
\subfloat[]{
\includegraphics[width=0.32\columnwidth]{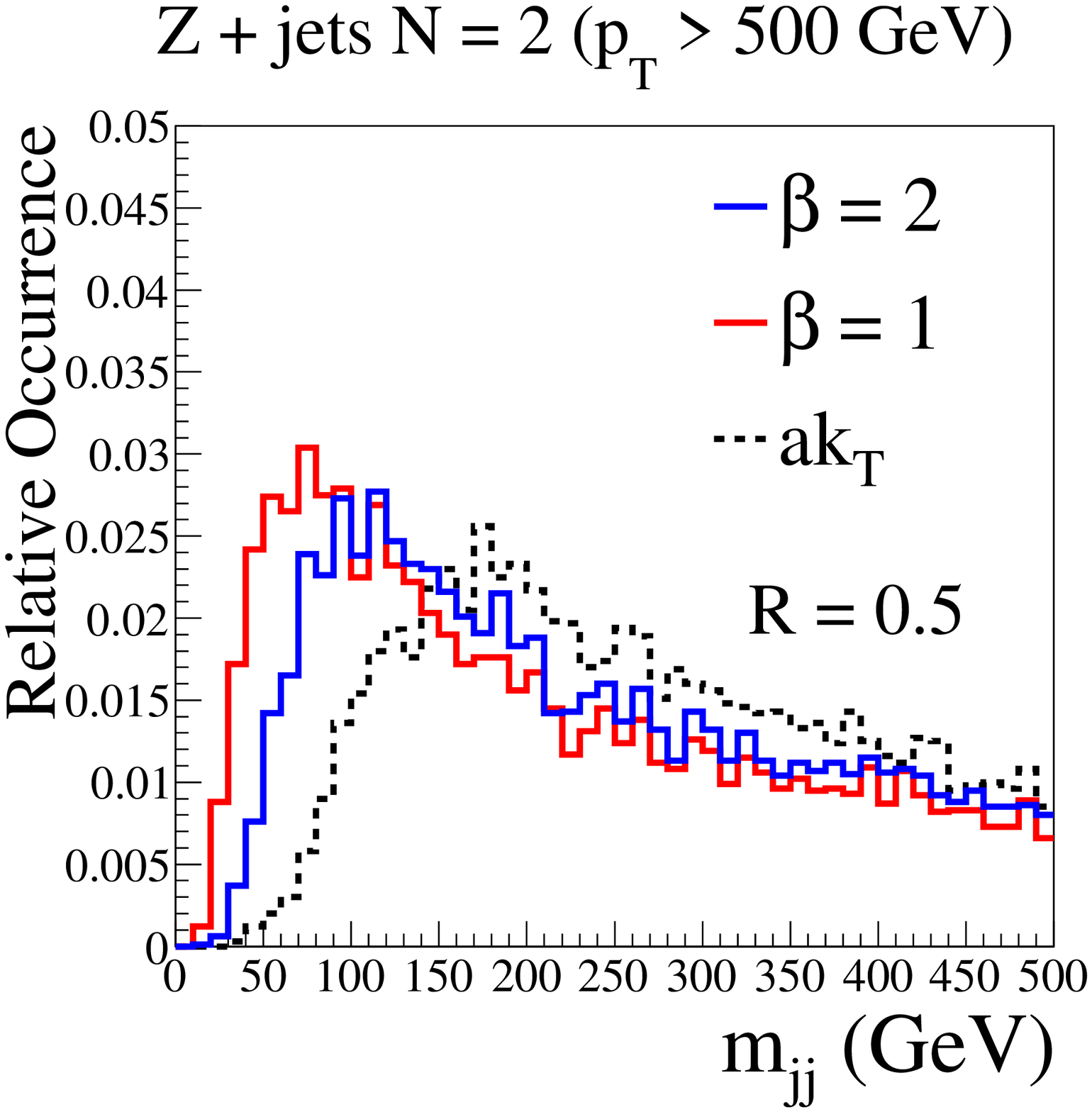}
\label{fig:back_higgs_2jets_mass_500}
}
\subfloat[]{
\includegraphics[width=0.32\columnwidth]{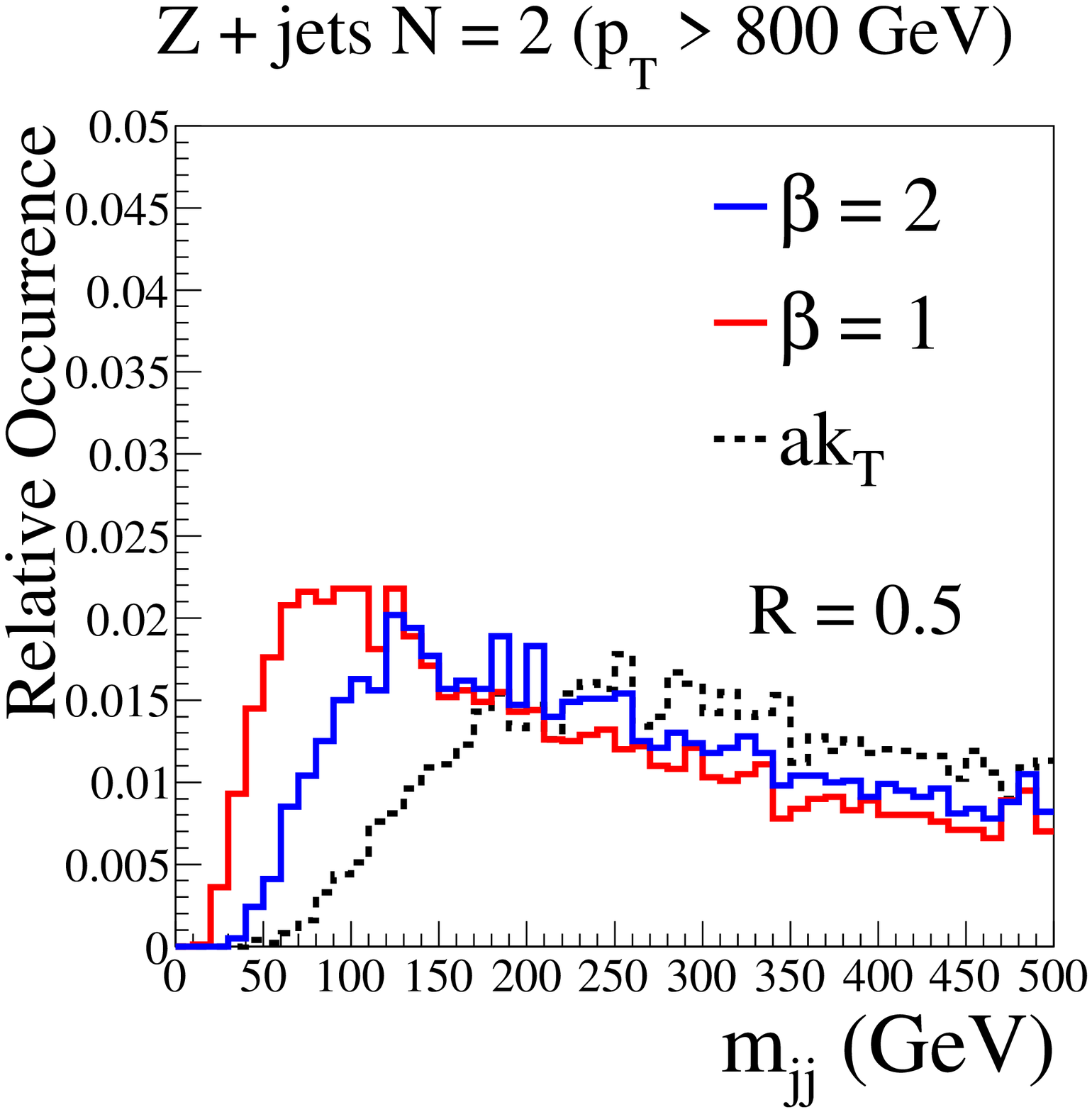}
\label{fig:back_higgs_2jets_mass_800}
}
\caption{Same as \Fig{fig:back_higgs_1jet_study}, but for $N=2$.  As the recoil $Z$ $p_T$ increases from (a) 200 GeV to (b) 500 GeV to (c) 800 GeV, the dijet invariant mass increases, since the two identified jets are unlikely to be proximate in phase space. Note that the scale on the $x$-axis has been increased compared to \Fig{fig:back_higgs_1jet_study} to include larger jet masses.  See \Fig{fig:higgs_2jet_study} for the corresponding signal study.}
\label{fig:back_higgs_2jets_study}
\end{figure}

\begin{figure}
\subfloat[]{
\includegraphics[width = .5\columnwidth]{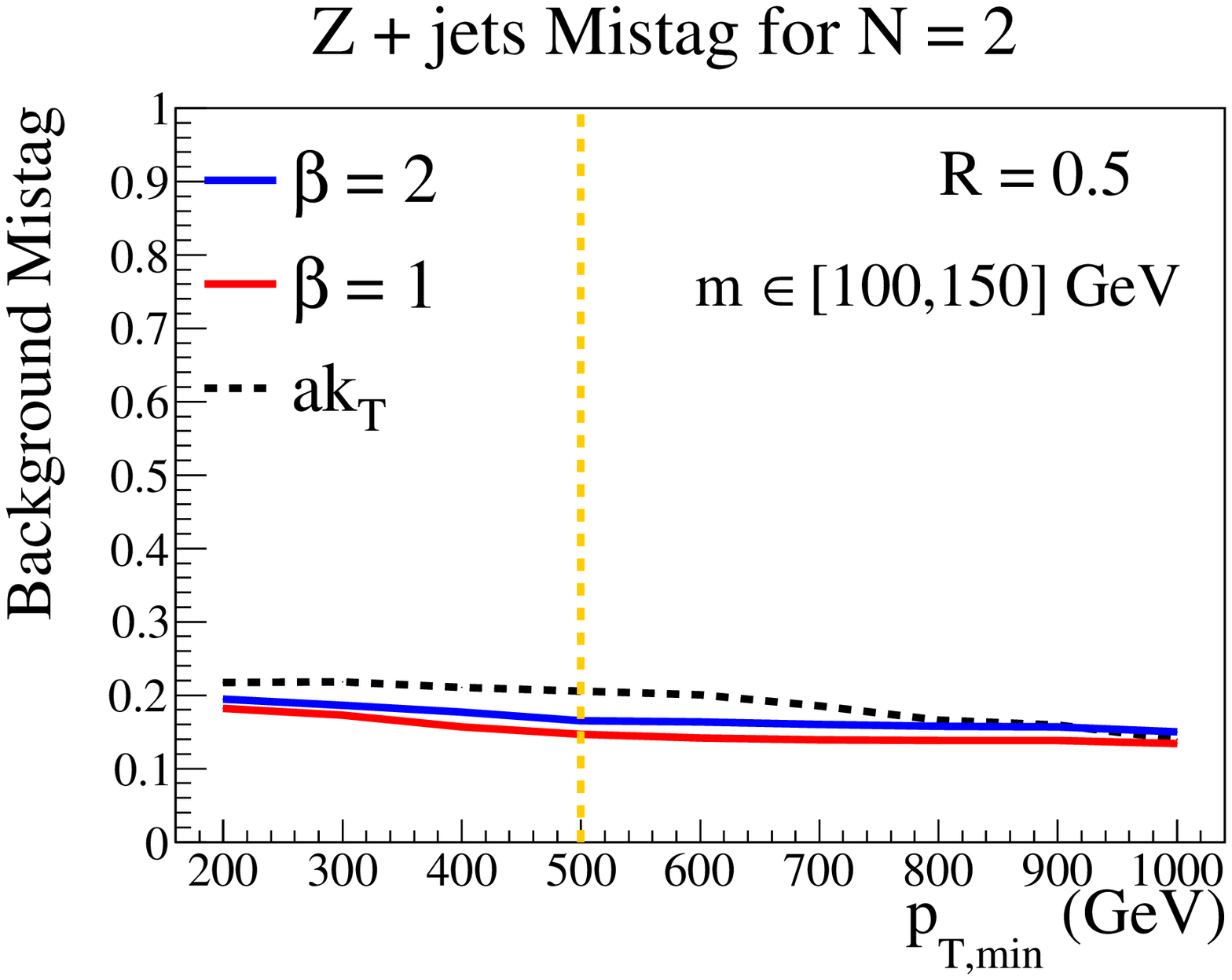}
\label{fig:back_higgs_2jet_eff}
}
\subfloat[]{
\includegraphics[width = .5\columnwidth]{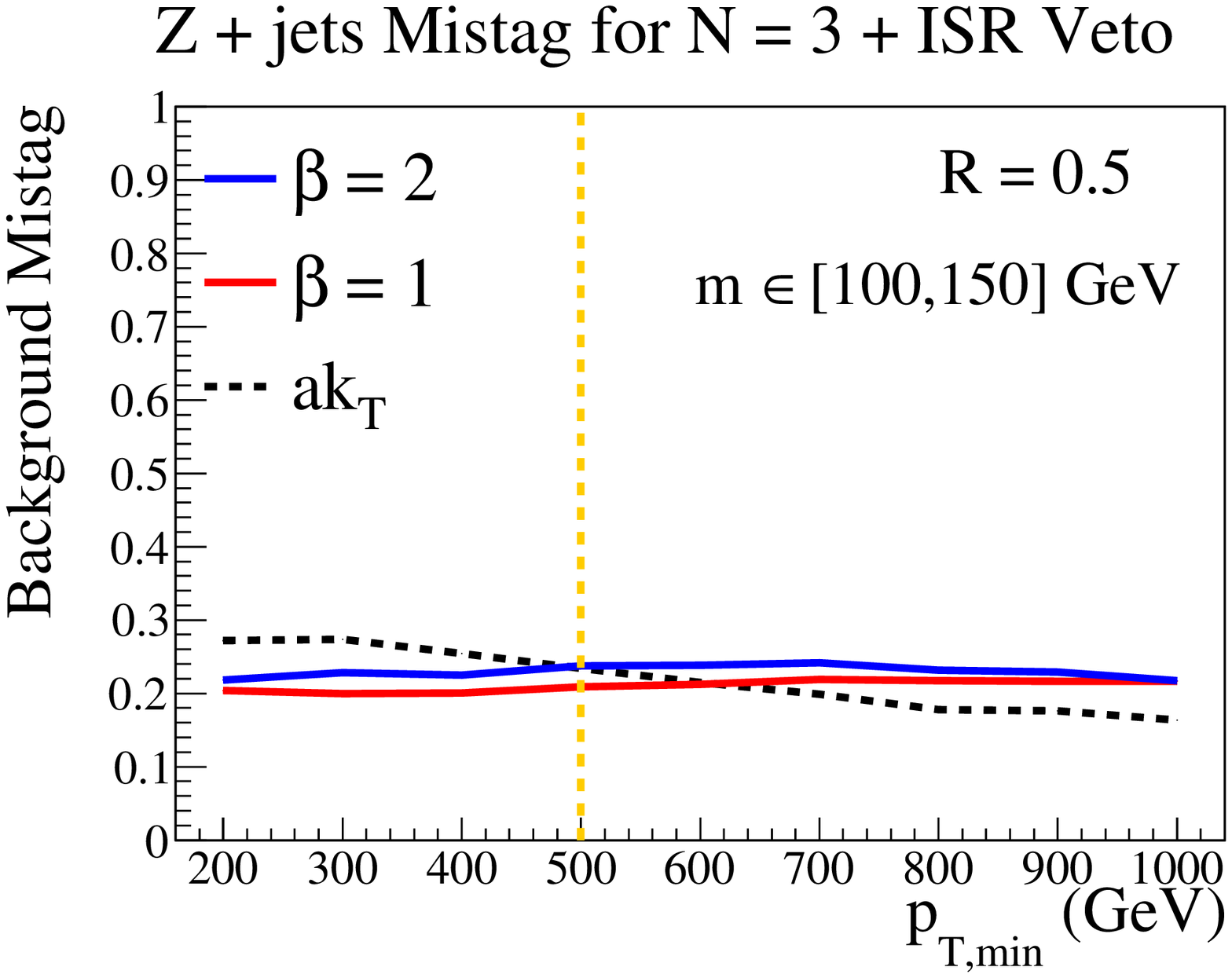}
\label{fig:back_higgs_2jet_eff_improved}
}
\caption{(a) $Z + \text{jets}$ mistag rate for $N = 2$ resolved dijet Higgs reconstruction as a function of $Z$ recoil $p_T$, with the mass window $m_{j} \in [100,150]~\GeV$.  The background rate is nearly constant as a function of $p_T$, and similar between anti-$k_T$ and XCone, showing that XCone does not unnecessarily sculpt background distributions.  (b)  Same as \Fig{fig:higgs_2jet_eff}, but combining $N = 2$ with $N=3$, using the minimum pairwise mass to veto ISR.  Again, we see a roughly flat mistag rate.   See \Fig{fig:higgs_2jet_eff_both} for the corresponding signal study.}
\label{fig:back_higgs_2jet_eff_both}
\end{figure}

Turning to $N=2$ in \Fig{fig:back_higgs_2jets_study}, the dijet invariant mass spectrum increases with recoil $Z$ $p_T$, both for XCone and anti-$k_T$. The reason is that there is no genuine 2-prong structure in the background, so it is far more likely for these algorithms to identify two widely separated jets at large invariant mass rather than two proximate jets at small invariant mass.  In \Fig{fig:back_higgs_2jet_eff}, this leads to an approximately flat mistag rate as function of $p_T$ of around 10-15\%.  As anticipated in the text, the combined $N=2,3$ strategy in \Fig{fig:back_higgs_2jet_eff_improved} has a higher mistag rate, though it still remains roughly constant as a function of $p_T$.   Though anti-$k_T$ has somewhat better mistag rates than XCone at high $p_T$, this is also where anti-$k_T$ has low signal efficiencies (recall \Fig{fig:higgs_2jet_eff_both}).

Note that the above analysis did not include further attempts to mitigate backgrounds through $b$-tagging, $N$-subjettiness, or other substructure discriminant variables.   Since the mistag rates for XCone and anti-$k_T$ are somewhat similar, we suspect that background mitigation techniques used in current Higgs analyses can be adapted to XCone, though we leave a detailed study to future work.

\bibliographystyle{JHEP}
\bibliography{njettiness}

\providecommand{\href}[2]{#2}\begingroup\raggedright\begin{thebibliography}{10}

\bibitem{Stewart:2015waa}
I.~W. Stewart, F.~J. Tackmann, J.~Thaler, C.~K. Vermilion, and T.~F. Wilkason,
  {\it {XCone: N-jettiness as an Exclusive Cone Jet Algorithm}},
  \href{http://arxiv.org/abs/1508.01516}{{\tt arXiv:1508.01516}}.

\bibitem{Abdesselam:2010pt}
A.~Abdesselam, E.~B. Kuutmann, U.~Bitenc, G.~Brooijmans, J.~Butterworth,
  et~al., {\it {Boosted objects: A Probe of beyond the Standard Model
  physics}},  {\em Eur.Phys.J.} {\bf C71} (2011) 1661,
  [\href{http://arxiv.org/abs/1012.5412}{{\tt arXiv:1012.5412}}].

\bibitem{Altheimer:2012mn}
A.~Altheimer, S.~Arora, L.~Asquith, G.~Brooijmans, J.~Butterworth, et~al., {\it
  {Jet Substructure at the Tevatron and LHC: New results, new tools, new
  benchmarks}},  {\em J.Phys.} {\bf G39} (2012) 063001,
  [\href{http://arxiv.org/abs/1201.0008}{{\tt arXiv:1201.0008}}].

\bibitem{Altheimer:2013yza}
A.~Altheimer, A.~Arce, L.~Asquith, J.~Backus~Mayes, E.~Bergeaas~Kuutmann,
  et~al., {\it {Boosted objects and jet substructure at the LHC}},
  \href{http://arxiv.org/abs/1311.2708}{{\tt arXiv:1311.2708}}.

\bibitem{Adams:2015hiv}
D.~Adams, A.~Arce, L.~Asquith, M.~Backovic, T.~Barillari, et~al., {\it {Towards
  an Understanding of the Correlations in Jet Substructure}},
  \href{http://arxiv.org/abs/1504.00679}{{\tt arXiv:1504.00679}}.

\bibitem{Chatrchyan:2013lca}
{\bf CMS} Collaboration, S.~Chatrchyan et~al., {\it {Searches for new physics
  using the $t\bar{t}$ invariant mass distribution in pp collisions at
  $\sqrt{s}$=8  TeV}},  {\em Phys.Rev.Lett.} {\bf 111} (2013), no.~21
  211804, [\href{http://arxiv.org/abs/1309.2030}{{\tt arXiv:1309.2030}}].

\bibitem{Aad:2012raa}
{\bf ATLAS Collaboration} Collaboration, G.~Aad et~al., {\it {Search for
  resonances decaying into top-quark pairs using fully hadronic decays in $pp$
  collisions with ATLAS at $\sqrt{s}=7$ TeV}},  {\em JHEP} {\bf 1301} (2013)
  116, [\href{http://arxiv.org/abs/1211.2202}{{\tt arXiv:1211.2202}}].

\bibitem{Catani:1993hr}
S.~Catani, Y.~L. Dokshitzer, M.~Seymour, and B.~Webber, {\it {Longitudinally
  invariant $K_t$ clustering algorithms for hadron hadron collisions}},  {\em
  Nucl.Phys.} {\bf B406} (1993) 187--224.

\bibitem{Cacciari:2008gp}
M.~Cacciari, G.~P. Salam, and G.~Soyez, {\it {The Anti-k(t) jet clustering
  algorithm}},  {\em JHEP} {\bf 0804} (2008) 063,
  [\href{http://arxiv.org/abs/0802.1189}{{\tt arXiv:0802.1189}}].

\bibitem{Gouzevitch:2013qca}
M.~Gouzevitch, A.~Oliveira, J.~Rojo, R.~Rosenfeld, G.~P. Salam, et~al., {\it
  {Scale-invariant resonance tagging in multijet events and new physics in
  Higgs pair production}},  {\em JHEP} {\bf 1307} (2013) 148,
  [\href{http://arxiv.org/abs/1303.6636}{{\tt arXiv:1303.6636}}].

\bibitem{Salam:2007xv}
G.~P. Salam and G.~Soyez, {\it {A Practical Seedless Infrared-Safe Cone jet
  algorithm}},  {\em JHEP} {\bf 0705} (2007) 086,
  [\href{http://arxiv.org/abs/0704.0292}{{\tt arXiv:0704.0292}}].

\bibitem{Stoll:2014hsa}
M.~Stoll, {\it {Vetoed jet clustering: The mass-jump algorithm}},  {\em JHEP}
  {\bf 1504} (2015) 111, [\href{http://arxiv.org/abs/1410.4637}{{\tt
  arXiv:1410.4637}}].

\bibitem{Hamaguchi:2015uqa}
K.~Hamaguchi, S.~P. Liew, and M.~Stoll, {\it {Jumping into buckets, or How to
  decontaminate overlapping fat jets}},
  \href{http://arxiv.org/abs/1505.02930}{{\tt arXiv:1505.02930}}.

\bibitem{Larkoski:2015yqa}
A.~J. Larkoski, F.~Maltoni, and M.~Selvaggi, {\it {Tracking down hyper-boosted
  top quarks}},  \href{http://arxiv.org/abs/1503.03347}{{\tt
  arXiv:1503.03347}}.

\bibitem{Spannowsky:2015eba}
M.~Spannowsky and M.~Stoll, {\it {Tracking New Physics at the LHC and beyond}},
   \href{http://arxiv.org/abs/1505.01921}{{\tt arXiv:1505.01921}}.

\bibitem{Cacciari:2007fd}
M.~Cacciari and G.~P. Salam, {\it {Pileup subtraction using jet areas}},  {\em
  Phys.Lett.} {\bf B659} (2008) 119--126,
  [\href{http://arxiv.org/abs/0707.1378}{{\tt arXiv:0707.1378}}].

\bibitem{Cacciari:2008gn}
M.~Cacciari, G.~P. Salam, and G.~Soyez, {\it {The Catchment Area of Jets}},
  {\em JHEP} {\bf 0804} (2008) 005, [\href{http://arxiv.org/abs/0802.1188}{{\tt
  arXiv:0802.1188}}].

\bibitem{Stewart:2010tn}
I.~W. Stewart, F.~J. Tackmann, and W.~J. Waalewijn, {\it {N-Jettiness: An
  Inclusive Event Shape to Veto Jets}},  {\em Phys.Rev.Lett.} {\bf 105} (2010)
  092002, [\href{http://arxiv.org/abs/1004.2489}{{\tt arXiv:1004.2489}}].

\bibitem{Thaler:2010tr}
J.~Thaler and K.~Van~Tilburg, {\it {Identifying Boosted Objects with
  N-subjettiness}},  {\em JHEP} {\bf 1103} (2011) 015,
  [\href{http://arxiv.org/abs/1011.2268}{{\tt arXiv:1011.2268}}].

\bibitem{Thaler:2011gf}
J.~Thaler and K.~Van~Tilburg, {\it {Maximizing Boosted Top Identification by
  Minimizing N-subjettiness}},  {\em JHEP} {\bf 1202} (2012) 093,
  [\href{http://arxiv.org/abs/1108.2701}{{\tt arXiv:1108.2701}}].

\bibitem{Jouttenus:2013hs}
T.~T. Jouttenus, I.~W. Stewart, F.~J. Tackmann, and W.~J. Waalewijn, {\it {Jet
  Mass Spectra in Higgs $+$ One Jet at NNLL}},  {\em Phys.Rev.} {\bf D88}
  (2013) 054031, [\href{http://arxiv.org/abs/1302.0846}{{\tt
  arXiv:1302.0846}}].

\bibitem{Kim:2010uj}
J.-H. Kim, {\it {Rest Frame Subjet Algorithm With SISCone Jet For Fully
  Hadronic Decaying Higgs Search}},  {\em Phys.Rev.} {\bf D83} (2011) 011502,
  [\href{http://arxiv.org/abs/1011.1493}{{\tt arXiv:1011.1493}}].

\bibitem{Thaler:2015uja}
J.~Thaler, {\it {Separated at Birth: Jet Maximization, Axis Minimization, and
  Stable Cone Finding}},  \href{http://arxiv.org/abs/1506.07876}{{\tt
  arXiv:1506.07876}}.

\bibitem{Ellis:2001aa}
S.~Ellis, J.~Huston, and M.~Tonnesmann, {\it {On building better cone jet
  algorithms}},  \href{http://arxiv.org/abs/hep-ph/0111434}{{\tt
  hep-ph/0111434}}.

\bibitem{Larkoski:2014uqa}
A.~J. Larkoski, D.~Neill, and J.~Thaler, {\it {Jet Shapes with the Broadening
  Axis}},  {\em JHEP} {\bf 1404} (2014) 017,
  [\href{http://arxiv.org/abs/1401.2158}{{\tt arXiv:1401.2158}}].

\bibitem{Larkoski:2014bia}
A.~J. Larkoski and J.~Thaler, {\it {Aspects of jets at 100 TeV}},  {\em
  Phys.Rev.} {\bf D90} (2014), no.~3 034010,
  [\href{http://arxiv.org/abs/1406.7011}{{\tt arXiv:1406.7011}}].

\bibitem{Sjostrand:2006za}
T.~Sjostrand, S.~Mrenna, and P.~Z. Skands, {\it {PYTHIA 6.4 Physics and
  Manual}},  {\em JHEP} {\bf 0605} (2006) 026,
  [\href{http://arxiv.org/abs/hep-ph/0603175}{{\tt hep-ph/0603175}}].

\bibitem{Sjostrand:2007gs}
T.~Sjostrand, S.~Mrenna, and P.~Z. Skands, {\it {A Brief Introduction to PYTHIA
  8.1}},  {\em Comput.Phys.Commun.} {\bf 178} (2008) 852--867,
  [\href{http://arxiv.org/abs/0710.3820}{{\tt arXiv:0710.3820}}].

\bibitem{Cacciari:2011ma}
M.~Cacciari, G.~P. Salam, and G.~Soyez, {\it {FastJet User Manual}},  {\em
  Eur.Phys.J.} {\bf C72} (2012) 1896,
  [\href{http://arxiv.org/abs/1111.6097}{{\tt arXiv:1111.6097}}].

\bibitem{fjcontrib}
``Fastjet contrib.'' \url{http://fastjet.hepforge.org/contrib/}.

\bibitem{Dasgupta:2007wa}
M.~Dasgupta, L.~Magnea, and G.~P. Salam, {\it {Non-perturbative QCD effects in
  jets at hadron colliders}},  {\em JHEP} {\bf 0802} (2008) 055,
  [\href{http://arxiv.org/abs/0712.3014}{{\tt arXiv:0712.3014}}].

\bibitem{Chatrchyan:2013zna}
{\bf CMS} Collaboration, S.~Chatrchyan et~al., {\it {Search for the standard
  model Higgs boson produced in association with a W or a Z boson and decaying
  to bottom quarks}},  {\em Phys.Rev.} {\bf D89} (2014), no.~1 012003,
  [\href{http://arxiv.org/abs/1310.3687}{{\tt arXiv:1310.3687}}].

\bibitem{Aad:2014xzb}
{\bf ATLAS} Collaboration, G.~Aad et~al., {\it {Search for the $b\bar{b}$ decay
  of the Standard Model Higgs boson in associated $(W/Z)H$ production with the
  ATLAS detector}},  {\em JHEP} {\bf 1501} (2015) 069,
  [\href{http://arxiv.org/abs/1409.6212}{{\tt arXiv:1409.6212}}].

\bibitem{Gallicchio:2010dq}
J.~Gallicchio, J.~Huth, M.~Kagan, M.~D. Schwartz, K.~Black, et~al., {\it
  {Multivariate discrimination and the Higgs + W/Z search}},  {\em JHEP} {\bf
  1104} (2011) 069, [\href{http://arxiv.org/abs/1010.3698}{{\tt
  arXiv:1010.3698}}].

\bibitem{Butterworth:2008iy}
J.~M. Butterworth, A.~R. Davison, M.~Rubin, and G.~P. Salam, {\it {Jet
  substructure as a new Higgs search channel at the LHC}},  {\em
  Phys.Rev.Lett.} {\bf 100} (2008) 242001,
  [\href{http://arxiv.org/abs/0802.2470}{{\tt arXiv:0802.2470}}].

\bibitem{Butterworth:2015bya}
J.~M. Butterworth, I.~Ochoa, and T.~Scanlon, {\it {Boosted Higgs $\rightarrow
  b\bar{b}$ in vector-boson associated production at 14 TeV}},  {\em Eur. Phys.
  J.} {\bf C75} (2015), no.~8 366, [\href{http://arxiv.org/abs/1506.04973}{{\tt
  arXiv:1506.04973}}].

\bibitem{Kribs:2009yh}
G.~D. Kribs, A.~Martin, T.~S. Roy, and M.~Spannowsky, {\it {Discovering the
  Higgs Boson in New Physics Events using Jet Substructure}},  {\em Phys.Rev.}
  {\bf D81} (2010) 111501, [\href{http://arxiv.org/abs/0912.4731}{{\tt
  arXiv:0912.4731}}].

\bibitem{Dokshitzer:1997in}
Y.~L. Dokshitzer, G.~Leder, S.~Moretti, and B.~Webber, {\it {Better jet
  clustering algorithms}},  {\em JHEP} {\bf 9708} (1997) 001,
  [\href{http://arxiv.org/abs/hep-ph/9707323}{{\tt hep-ph/9707323}}].

\bibitem{Wobisch:1998wt}
M.~Wobisch and T.~Wengler, {\it {Hadronization corrections to jet
  cross-sections in deep inelastic scattering}},
  \href{http://arxiv.org/abs/hep-ph/9907280}{{\tt hep-ph/9907280}}.

\bibitem{Wobisch:2000dk}
M.~Wobisch, {\it {Measurement and QCD analysis of jet cross-sections in deep
  inelastic positron proton collisions at $\sqrt{s} = 300$~GeV}},  {\em
  DESY-THESIS-2000-049} (2000).

\bibitem{Blazey:2000qt}
G.~C. Blazey, J.~R. Dittmann, S.~D. Ellis, V.~D. Elvira, K.~Frame, et~al., {\it
  {Run II jet physics}},  \href{http://arxiv.org/abs/hep-ex/0005012}{{\tt
  hep-ex/0005012}}.

\bibitem{Brooijmans:1077731}
G.~Brooijmans, {\it High pt hadronic top quark identification},  Tech. Rep.
  ATL-PHYS-CONF-2008-008. ATL-COM-PHYS-2008-001, CERN, Geneva, Jan, 2008.

\bibitem{Thaler:2008ju}
J.~Thaler and L.-T. Wang, {\it {Strategies to Identify Boosted Tops}},  {\em
  JHEP} {\bf 0807} (2008) 092, [\href{http://arxiv.org/abs/0806.0023}{{\tt
  arXiv:0806.0023}}].

\bibitem{Kaplan:2008ie}
D.~E. Kaplan, K.~Rehermann, M.~D. Schwartz, and B.~Tweedie, {\it {Top Tagging:
  A Method for Identifying Boosted Hadronically Decaying Top Quarks}},  {\em
  Phys.Rev.Lett.} {\bf 101} (2008) 142001,
  [\href{http://arxiv.org/abs/0806.0848}{{\tt arXiv:0806.0848}}].

\bibitem{Almeida:2008yp}
L.~G. Almeida, S.~J. Lee, G.~Perez, G.~F. Sterman, I.~Sung, et~al., {\it
  {Substructure of high-$p_T$ Jets at the LHC}},  {\em Phys.Rev.} {\bf D79}
  (2009) 074017, [\href{http://arxiv.org/abs/0807.0234}{{\tt
  arXiv:0807.0234}}].

\bibitem{CMS:2013vea}
{\bf CMS Collaboration} Collaboration, C.~Collaboration, {\it {Performance of b
  tagging at sqrt(s)=8 TeV in multijet, ttbar and boosted topology events}},
  Tech. Rep. CMS-PAS-BTV-13-001, 2013.

\bibitem{ATL-PHYS-PUB-2014-013}
{\it {Flavor Tagging with Track Jets in Boosted Topologies with the ATLAS
  Detector}},  Tech. Rep. ATL-PHYS-PUB-2014-013, CERN, Geneva, Aug, 2014.

\bibitem{Curtin:2012rm}
D.~Curtin, R.~Essig, and B.~Shuve, {\it {Boosted Multijet Resonances and New
  Color-Flow Variables}},  {\em Phys.Rev.} {\bf D88} (2013) 034019,
  [\href{http://arxiv.org/abs/1210.5523}{{\tt arXiv:1210.5523}}].

\bibitem{CMS:2014fya}
{\bf CMS} Collaboration, C.~Collaboration, {\it {Boosted Top Jet Tagging at
  CMS}}, .

\bibitem{Plehn:2010st}
T.~Plehn, M.~Spannowsky, M.~Takeuchi, and D.~Zerwas, {\it {Stop Reconstruction
  with Tagged Tops}},  {\em JHEP} {\bf 1010} (2010) 078,
  [\href{http://arxiv.org/abs/1006.2833}{{\tt arXiv:1006.2833}}].

\bibitem{Plehn:2009rk}
T.~Plehn, G.~P. Salam, and M.~Spannowsky, {\it {Fat Jets for a Light Higgs}},
  {\em Phys.Rev.Lett.} {\bf 104} (2010) 111801,
  [\href{http://arxiv.org/abs/0910.5472}{{\tt arXiv:0910.5472}}].

\bibitem{Fleming:2007xt}
S.~Fleming, A.~H. Hoang, S.~Mantry, and I.~W. Stewart, {\it {Top Jets in the
  Peak Region: Factorization Analysis with NLL Resummation}},  {\em Phys. Rev.}
  {\bf D77} (2008) 114003, [\href{http://arxiv.org/abs/0711.2079}{{\tt
  arXiv:0711.2079}}].

\bibitem{Fleming:2007qr}
S.~Fleming, A.~H. Hoang, S.~Mantry, and I.~W. Stewart, {\it {Jets from massive
  unstable particles: Top-mass determination}},  {\em Phys. Rev.} {\bf D77}
  (2008) 074010, [\href{http://arxiv.org/abs/hep-ph/0703207}{{\tt
  hep-ph/0703207}}].

\bibitem{Freytsis:2014hpa}
M.~Freytsis, T.~Volansky, and J.~R. Walsh, {\it {Tagging Partially
  Reconstructed Objects with Jet Substructure}},
  \href{http://arxiv.org/abs/1412.7540}{{\tt arXiv:1412.7540}}.

\bibitem{Krohn:2009th}
D.~Krohn, J.~Thaler, and L.-T. Wang, {\it {Jet Trimming}},  {\em JHEP} {\bf
  1002} (2010) 084, [\href{http://arxiv.org/abs/0912.1342}{{\tt
  arXiv:0912.1342}}].

\bibitem{Krohn:2009zg}
D.~Krohn, J.~Thaler, and L.-T. Wang, {\it {Jets with Variable R}},  {\em JHEP}
  {\bf 06} (2009) 059, [\href{http://arxiv.org/abs/0903.0392}{{\tt
  arXiv:0903.0392}}].

\bibitem{Nachman:2014kla}
B.~Nachman, P.~Nef, A.~Schwartzman, M.~Swiatlowski, and C.~Wanotayaroj, {\it
  {Jets from Jets: Re-clustering as a tool for large radius jet reconstruction
  and grooming at the LHC}},  {\em JHEP} {\bf 02} (2015) 075,
  [\href{http://arxiv.org/abs/1407.2922}{{\tt arXiv:1407.2922}}].

\bibitem{Ellis:1993tq}
S.~D. Ellis and D.~E. Soper, {\it {Successive combination jet algorithm for
  hadron collisions}},  {\em Phys.Rev.} {\bf D48} (1993) 3160--3166,
  [\href{http://arxiv.org/abs/hep-ph/9305266}{{\tt hep-ph/9305266}}].

\bibitem{Bertolini:2013iqa}
D.~Bertolini, T.~Chan, and J.~Thaler, {\it {Jet Observables Without Jet
  Algorithms}},  {\em JHEP} {\bf 1404} (2014) 013,
  [\href{http://arxiv.org/abs/1310.7584}{{\tt arXiv:1310.7584}}].

\bibitem{Salambroadening}
G.~Salam, {\it {$E_t^\infty$ Scheme}},  {\em Unpublished}.

\end{thebibliography}\endgroup
\end{document}